\documentclass[onecolumn]{aastex631}
\usepackage{epstopdf}
\usepackage{epstopdf}
\usepackage{hyperref}

\newcommand{\farcsec}{\hbox{$.\!\!^{\prime\prime}$}}

\submitjournal{ApJ}

\shorttitle{B335 outflow}
\shortauthors{Hodapp et al.}

\begin{document}
\turnoffedit
\title{The Outflow of the B335 Protostar II: After the Outburst}

\correspondingauthor{Klaus Hodapp}
\email{hodapp@hawaii.edu}

\author[0000-0003-0786-2140]{Klaus W. Hodapp}
\affil{University of Hawaii, Institute for Astronomy, 640 N. Aohoku Place, Hilo, HI 96720, USA}

\author[0000-0001-9344-0096]{Adwin Boogert}
\affil{University of Hawaii, Institute for Astronomy, 2680 Woodlawn Drive, Honolulu, HI 96822, USA}

\author[0000-0002-6773-459X]{Doug Johnstone}
\affiliation{NRC Herzberg Astronomy and Astrophysics, 5071 West Saanich Road, Victoria, BC, V9E 2E7, Canada}
\affiliation{Department of Physics and Astronomy, University of Victoria, Victoria, BC, V8P 5C2, Canada}

\author[0000-0002-5714-799X]{Valentin J. M. Le Gouellec}
\affil{Institut de Cienci\`es de l’Espai (ICE-CSIC), Campus UAB, Carrer de Can Magrans S/N, E-08193 Cerdanyola del Vall\`es, Spain}
\affil{Institut d’Estudis Espacials de Catalunya (IEEC), c/ Gran Capitá, 2-4, 08034 Barcelona, Spain}

\author[0009-0001-2707-8903]{Eleni Tsiakaliari}
\affil{School of Physical Sciences, Walton hall, The Open University, Milton Keynes, MK7 6AA, UK}

\author[0000-0003-0972-1595]{Helen J. Fraser}
\affil{School of Physical Sciences, Walton hall, The Open University, Milton Keynes, MK7 6AA, UK}

\author[0000-0002-1437-4463]{Laurie L. Chu}
\affil{National Optical-Infrared Astronomy Research Laboratory, 950 N Cherry Ave, Tucson, AZ 85719}

\author[0000-0002-8963-8056]{Thomas Greene}
\affil{Caltech/IPAC, 1200 East California Boulevard, Pasadena, CA 91125, USA}

\author[0000-0002-7893-6170]{Marcia J. Rieke}
\affil{Steward Observatory, University of Arizona, Tucson, AZ 85721, USA}




\begin{abstract}

The B335 protostar has recently undergone a major, still ongoing, outburst detected in scattered light from its outflow cavity, offering a rare opportunity to study its impact on a protostellar jet. We use JWST/NIRCam photometry of background stars behind B335 from 2.7 to 4.4 $\mu$m to map extinction and H$_2$O ice absorption, showing that the outflow has carved a cavity in the molecular core. We measure proper motions of shock fronts emerging from the protostar of 131--227 km s$^{-1}$. The kinematic age of the most prominent shock front, 3E, corresponds to the early phase of the current outburst. JWST/NIRSpec IFU data show that the youngest shock, 2E, exhibits ionic lines but no molecular emission. Shock 3E shows strong CO emission together with H$_2$ and [\ion{Fe}{2}], whereas older shocks show weaker CO and are dominated by H$_2$ and [\ion{Fe}{2}]. The feature 0E, closest to the protostar, appears to be a stationary shock. CO-line-removed spectra near the protostar show that the unsaturated absorption features of $^{13}$CO$_2$, OCN$^-$, and OCS increase strongly toward the source. The ice properties are otherwise similar to those along lower-extinction sight lines. In the central bipolar reflection nebula, CO gas is seen in scattered emission from the immediate protostellar surroundings, but a few arcsec farther out, absorption by cooler CO gas in the outflow cavity is detected.

\end{abstract}







\section{Introduction} \label{sec:intro}

The accumulation of mass from a molecular core onto a protostar
proceeds initially by infall of mass from the molecular core onto
a circumstellar disk, from which mass is then finally accreted onto
the protostar. The accretion process requires a shedding of excess angular
momentum via a disk wind and a faster jet, which carve out an outflow cavity
into the surrounding cloud and form a bipolar outflow. This outflow is generally detected
both at radio wavelengths, most readily in the CO rotation lines, and at 
near-infrared wavelengths through shock-excited H$_2$ emission. 
The mass accretion onto the protostar is, in general, not a steady process,
but is characterized by fluctuations in the accretion rate, and consequently
the accretion luminosity, observable as photometric variability of the
FUor, EXor or similar types. A list of all known such outbursts has recently been published by
\citet{ContrerasPena.2025.JKAS.58.209.OYCAT} and the impact of these outbursts on the formation
of the star was discussed. As a byproduct of the variations in accretion,
the jet velocity is modulated, so that the jets emerging from protostars usually
appear as a string of individual internal shock fronts. 

The globule \object{Barnard 335} (B335 in short), discovered by  
\citet{Barnard.1927.pasr.book.B} on wide-field photographs, is part
of a larger system of similar globules, but is itself isolated from
those other clouds. The protostar in B335 was discovered by 
\citet{Keene.1983.ApJ.274L.43.B335.protostar} as a far-infrared source.
Its isolation and the presence of only one protostar in the globule
has made B335 a frequently studied example of isolated star formation.
A summary of ``Star Formation in Bok Globules and Small Clouds''
and B335 specifically
was published by \citet{Reipurth.2008.Handbook.B335}.
Using purely photometric methods, 
\citet{Olofsson.2009.AA.498.455O.B335.distance}
measured a distance to B335 in the range of 90 - 120 pc. 
They had already mentioned a possible association of the B335 cloud 
with the star HD 184982 on the basis that this star appears to be illuminating 
the cloud in addition to the general interstellar radiation field. 
\citet{Watson.2020.RNAAS.4.88.B335.distance}
obtained new images of this reflection nebulosity 
and used the Gaia DR2 distance to HD 184982 of 164.5 pc as a proxy 
for the distance of the B335 molecular core.
Recently, however, Neal Evans and Steve Federman have communicated
(personal communication, 2026)
yet unpublished results pointing out that HD 184982 does not show
interstellar absorption lines at the velocity of the B335 cloud.
Rather, interstellar absorption lines at the velocity matching B335
were observed toward HD 185176 at a distance of 253 pc. 
Moreover, extinction plots for the lines of sight towards these
two stars show a prominent peak at $\approx$ 210 pc. 
Evans and Federman conclude
that the distance to B335 is well constrained between 163 pc, the
newer Gaia DR3 distance to HD 184982, and 253 pc.
Since these latest results are not yet published, we will use the rounded distance value of 165 pc from \citet{Watson.2020.RNAAS.4.88.B335.distance} in this paper.

The molecular cloud surrounding the Bok globule
B335 was discovered in CO emission by 
\citep{Frerking.1982.ApJ.256.523.B335.protostar}
and mapped in detail by
\citet{Frerking.1987.ApJ.313.320.B335_CO_map}.
\citet{Goldsmith.1984.ApJ.286.599.B335.outflow}
discovered the molecular outflow associated with the embedded
protostar.
\citet{Hirano.1988.ApJ.327L.69.B335.inclination}
mapped it in more detail and
determined that the axis of the large-scale outflow on
scales of arcminutes is inclined only by
10$\arcdeg$ from the plane of the sky, i.e., the bipolar
outflow is seen nearly, but not precisely, edge-on.
The bipolar outflow emerging from the single protostar
in B335 was further studied by
\citet{Stutz.2008.ApJ.687.389.B335.Spitzer}
based on Spitzer Space Telescope imaging data.
Out to a wavelength of 8 $\mu$m  B335 appears as a 
bipolar nebula. Only at wavelengths of 24 $\mu$m and longer
does it appear as an unresolved point source,
probably direct light from 
the central embedded protostar. 

The protostar in B335 is a low luminosity object, with
\citet{Green.2013.ApJ.770.123.Lbol} deriving a luminosity,
based on Herschel data, of only 0.68 L$_\odot$, with an effective
T$_{bol}$ of 39K, overall being classified as SED Class 0.
Based on more elaborate models, \citet{Evans.2023.ApJ.943.90.B335.variability} 
derives a quiescent luminosity of 3 L$_\odot$ with a peak outburst
luminosity of 23 L$_\odot$.
The B335 protostar is a low mass object, with \citet{Federman.2026.ApJ...998..282.MIRIjet}
listing it as 0.25 M$_\odot$ as an approximate consensus value
from a large number of individual measurements.
Age estimates, discussed in \citet{Evans.2023.ApJ.943.90.B335.variability},  
are in the range of 3 to 4 $\times10^4$ yrs.
\citet{Cabedo.2021.AA...653.166.infall} and
\citet{Bjerkeli.2023.AA.677.62.B335.episotic.accretion}
observed infalling gas towards B335 with evidence for episodic accretion.
In ALMA observations early in the present outburst,
\citet{Cabedo.2023.AA...669A..90.ionization} showed an unusually high ionization fraction near the protostar,
indicative of on-going strong accretion.

The outflow cavity contains numerous outflow shock fronts. 
High proper motion shock fronts are found mostly along
a fairly narrow jet near the center of the outflow cavity,
while low proper motion shock fronts are less concentrated
and may have formed by interaction of the outflow with the
cavity walls.
Optical Herbig-Haro shock fronts emerging from the B335 globule 
were found first by 
\citet{Vrba.1986.AJ.92.633.HH119}
and then confirmed and added to by 
\citet{Reipurth.1992.AA.256.225.B335.HH}
who labeled them as HH 119 A-C. 
In the infrared,  
\citet{Hodapp.1998.ApJ.500L.183.B335} did the first major imaging study and
labeled additional shock fronts found in the H$_2~1{-}0~ S(1)$ emission line, 
but not confirmed by optical spectroscopy as HH 119 IR 1-5.
\citet{Galfalk.2007.A&A.475.281.B335.outflows} expanded this naming scheme
to include newly identified shocks
and measured the proper motion of the
distant shocks detectable at near-infrared wavelengths.

B335 has, in the past decade, undergone a major accretion outburst that has not yet completely ended
at the time of writing this paper. It offers the
rare opportunity to study the effects of an accretion outburst on the 
outflow and jet emerging from a protostar. Fortuitously, B335 was also
included in several JWST observing programs of small samples of star-forming
regions, so that a uniquely comprehensive data set is available to study the
effects of the accretion outburst on the outflow, down to timescales of a
few years.
\citet{Hodapp.2024.AJ.167.102.B335} have published a first paper, hereafter referred to as Paper I, 
on the JWST observations of the
B335 outflow based on one epoch of observations obtained in 2023. They identified 
additional highly obscured shock fronts and named them HH119 JWST 0E through 8E for the eastern
outflow lobe, and correspondingly for the western lobe. These shock fronts are usually abbreviated
to just, for example, 3E,  when the context is clear. The naming scheme used in Paper I implies that
the shocks appear to occur in east-west pairs, suggesting that they were ejected at the same time.
This second paper is based on two epochs, in 2023 and 2024, of JWST/NIRCam imaging in the F444W filter,
and one epoch of JWST NIRSpec IFU observations of the central region of this outflow, done in
2022 (program 1802, P.I.: T. Megeath ).
Initial results from that NIRSpec IFU program were published by
\citet{Rubinstein.2024.ApJ.974.112.B335.IFU} and
\citet{Federman.2024.ApJ.966.41}.
A detailed analysis of the excitation conditions in the shock fronts, in particular shock 3E,
is beyond the scope of this paper and will be published in a separate paper in preparation
by \citet{Kim.2026.inpreparation.B335CO}. A future paper by \citet{LeGouellec.2027.in.prep.COexcitation}
will further study the excitation conditions based on additional NIRSpec IFU data.

This paper is structured as follows:
The observing and analysis methods are outlined in Section 2. The NIRCam imaging observations
are described in subsection 2.1, with sub-subsections on the Astrometry in general (2.1.1), the shock proper motion measurements specifically (2.1.2), and the processing of the images to reveal faint extended emission (2.1.3). Subsection 2.2 explains
the photometry of background stars, with subsections on the Continuum Extinction (2.2.1)
and the H$_2$O ice column density (2.2.2). Subsection 2.3 briefly discusses the
public NIRSpec data that we are using.

The Discussion in Section 3 starts with the overall structure of the B335 cloud core
from scattered light (coreshine) in 3.1, and extinction and photometric H$_2$O ice column density in 3.2.
Section 3.3 discusses the photometric variability and local shadow effects in the B335 reflection nebula.
Section 3.4 discusses the likely repetitive nature of the accretion outbursts.
Section 3.5 presents the results on the B335 outflow and section 3.6 and its sub-sections discuss the excitation
conditions in the individual shock fronts of the jet, summarized in a synoptic view in
sub-section 3.6.7.
The outflow cavity, outside of the jet, is discussed in section 3.7 with sub-sections
on extinction in 3.7.1, ice absorption in 3.7.2, and CO gas emission and absorption
in 3.7.3. Section 4 are the summary and conclusions.
The Appendix A presents results on the proper motion of the B335 protostar from public
ALMA observations that are used to identify the position of the protostar at the epochs
of the JWST observations presented here.
Appendix B outlines the photometric measurement method used to produce the H$_2$O ice column density map.
Appendix C presents results of a component analysis of the ice features in the B335 reflection nebula.

\section{Observations and Analysis Methods}

\subsection{JWST NIRCam Observations} 
The results reported here are based on guaranteed time (GTO) observations 
with the James Webb Space Telescope (JWST) 
\citep{Gardner.2023.PASP.135f8001.JWST.mission}.
The main goal of our 
JWST/NIRCam 
\citep{Rieke.2023.PASP.135b8001.NIRCam} program 1187 was
Wide Field Slitless Spectroscopy (WFSS) 
of background stars behind B335 aimed at mapping the 
column density of H$_2$O, CO$_2$, and CO ice. These WFSS results will be published
in a separate series of papers,
the first paper being \citet{Tsiakaliari.2026.inpreparation.B335COmap}.
At each position where WFSS data were taken, we also obtained a small set
of direct images with the long module of NIRCam without the grism,
originally to have a reference for
WFSS wavelength calibration. These direct images were taken in the F277W, F300M, F356W, and F444W filters. In addition, we took parallel direct imaging data
with the short module of NIRCam in the F070W, F090W, F115W, F150W, and F200W filter.
We will only use and discuss these direct images in this paper.

While this program on B335 was originally planned to be done in a single
epoch, a guide star acquisition error during visit 15:1 in 2023
necessitated, or better, allowed, a repeat observation of the failed visit one year later.
Fortuitously, both the 2023 and 2024 visits covered the region around the B335 protostar in
the F444W filter,
the only filter to show the heavily 
obscured inner, and youngest, shock fronts of the B335 outflow.
The resulting two-epoch F444W dataset is therefore ideally suited
for proper motion measurements of the inner, youngest shock fronts, whose
proper motion have not been measured previously.
However, the more distant shock fronts to the east of B335, HH~119 IR 3, 4, and 5
in the terminology of 
\citet{Hodapp.1998.ApJ.500L.183.B335} were not covered in the second epoch and no
new proper motion data were obtained.
The imaging observations presented here were carried out on 2023 April 25 UTC (MJD 60059.080)
and on 2024 April 24 UTC (MJD 60424.258), quite precisely one year apart. 
The data are available at MAST:
\dataset[doi:10.17909/etar-9w74]{10.17909/etar-9w74}.
Results from the 2023 imaging WFSS observations with some proper motion information based on
older ground-based observations
were  presented in the first paper
in this series on B335 by
\citet{Hodapp.2024.AJ.167.102.B335}, hereafter referred to as Paper~I. 

\subsubsection{Astrometry}
The reduced images from MAST are nominally astrometrically calibrated,
but that calibration was found to be of insufficient quality for a 
comparison of shock proper motions, probably because of the scarcity
of Gaia catalog stars in and near the dark cloud.
The NIRCam F444W images at the two epochs were not taken at exactly the same position 
on the sky. As a first step in aligning these two images, approximately matching cutouts containing
the central region of the outflow were produced. 
We used two methods for fine-aligning these two cutout images. A set of reference stars
visible on both epoch F444W images was selected, excluding saturated,
very faint, and overlapping stars. 
There are no foreground
stars toward the B335 globule, so this sample consists of stars at a distance
greater than that of B335. Due to the extinction in the B335 cloud, none of these stars are optically visible and therefore, none of
them have Gaia positions or proper motions. 
It is reasonable to assume that such a
large ensemble of stars 
at various, but unknown, distances 
is a close approximation of the celestial reference
system, but systematic effects of streaming motions in the background star field
can not be excluded. 
For the second method, we used a sample of small extended sources not looking
like shock-excited outflow features. The expectation is that this is a sample
of somewhat compact extragalactic sources that have negligible proper motions.

For the purpose of fine-aligning the two images, we magnified them by a zoom factor of 30.
We measured the correlation of both images on both sets of objects, varying rotation and
spatial offsets in integer steps in the zoomed image.
The background star sample contains more objects, and for each, the position can be correlated  with better precision.
The correlation of the extragalactic sources is less precise
than for stars, but the sample is expected to be a more accurate representation of
the reference frame.
Both background object samples gave closely similar results and in the end, we used the
mean of the two measurements of the
spatial offsets to match the two images. The correction in row and column direction
26 $\pm$ 2 and 22 $\pm$ 2 mas, respectively, i.e., less than half of one pixel in the original data.
The field rotation was found to be zero.
This procedure results in the two images having matching sky coordinates, even
though the absolute accuracy of these coordinates has not improved by matching
the background objects in the two epoch images.
The problem remains that the whole B335 cloud, and, in particular, the protostar
embedded in it, cannot be expected to be stationary in front of the background stars.

We have measured the proper motion of the protostar itself, which is not detected
at near-infrared NIRCam wavelengths, on public ALMA data. The procedure
is detailed out in Appendix A.\\

\begin{flushleft}
$
\mathrm{RA}(\mathrm{MJD})\mathrm{[deg]}  =
\left( 294.2537586 \pm 1.1\times10^{-6} \right)\,\mathrm{[deg]}
+
\left( 7.1 \pm 0.5 \right)\times10^{-9}\,\mathrm{[deg\,day^{-1}]}\,
(\mathrm{MJD}-60000)\mathrm{[day]}
$
\end{flushleft}

\begin{flushleft}
$
\mathrm{DEC}(\mathrm{MJD})\mathrm{[deg]} =
\left( 7.5692829 \pm 1.7\times10^{-6} \right)\,\mathrm{[deg]}
-
\left( 15.0 \pm 0.7 \right)\times10^{-9}\,\mathrm{[deg\,day^{-1}]}\,
(\mathrm{MJD}-60000)\mathrm{[day]}
$
\end{flushleft}

\begin{flushleft}
RA PM on sky [mas/yr] =  9.3 $\pm$ 0.6, 
DEC PM in [mas/yr] =  -19.7 $\pm$ 0.9
\end{flushleft}

This measured proper motion of the B335 protostar in the one-year epoch difference
of 9 mas in RA and -20 mas in DEC was applied
as a final step in the alignment process to the
matched F444W images, so that, instead of being
referenced to the mean background star positions and extragalactic reference
positions, the final two images are instead registered with respect to the position
of the protostar.

As explained in detail in the Appendix A,
the measured projected velocity of the B335 protostar is 16.7 kms$^{-1}$ at a 
position angle of 154\arcdeg. It is an order of magnitude smaller than the
proper motions of shock fronts in the jet emanating from the protostar.
We only use these relatively aligned images for the purpose of shock front
proper motion measurements.
For the purpose of displaying the absolute coordinate system in the various
figures in this paper, we rely on the world coordinate system (WCS) produced by the
STScI pipeline. The comparison of the two epochs of imaging data indicates that this
WCS is accurate to better than one half of a NIRCam pixel.
The WCS of the NIRSpec IFU data cube is less precise, as was noted, for example, in
\citet{Wang.2025.arXiv.NIRSpecAlign}.
Fortunately, there is one background star detected in the NIRSpec IFU data that is also
detected in the NIRCam F444W image. Assuming that the NIRCam data are more accurate, and because the
star has negligible proper motion in the $\approx$ 9 months between the NIRSpec and the 2023 NIRCam
images, we have used this star to measure the NIRSpec IFU offset relative to NIRCam of 0\farcsec24, and applied
this correction. The resulting corrected protostar position in the NIRSpec data cube, calculated for the
epoch of the NIRSpec IFU observations, is astrophysically plausible and is shown in all figures based on the NIRSpec data.

\subsubsection{Shock Proper Motion Measurements}
The individual shock fronts were labeled in Paper 1 (their Figure~1), and are again
identified here in Section 3.4.
The emission knot closest to the protostar, 0E, does show some motion between
the 2023 and 2024 images. It shows CO, but little shock-excited H$_2$ and [\ion{Fe}{2}] emission.
On the western side of the outflow, knot 0Wn did not move measurably between
2023 and 2024, but it does show CO, H$_2$ and [\ion{Fe}{2}] line emission.
The shock front identified as 1E in the 2023 data (Paper 1) changed its appearance
so much in 2024 that we could not measure a proper motion with confidence.
The proper motions of most individual knots of shock-exited emission were
measured
by computing the correlation of the 2023 and 2024 images in selected sections
containing only the knot in question, while varying the spatial offset
between the two image sections. 
The maximum of the correlation matrix gives the shift of the knot between
the two images, i.e., the proper motion. 
For the complex bow shocks at larger distance from the protostar, we have
tried to divide the shock front up into smaller sub-regions.
For the faint shock fronts 2E, and the knots immediately next to the
protostar, 0E, 0Wn, and 0Ws, we had to resort to measuring the proper
motion by tracking the flux maximum of the knots on contour maps.
The knots in this region close to the protostar were not sufficiently
separated spatially for the cross-correlation technique to work reliably.
The errors of the proper motion measurements are dominated by the
complexity of the flux distribution and the changes in the shape of
the shock-excited knots over the course of the year between the observations.
We estimate that in the best case the isolated, small and bright shock front
3E can be measured with one pixel error in the zoomed image, i.e., approximately
2 mas. On the other hand, shock 2E, appearing faint in the F444W image and being situated in complex extended reflection
nebulosity, can probably only be measured to a precision of 20 mas, and for this reason,
we do not include its proper motion in Table I in Section 3.5.
The variability of the reflection nebula and the measured proper motion vectors will be
discussed in sections 3.4 and 3.5.

\subsubsection{Faint Extended Emission: Coreshine}
The coreshine effect in B335 was studied on the basis of a set images with uniform pixel scale of 0.030 arcsec pixel$^{-1}$
in all the filters used by us for B335 imaging: F070W, F090W, F115W, F150W, F200W,
F277W, F300M, F356W, and F444W. The shortest wavelengths show the B335 cloud as a featureless
area of absorption against the background, while at the longest wavelengths, scattered 
radiation from the protostar at the outflow cavity dominates.
Figure\ \ref{cloudshine-image} shows a false-color representation of three of the shorter
wavelength (F090W, F150W, and F277W) images where scattering reveals useful information about the cloud structure.
The cross sections through our images shown in Figure~\ref{cloudshine-crosscuts} were extracted in broad stripes of 400 pixel width in RA,
and 4200 pixels length in DEC. All pixels above a certain small threshold were
excluded  from the computation of the sky level to eliminate images
of stars and the diffraction spikes of bright stars located outside of the stripes. Fainter extended
features from the reflection nebula and shock-excited regions are retained in the stripes so processed. 
In each row (400 pixel extent), we sorted all remaining pixels, and computed the median of
the 5th to 16th faintest pixel in this sorted set. Essentially, this is close to a noise-clipped minimum signal.
This process further suppresses the signal from stars and star diffraction
spikes, but also eliminates low signal noise spikes.

\clearpage
\subsection{Photometry of Background Stars}

Generating the maps of extinction and H$_2$O ice column density starts
with a flux-limited list of stars in the field observed with NIRCam in B335. This star list
was carefully visually inspected and detector defects, artifacts from diffraction
spikes, extragalactic extended objects, closely spaced stars and stars behind shock features were excluded.
For all stars in this curated list, aperture photometry
was performed in the F277W, F300M, F356W, and F444W filters. Since the signal
in the F300M filter, designed to match the H$_2$O ice absorption band, was the
faintest of these filters, all stars with F300M photometry fainter than 1 $\mu$Jy
had their photometry marked as invalid.

Each star with valid photometry in all those four filters represents a column
density (or extinction) measurement in a line of sight through the molecular
cloud. This set of randomly distributed line-of-sight measurements can only be
processed into a contiguous map with some assumptions.
We have chosen a technique minimizing assumptions about the structure of the
molecular cloud. For each position on the sky within the images, the assigned
water ice column density is the median column density in the closest seven lines
of sight (stars) with valid photometry. 
We found that the median combining of several lines of sight is essential
for eliminating noise spikes produced by low S/N individual stars.
The resulting column density map is
therefore a
Voronoi-like irregular tessellation
of small areas of constant column density.
It is an unfortunate side effect that those tessellation elements of constant column density
are largest in areas of high column density, i.e., close to the protostar, where
detectable background stars are scarce. 
The extinction near the protostar position is therefore
systematically underestimated.

The color of each star is a combination of its intrinsic color, the interstellar
extinction in the background and unrelated to B335, and the extinction in B335 
itself, the latter being the quantity of interest. While the intrinsic color
of each background star could, in principle, be determined from spectroscopy,
this is not a realistic option for most of these faint background stars.
We rely instead on the fact that the intrinsic colors of stars in the 
infrared vary much less than the expected effects of many magnitudes of
A$_V$. This same assumption was originally made by 
\citet{Lada.1994.ApJ.429.694.NICE} in their seminal paper
on infrared extinction measurements.
The small contribution of the average intrinsic color of the background stars
contributes a nominal extinction value as a baseline in the map.
We have computed the mean of the lowest 10\% of extinction values as
an estimate for this combined effect of intrinsic star color and 
average background interstellar extinction, 
and subtracted this
average from the B335 field.

\subsubsection{Continuum Extinction}

For extinction mapping, the color based on the synthetic ``F277W-F300M'' (Appendix B)
and the F444W AB magnitudes has the advantage that it
is not affected by H$_2$O ice absorption. However, its severe disadvantage is
that by using the F277W and F300M data, the shortest wavelength filters of the four
that we discuss here, the high extinction in the immediate vicinity of the protostar leads to a 
scarcity of stars, i.e., lines of sight. The map value including the position of
the protostar is then the median of 7 nearest detected stars, all of which lie in
the relatively low extinction of the outflow cavities. 
This map, shown in Section 3.2,
thus shows relatively
low extinction at the position of the protostar. This is not consistent with the
extremely high extinction and ice column densities that we actually measure at
this position with spectroscopy of the scattered light, as we will discuss in
Sections 3.7.1 and 3.7.2. 

To avoid this problem to the maximum degree possible with our imaging data,
we produced a higher spatial resolution extinction map on the basis
of the F356W and F444W filters. 
This color does not separate continuum extinction
and H$_2$O ice absorption, but the F356W image goes substantially deeper.
We used the near-infrared extinction law by 
\citet{Wang.2019.ApJ.877.116.Extinction}, their equation 10, with
a power-law index of -2.07
to convert the measured color excess to the usual unit of A$_V$.

Our continuum extinction maps have better spatial resolution, but a much smaller
area coverage than the extinction map by 
\citet{Chu.2021.ApJ.918.2.extinction.maps}, which was based on ground-based
J, H, and K photometry with UKIRT and Spitzer channel 1 and 2 data.
That larger area map also shows the reduced extinction in the outflow cavities
at larger distances from the protostar and in regions of overall much lower
extinction values. 
The resulting tessellated  extinction map is certainly not the true structure of the molecular cloud, but it is
the closest approximation that our data directly support.

\subsubsection{H$_2$O Ice Column Density from Photometry}

The optical depth of the broad absorption feature of H$_2$O ice can be approximately
measured using filter photometry of background stars behind the B335 molecular core. 
We used photometry in the F277W, F300M, F356W, and F444W filters.
Our method of calculating an approximate H$_2$O ice optical depth and column density
from this filter set is explained in the Appendix B.
In this paper, we will use this photometric
H$_2$O ice column density map only for qualitative and morphological arguments. A more detailed
study of the ice features in the B335 cloud on the basis of more NIRCam WFSS spectra will be
published in a separate paper in the future \citep{Tsiakaliari.2026.inpreparation.B335COmap}. 
We used the band strength of A = $2.0 \times 10^{-16}$ cm molecule$^{-1}$  for amorphous water ice at T = 25K
recommended by \citet{Bouilloud.2015.MNRAS.451.2145.waterice.A} as the presently best consensus value.
The resulting map of the H$_2$O ice distribution is thus calibrated in units of molecules cm$^{-2}$.
We display this photometric H$_2$O ice column density
map in greyscale in the middle panels of Figure~\ref{H2O-AV-map}. 
For the contours overlayed in the lower left panel of Figure \ref{H2O-AV-map} , we have smoothed this
map with a Gaussian of sigma = 15 pixels.
As will be discussed in Section 3.2 the ice column density map shows
reduced column density near the protostar. This is an artifact caused by the lack of stars detected in the F277W,
and most problematically, in the F300M filter.

\subsection{NIRSPEC IFU }
We use the publicly available NIRSpec IFU spectroscopy obtained under GO program 1802 (P.I. T. Megeath)
to extract spectra of selected knots, and to try to determine proper motions.
Program 1802 observed B335 with the G395M medium resolution granting and F290LP order sorting filter,
covering the wavelength range 2.87–5.27 $\mu$m with a spectral resolution of $\approx$ 1000. Each IFU pointing
covers a 3\arcsec$\times$3\arcsec field with 100 mas spaxels, and the total dataset was a mosaic of several such exposures. The
data acquisition process was described in full detail in \citet{Federman.2024.ApJ.966.41}.
The data are available at MAST:
\dataset[doi:10.17909/z839-ka94]{\doi{10.17909/z839-ka94}}.
The data on B335 were obtained on MJD 59838.0 (2022 Sept. 16).
Spectral images based on this data set have already been published by
\citet{Federman.2024.ApJ.966.41} and extracted spectra in the east and west outflow
lobe have been presented and discussed by
\citet{Rubinstein.2024.ApJ.974.112.B335.IFU}.
For our analysis, we used the data cube 
produced by the STScI pipeline.
From these, we extract the flux spectra in 
specifically selected smaller spatial regions of the outflow lobes so that individual shock fronts
and changes in the scattered continuum light can be studied and we extract images of line
flux in emission lines of H$_2$, [\ion{Fe}{2}], and CO, as will be presented in
the Discussion Section.

\section {Discussion}

\subsection{Overall Structure of the B335 Core from Scattered Light, Coreshine}
At optical and near-infrared wavelengths, B335 appears as a small dark cloud slightly 
elongated in the north-south direction, or, as we will discuss in the following, better
described as flattened in the east-west direction.
A ``shadow'' minimum in the extended flux seen by
\citet{Stutz.2008.ApJ.687.389.B335.Spitzer}
at 3.6$\mu$m and 8.0$\mu$m to the south of the protostar, and also pointed out in our Paper I,
is clearly associated with the nearly complete absence
of background stars in this area of the cloud.
This finding at longer infrared wavelengths confirms the 
flattened shape of the B335 core at higher column densities.

Infrared polarimetry of background stars was used by 
\citet{Hodapp.1987.ApJ.319.842.B335.mag}
to map the magnetic field structure  in B335. 
A similar study over a larger field by \citet{Kandori.2020.ApJ.891.55.magmap}
obtained very similar results in the core, but also measured the field star polarization
in the off-core, low-extinction field, which showed similar degrees of polarization
than the on-core field. Subtracting the off-core polarization, caused by the anisotropic general interstellar
extinction, revealed a cleaner
``hourglass'' pattern of the in-core polarization vectors. 
From these two studies, and the even wider
field result by
\citet{Bertrang.2014.AA.565.94.B335.magmap}, it is clear that
the magnetic field traced by near-infrared polarization is parallel to the axis
of the outflow, and that the core of the molecular cloud is flattened along that same
direction, nearly east-west.
Much closer to the protostar, ALMA dust emission polarization measurements by 
\citet{Maury.2018.MNRAS.477.2760.B335.ALMA} and
\citet{LeGouellec.2023.AA...671.167.magfield} confirm that the outflow axis is aligned with
the magnetic field and find that the magnetic field in the disk around the protostar is
in north-south direction, i.e., perpendicular to the outflow axis. 

The B335 globule is larger than the field of view of a single module of NIRCam, in particular
at short wavelengths where the lower extinction regions show cloudshine. We therefore used a
reference position farther away from B335 in NIRCam Module B as the blank sky reference region. This
region was carefully selected to be free of bright stars and diffraction spikes from bright
stars. Using effectively the minimum of the signal measured in this region as the sky value,
we subtract this from the crosscuts shown in Figure~ \ref{cloudshine-crosscuts}.
The crosscuts in the F150W (blue) and F200W (green) filters begin and end
with signal above zero, indicating the field of the NIRCam Module A was not sufficient
to cover the full spatial extent of the cloudshine interstellar scattered light. 
At the longer wavelengths covered by the NIRCam long-wave channel,
i.e., in filters F277W (yellow) and  F356W (red), the crosscut
signal starts and ends near zero. At the short wavelengths, shown blue and green, the
crosscuts show a minimum of cloudshine flux near the position of the protostar (left panel),
in other words, the scattered light is concentrated in a rim around the cloud center. This 
minimum in the scattered light gets narrower with increasing wavelength.
In the F277W filter (yellow), light scattered off the walls of the outflow cavity, i.e., the reflection nebula, produces
a maximum of flux in the crosscuts.

\begin{figure*}[h]
\begin{center}
	\includegraphics[angle=0.,scale=0.50]{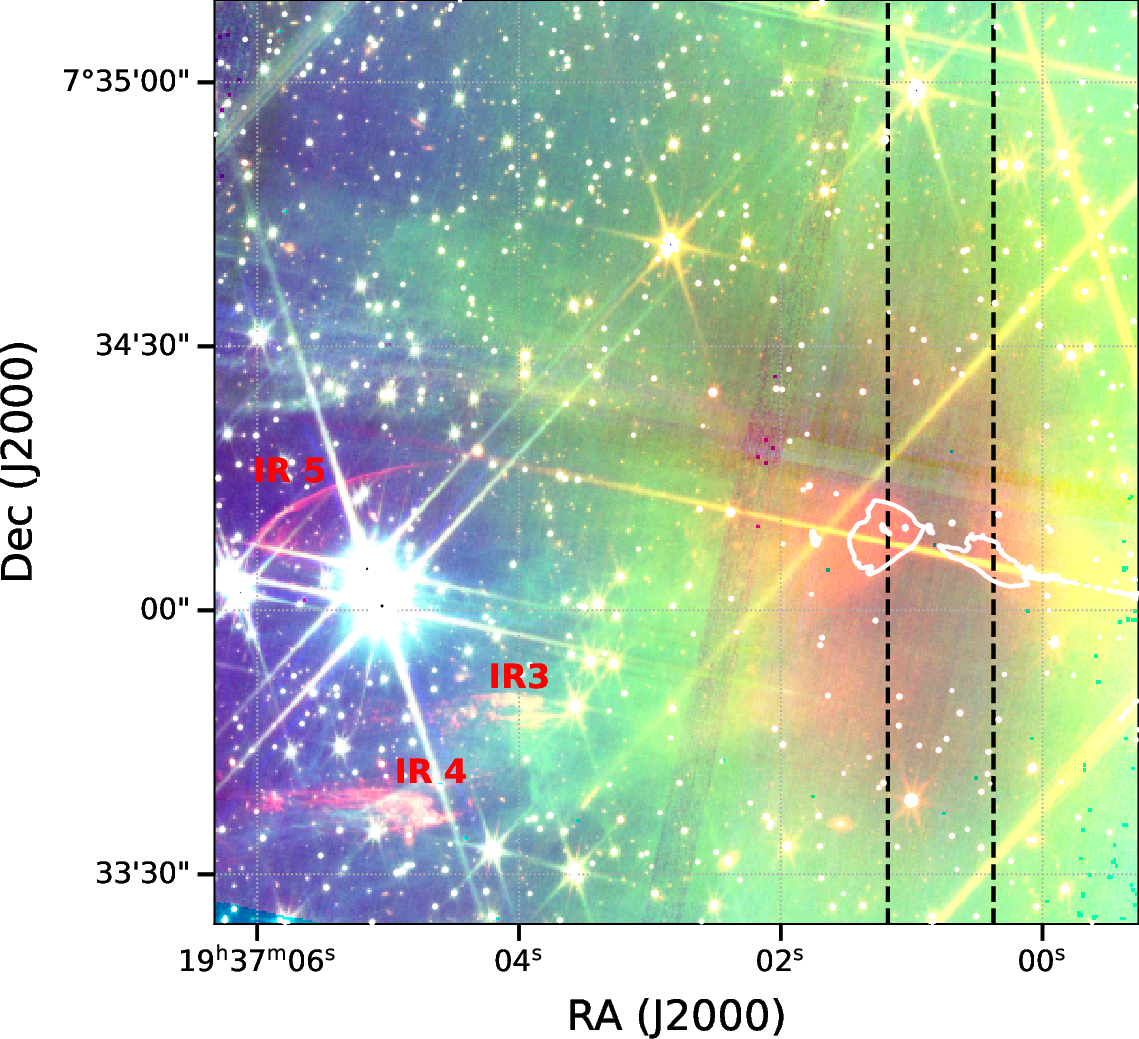}
	\caption{RGB color composite sky-subtracted image of the B335 outflow: F277W, F150W, F090W.
                 This image clearly shows the ``core shine'' scattered light from the interstellar radiation field outlining
                 the B335 cloud and core. 
                 Since the scattering is strongest at the shorter of the 
                 wavelengths shown here, the outer regions of the B335 cloud appear
                 in blue-green color, and increasingly dense regions in yellow and orange.
                 The densest parts of the core,
                 where the protostar is located, are seen as a depression in the scattered interstellar light, showing in a brownish color.
                 In the center of this dark region, in the F277W filter images displayed in red, the bipolar reflection nebula, illuminated by the protostar appears in reddish color. 
                 A few contours (white lines) of the flux in the F444W filter
                 are included to show the knots in the inner jet and the outline of the inner outflow cavity.
                 The distant parts of the outflow cavity appear as a region of reduced scattered light in
                 the eastern part of the image. The B335 cloud only contains one flattened dense molecular core.
                 The extent of the stripe used for extracting the crosscuts shown in Figure~\ref{cloudshine-crosscuts} is indicated by dashed lines. 
                 We have added labels (in red) for the distant shock fronts (appearing pink in this RGB image) in the
                 eastern lobe, since this figure is the only one in this paper showing these shock fronts.
                 \label{cloudshine-image}}
	\end{center}
\end{figure*}

\begin{figure*}[h]
\begin{center}
	\includegraphics[angle=0.,scale=0.40]{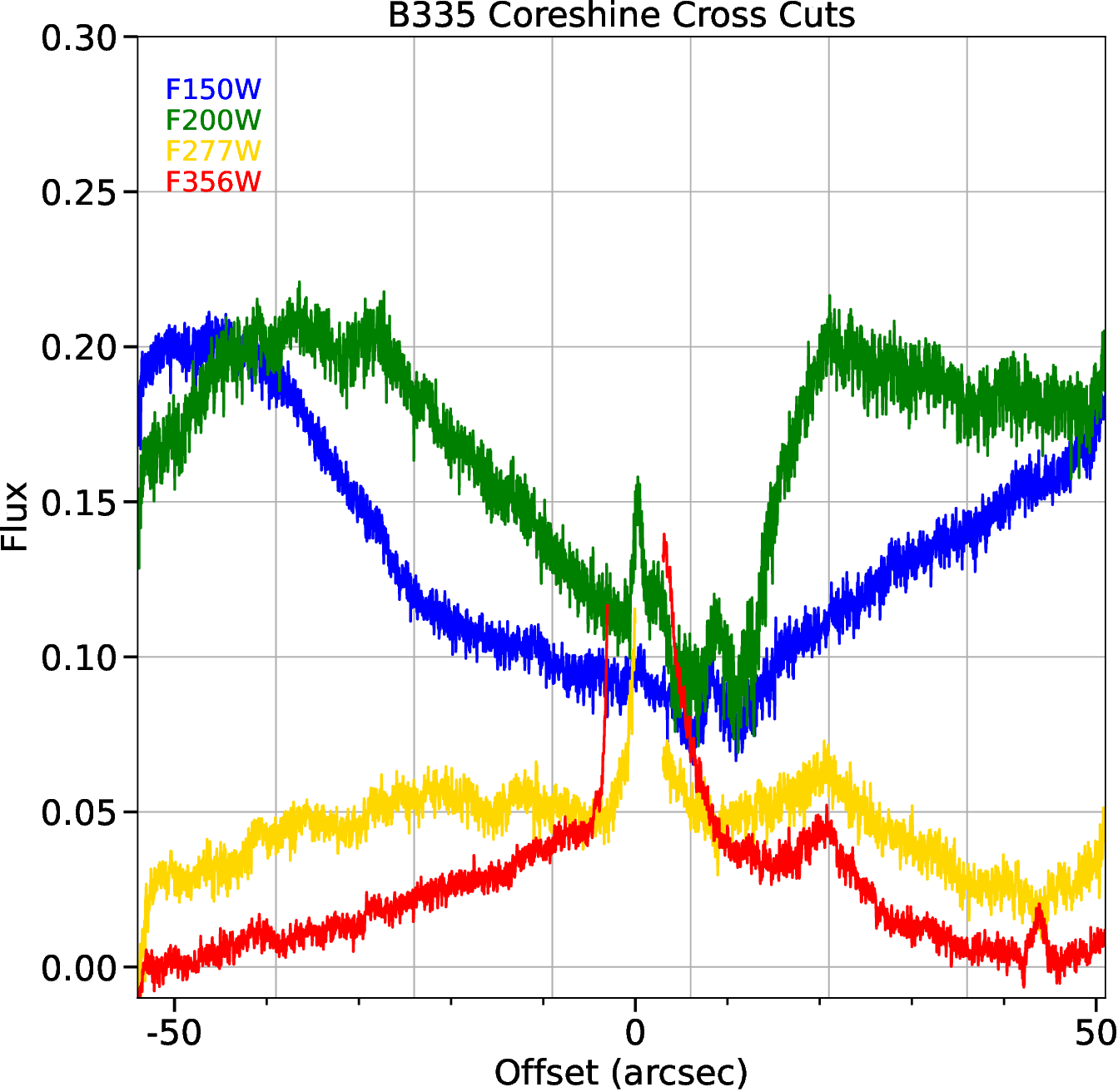}
	\caption{Crosscuts in declination direction centered on the protostellar source (offset 0), between the dashed lines
    in Figure~\ref{cloudshine-image}. The crosscuts are colored in wavelength order and the filters used are indicated. At the shortest wavelength (F150W, blue), B335 appears as a dark cloud against the background, the apparent size of which gets smaller with increasing wavelength (F200W, green). At even longer wavelengths, the reflection nebula illuminated by the protostar begins to be detectable (F277W, yellow), and becomes dominant in the F356W filter (red) \label{cloudshine-crosscuts}}
	\end{center}
\end{figure*}

The image in Figure~\ref{cloudshine-image} also shows that the
scattered light in the F090W (blue) and F150W (green) filters
shows a minimum in the area of the eastern (blue-shifted) outflow cavity. 
The eastern outflow cavity contains the bow shock front IR5 on its symmetry axis,
and to the south is limited by the shock fronts IR3 and IR4. The northern edge is 
indicated by an increase in coreshine scattered light (green) roughly at $7\arcdeg 34\arcmin 28\arcsec$.
We note that the two brightest stars in the image appear roughly in the center of this eastern outflow cavity. However, these stars, 
Gaia DR3 4295203523534207744 with Gaia g  = 13.68 mag at a distance of 535$\pm$6 pc and
Gaia DR3 4295203519210994816 with Gaia g 14.50 mag at a distance of 529$\pm$6 pc are 
far behind B335 and not physically associated with it.
In comparison to the extinction and column density maps presented in the next subsection,
the cloudshine and coreshine data extend to regions of lower density in the perimeter of the
molecular core and demonstrate that the outflow cavity can be traced into those peripheral regions
of the molecular core.
\clearpage

\subsection{Extinction and H$_2$O Ice Column Density Maps}
\begin{figure*}[h]
\begin{center}
	\includegraphics[angle=0.,scale=0.60]{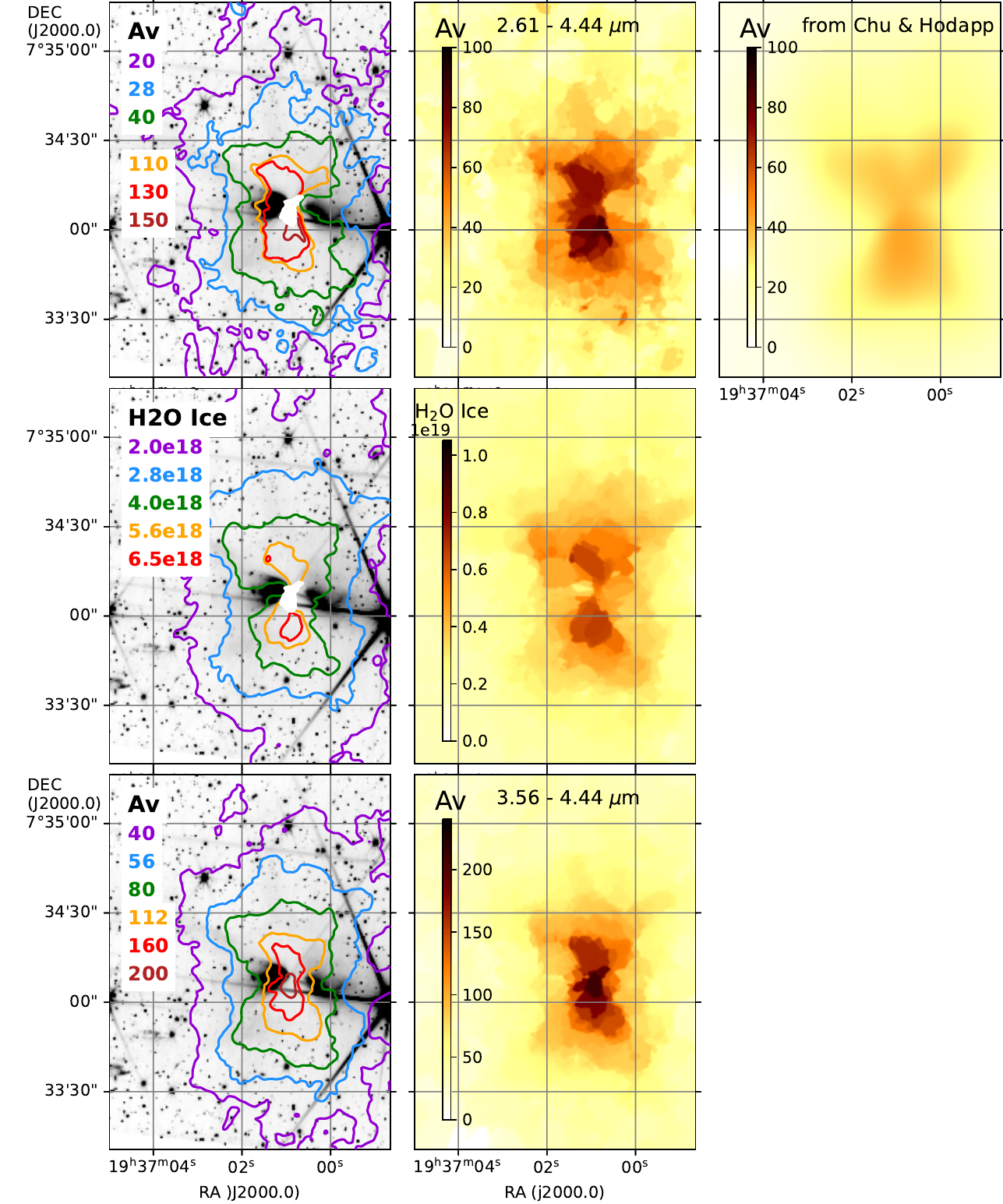}
	\caption{
    Left Column: Cutout of the F356W image with smoothed ($\sigma$ = 15 pixels) contours of the extinction or column density overlaid, to demonstrate the relation between the outflow cavity and the cloud structure. 
    The outflow cavity is traced by the bipolar reflection nebula. 
    In the two top rows, areas of high extinction near the protostar
    are indicated by a white patch to indicate the area where those
    maps are invalid due to the lack of detected stars.
    Center Column: The respective extinction and column density maps in halftone scaling.
    Top Row: The ''Continuum'' extinction map from the synthetic 2.61$\mu$m and F444W (4.44 $\mu$m) photometry.
    Center Row: The photometric H$_2$O column density map.
    Bottom Row: The extinction map from F356W - F444W photometry, which reaches the faintest stars and is therefore best sampled.
    Top Right: The data of the
    A$_V$ map by \citep{Chu.2021.ApJ.918.2.extinction.maps} with identical cutout and signal scaling for comparison.
    The extinction and column density data shown in the central column are
    available as FITS files in the data-behind-the-figure.
    }
    \label{H2O-AV-map} 
	\end{center}
\end{figure*}

Figure \ref{H2O-AV-map} shows maps
of the continuum extinction in units of A$_V$ in the top panels 
the H$_2$O ice column density in units of cm$^{-2}$
in the middle panel, and the extinction map based on the broad-band
F356W and F444W filters in the bottom row.
In the middle
left panel, we mark the area of highest extinction from the
broadband extinction map as a white area where the H$_2$O map is incomplete and not reliable.

The B335 core appears as a roughly ellipsoid shape, with the long axis approximately in the NS direction and, more to the point, the short axis lying in the EW direction.
As mentioned before, in B335, the magnetic field is roughly oriented
in EW direction near the position of the protostar \citep{Kandori.2020.ApJ.891.55.magmap}.
This suggests that the core was initially formed by contraction along the field lines of the local magnetic field, 
as was first theoretically predicted by 
\citet{1956.MestelSpitzer.MNRAS.116.503.magfield} and further developed by
\citet{1987.Shu.ARAA.25.23}. The result of this contraction is the flattened overall structure of the
B335 core, seen apparently nearly edge-on and therefore appearing elongated in the north-south direction
in the extinction maps.

The outflow activity originating at the single protostar has carved out bipolar outflow cavities.
Both in the extinction maps (top and bottom rows in Figure \ref{H2O-AV-map} and in the H$_2$O ice column density map (middle row), lines of sight through the outflow cavities
show lower extinction than other areas of the core, leading to an hourglass shape of the contours.
The hollowed-out outflow cavity was first formed by a wind from the protostar, pushing material out from
its surrounding molecular core. The outflow cavity at present contains a very hot jet and is limited
by a layer emitting in shock-excited H$_2$, so overall, it is too warm to allow the formation
of dust and ice mantles, and therefore appears as a region of lower extinction and ice column density.
The hollowed-out structure of the cavity follows closely the walls of this cavity
seen in ALMA molecular line emission, as is shown, for example, in Figure~1 of
\citet{LeGouellec.2023.AA...671.167.magfield}.

\clearpage

We note again that all the extinction maps underestimate the column density
in the immediate vicinity of the protostar, where even in the
F444W filter, no stars were detected. The mapping algorithm therefore
fills this region with the median extinction of stars detected
in the less obscured perimeter of this opaque region, and therefore underestimates
the extinction in this opaque region itself.

The comparison with the older pre-JWST extinction map
by \citet{Chu.2021.ApJ.918.2.extinction.maps}, shown here again in the top right panel of \ref{H2O-AV-map} illustrates
the limitations of this mapping technique. The older map,
based on UKIRT and Spitzer photometry with much shallower
limiting magnitude, does not show the massive rise in
extinction in the immediate vicinity of the protostar.
This fact was noted and those areas
had been marked as incomplete in that paper.
The lower extinction areas of the B335 cloud match our
new JWST extinction map quite well. The comparison with 
this earlier map illustrates the substantial 
advance in sensitivity and thereby spatial resolution of
the background star extinction mapping method enabled by JWST.
We also note that the H$_2$O Ice map similarly misses
the rise high column density near the protostar, which
is caused by the H$_2$O photometric ice index using the
shorter F277W and F300M filters, where fewer background stars were
detected near the protostar.

With the conversion factor of N$_H$ to A$_V$ of 2.0 $pm$ 0.2 10$^{21}$ cm$^{-2}$ mag$^{-1}$ from
\citet{Draine.2003.ARAA..41..241.NHvsAV}, our A$_V$ map can be compared to the
column density map by \citet{Launhardt.2013.AA...551A..98.Hcoldens}, their Figure B.10, where
they show the highest contour in their map at 20$\times$10$^{21}$ cm$^{-2}$, corresponding to
A$_V$ = 10. That highest contour is only slightly larger than their Herschel beam size and is
slightly elliptical in north-south direction, similar to the elongation in our maps.
The long axis of their highest contour is $\approx$ 1 \arcmin, similar to the A$_V$ = 40 (N(H) = 80 $\times$ 10$^{21}$ cm$^{-2}$) 
contour in our map in Figure \ref{H2O-AV-map}.

We have tried to compute a ratio map of the A$_V$ and 
H$_2$O ice data. This map did not show any significant
features outside the region near the protostar where
our mapping method is unreliable because of a lack
of detected star. We conclude that the broadband extinction
and the H$_2$O ice column density are closely coupled.
\subsection{Variability of the Reflection Nebula}

Bipolar nebulae associated with protostars are commonly understood to be primarily reflection
nebulae with an additional component of line emission, mostly shock-excited H$_2$. Infrared polarization observations for Class I and II objects typically show
the large degrees of polarization and an orientation perpendicular to the outflow axis as is expected
from scattering, as first pointed out by \citet{Hodapp.1984.AA.141.255.IRpol}, and shown in spatially
resolved detail
for many bipolar nebulae. Such polarimetric observations have not been done specifically for B335, due to its extreme extinction, but there is no doubt that B335 contains reflection nebulosity.

The variability of the B335 reflection nebula was found
and discussed in detail by
\citet{Evans.2023.ApJ.943.90.B335.variability}, on the basis of data from 
the WISE/NEOWISE space telescope
\citep{Wright.2010.AJ.140.1868.WISE}. At the wavelength of the WISE W1 and W2 filters,
only the reflection nebula is detected and variability of the protostar is indirectly inferred.
At millimeter radio wavelengths, based on ALMA CO data in various isotopologues,
\citet{Cabedo.2021.AA...653.166.infall} and
\citet{Bjerkeli.2023.AA.677.62.B335.episotic.accretion}
observed infalling gas towards B335 and concluded that a disk is being formed.
In ALMA observations between October 2016 to September 2017, i.e. in the early years of the present 
outburst,
\citet{Cabedo.2023.AA...669A..90.ionization} have found a high ionization fraction near the protostar,
indicative of on-going strong accretion.
\citet{Bjerkeli.2023.AA.677.62.B335.episotic.accretion} noted that
the infall kinematics suggest episodic infall.
It should also be noted that variability of B335 was also noted earlier at radio wavelengths
by \citet{Avila.2001.RMxAA.37.201.B335.variability}.
The CO outflow of B335 and its changes after the present outburst were discussed by
\citet{Kim.2024.ApJ.961.108.COoutflow}.
The fact that the B335 nebula can show strong variations on timescales of a few years
shows that it is largely a reflection nebula. At its distance of 165 pc, the light travel
time is 0.95 days for one arcsecond in the plane of the sky. Changes in the illumination from
the protostar during an accretion outburst therefore propagate across the nebula in approximately
one month, while, on the other hand, physical movement of material, even in the fast jet, takes
of order of a century, as we list in Table I. Variability due to changes in the shock excitation
would therefore be expected to be on such a longer timescale.

\begin{figure*}[h]
\begin{center}
	\includegraphics[angle=0.,scale=0.60]{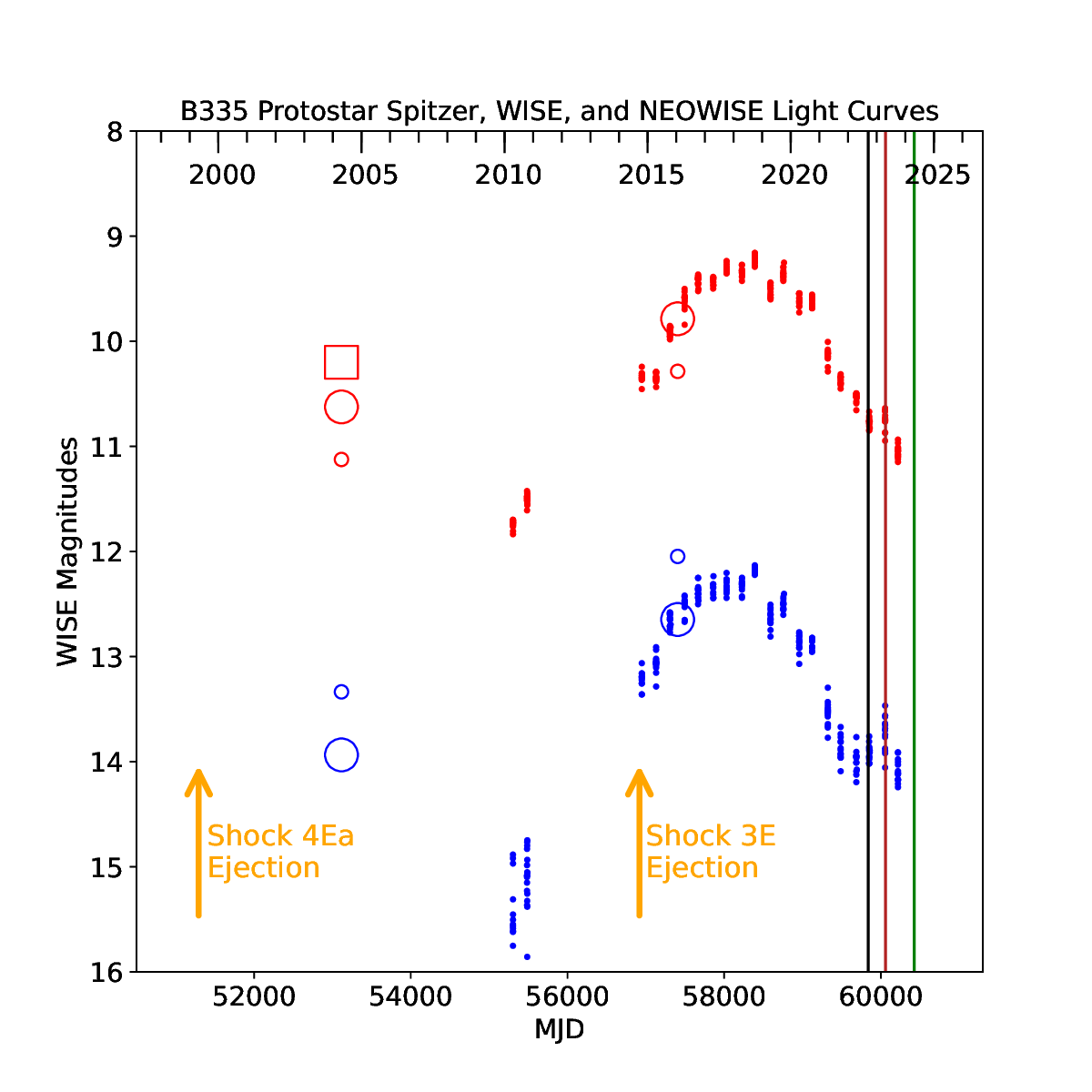}
	\caption{NEOWISE W1 and W2 light curve in Vega magnitudes. The epoch of
the NIRSpec IFU (MJD 59838.0, 2022 Sept. 16, black line)  and the two epochs of NIRCam imaging 
2023 04 25 UTC (MJD 60059.1, 2023 Apr. 25, brown line, and MJD 60424.3, 2024 Apr. 24, green line)
are indicated by vertical lines. We have also included the photometry based on Spitzer images in 2004 and 2016. The open red square is the large-aperture magnitude from 
\citet{Kim.2024.ApJ.961.108.COoutflow}, the open
circles are our raw and corrected r=4\arcsec magnitudes. We have indicated the
ejection times of shocks 4Ea and 3E. Shock 3E was ejected in the early phase of the presently on-going outburst. The wind causing shock 4Ea was plausibly ejected during the earlier outburst indicated by the 2004 Spitzer photometry.\label{NEOWISE-lightcurve}}
	\end{center}
\end{figure*}

We add to this discussion in Figure\ \ref{NEOWISE-lightcurve} by putting our two epochs of NIRCam imaging
into the context of the on-going decline of the lightcurve of B335. Both the light curves in the
WISE W1 (blue) and W2 (red) bands show that, compared to the original WISE data points between MJD 55000 and 56000,
the light curve has not returned to that level. In particular, the shorter wavelength W1 (blue) light curve (in magnitudes)
appears to have stabilized about halfway between the maximum and the earliest recorded W1 photometry.
The two earliest epochs of WISE data, in the cryogenic operations phase, already indicate an increasing brightness,
both in W1 and W2. We therefore have to assume that those early WISE measurements do not represent a
true quiescent condition of the protostar, but rather some early stage in the on-going outburst, or 
that a truly quiescent state does not exist in B335 so that low-level variability is always present.
The measured outburst amplitude is at least 3.5 mag in W1, and the outburst duration exceeds 5000 days,
indicating a duration and amplitude similar to the few classical FUor outbursts for
which the infrared outburst amplitudes have been compiled by
\citet{ContrerasPena.2025.JKAS.58.209.OYCAT}. 
B335 is a member of the still very small subgroup of very deeply embedded protostars
exhibiting long-duration outbursts of similar amplitude and duration as some classical
FU Orionis stars. The first example of such an object was OO Ser \citep{Hodapp.1996.ApJ.468.861.OOSer},
with a similar overall appearance to B335 \citep{Hodapp.2012A.pJ.744.56.OOSer}. A more recent example
is HOPS 383 discussed by \citet{Safron.2015.ApJ.800L.5.HOPS383}.

\begin{figure*}[h]
\begin{center}
	\includegraphics[angle=0.,scale=0.55]{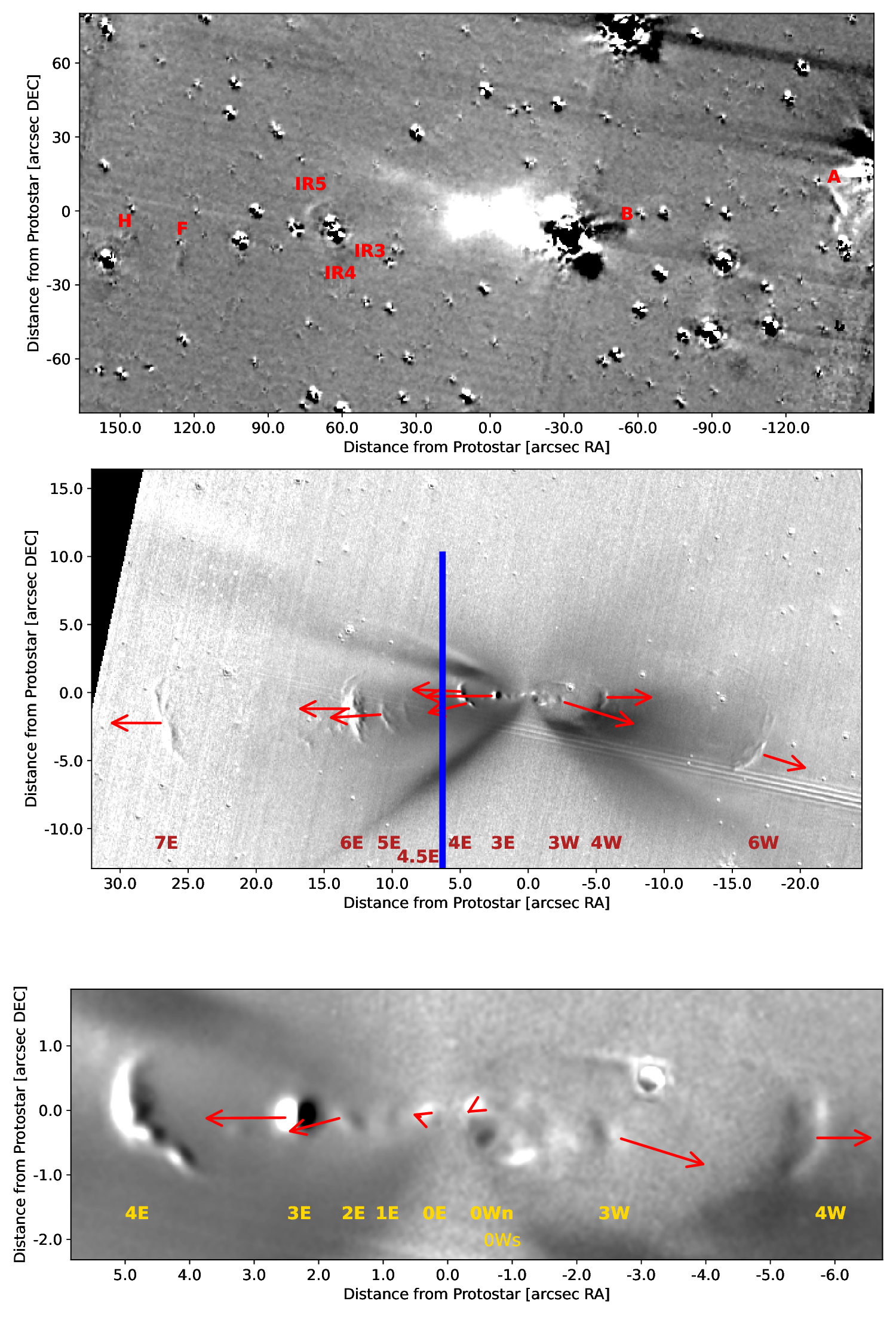}
	\caption{Top Panel: Wide field difference image of Spitzer channel 2 images of B335 taken in 2004 and 2016. The coordinates are offset from the position of the protostar. Moving shock fronts are seen adjacent pairs of black (2004) and white (2016) images of the same object. The previously known shock fronts are labeled. Shock fronts IR3 and IR4 do not show noticeable shifts in the images. We conclude that these are nearly stationary shocks, and not part of the fast jet. 
    Middle Panel: Shock proper motion vectors overlaid on the ratio of the registered NIRCam F444W
images at the two epochs: April 25, 2023 and April 24, 2024.
The grayscale was chosen so that local changes in the illumination
pattern in the reflection nebula are apparent. The middle  panel is the overall view of
the B335 reflection nebula and shock front, the lower panel is zoomed in on the innermost
shock fronts. The images at the two epochs are registered to a common
position for the protostar. Due to the proper motion of the protostar, all background objects
appear to have slightly moved in the year between the two epochs. The proper motion values
are listed in Table I.\label{Shock-PM}}
	\end{center}
\end{figure*}

Figure\ \ref{Shock-PM} shows the ratio of the 2023 and 2024 F444W images, so that local changes
on top of the overall variability are visible. There are a number of straight features in the ratio image pointing back at the position of the protostar (coordinate origin in 
Figure\ \ref{Shock-PM}), most noticeably at the northern edge of the blueshifted (eastern) 
lobe of the bipolar nebula.
These straight features cannot be explained by
shock fronts or thermal effects.
We interpret these features in our ratio images as shadows cast into the reflection nebula by local variations in the illumination and absorption
from the immediate surroundings of the protostar. They indicate that the light
path illuminating the reflection nebula contains absorbing dust condensations moving
substantially on a timescale of one year.
We also have to assume that these absorbing
dust condensations are above the plane of the disk around the protostar, so that shadow
effects can be produced in the reflection nebula. 
A simple estimate on the basis of Keplerian motion
around an object of 0.25 M$_\odot$ gives an orbital period of 2.0 years at 1 AU
distance. While we cannot determine what orbital motion of a dust condensation is producing
the changes observed within our one-year epoch difference, a distance from the
protostar in the range of a few AUs appears as a reasonable estimate.
We note again that the light travel time for the extent of the reflection nebula is of order one month,
short compared to this estimate of orbital period,
so that the formation of straight shadows is expected.

Similar straight features in the temporal ratio or difference
images have been noted before in other variable reflection nebulae.
Hubble's Variable Nebula NGC 2261 \citep{Hubble.1916.ApJ.44.190.var.neb}, illuminated by the variable young star R Mon is arguably the best studied case and shows both the slow moving
straight shadow effects pivoting around the position of the illuminating star
\citep{Lightfoot.1989.MNRAS.239.665.RMon}
as well as the faster moving light echos
\citep{Lightfoot.2025.MNRAS.540.52.RMon}.
In the more deeply embedded L483, an object similar to B335 in that it is also a
very young low mass protostar in an isolated small globule,
\citet{Connelley.2009.AJ.137.3494.L483.var} found evidence for straight shadow effects. 
The phenomenon of straight shadows in variable reflection nebulae has also been observed in much larger and higher luminosity deeply embedded objects
like Cep A by \citet{Hodapp.2009.AJ.137.3501.CepA.var} and also, based on optical images, in a
less obscured YSO: the EXor variable V1647 Ori studied by \citet{Aspin.2009.ApJ.692L.67.V1647}.

In typical protostellar outflow objects, see, for example, the review
by \citet{Ray.2021.NewAR.9301615.jets}, and specifically in B335, a fast jet is propagating
along the central axis of the outflow cavity, which is carved out of the surrounding
molecular core by a slower disk wind.
Following the model proposed by \citet{Lightfoot.1989.MNRAS.239.665.RMon}, our finding of
shadows implies that somewhere between the jet and the outflow cavity wall, density enhancements (clumps)
in the rotating disk wind are launched onto spiraling paths, and partly obscure the scattering
optical path, leading to the observed shadow effects. The amplitude of the shadow fingers is of order
20\% in the F444W filter, i.e., the optical depth changes in the scattering light path are less than
unity wavelengths around 4.4 $\mu$m. 

\begin{figure*}[h]
\begin{center}
	\includegraphics[angle=0.,scale=0.60]{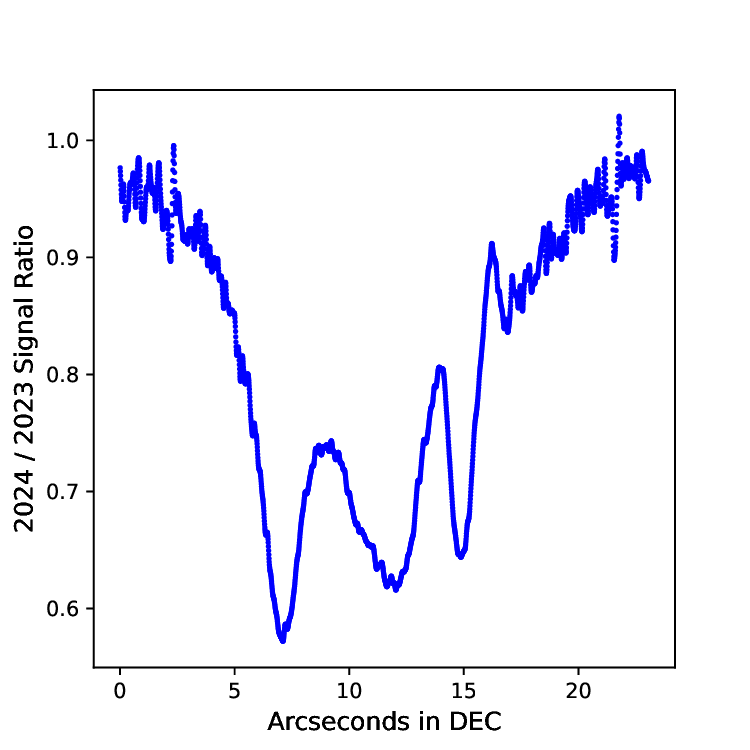}
	\caption{Crosscut from south to north through the B335 outflow cavity along the line indicated in blue in Figure~\ref{Shock-PM}.
    Shown is the flux ratio of the 2024 and 2023 F444W images. On average, the flux in 2024 was only $\approx$0.7 of the 2023 flux, but substantial local variations of order 20\% are seen, indicating strong shadowing variations on timescales of one year, i.e., an order of magnitude shorter than the timescale of the overall brightness outburst of B335 documented in Figure \ref{NEOWISE-lightcurve}.\label{Variation-Crosscut}}
	\end{center}
\end{figure*}

In Figure\ \ref{Variation-Crosscut}, we show a cross-cut of the ratio of the 2024 image to the 2023 image through the reflection nebula along the line
indicated in blue in Figure\ \ref{Shock-PM}.
At large distance from the outflow cavity, the ratio is near unity, even though noisy due to the faintness of the extended emission. In the outflow cavity, the ratio drops, on average, to 
$\approx$ 0.7, but with local variations of $\pm$ 0.1. 

\clearpage

\subsection{Are there Repetitive Outbursts ?}
The chain of shock fronts emerging from the B335 protostar
certainly suggests repeated ejection events.
There is only one set of Spitzer channel 1 and 2 images available that
were obtained prior to the WISE/NEOWISE missions, on MJD 53116, referred to
here simply as 2004. 
\citet{Kim.2024.ApJ.961.108.COoutflow}
have extracted photometry from the channel 2 image in a large 12\arcsec aperture
and have corrected for the difference in the bandpass between Spitzer and WISE.
Their channel 2 photometric point at 10.2 mag is indicated in 
Figure \ref{NEOWISE-lightcurve} by an open square. It indicates that the 
protostar was brighter in 2004 than in the earliest WISE observations.\\

We have analyzed the same archival Spitzer image with a different strategy.
We used a smaller, r=4\arcsec aperture centered on the blue-shifted outflow lobe.
We have also used a set of Spitzer channel 1 and 2 images obtained on MJD 57406 (2016) for the study of background star extinction in B335 by
\citet{Chu.2021.ApJ.918.2.extinction.maps}.
The difference of the Spitzer 2004 and 2016 images has already been shown
by \citet{Hodapp.2024.AJ.167.102.B335} to illustrate the local variations
in the brightness variations.
We show the uncorrected extracted photometry as small open circles in
Figure \ref{NEOWISE-lightcurve} and have used the 2016 photometry
to determine the offset to nearly contemporaneous NEOWISE measurements.
This offset, accounting for the different integrating apertures used,
was then applied to the 2004 data, the result being shown as large open circles
in the lightcurve.
While our data point lies 0.4 mag fainter than the larger aperture photometry by
\citet{Kim.2024.ApJ.961.108.COoutflow}
our data, in both Spitzer channel 1 and 2, confirm that in 2004, B335 was in a state
of brightness comparable to the average, not the maximal, brightness
of the present (2018 maximum) outburst. This is a tentative indication for
an outburst prior to the present one and makes it likely that shock front 4E,
with ejection dates of individual components around 2000 $\pm$ 2 (Table 1), was ejected during this prior outburst.

\subsection{The Outflow}
Our imaging in Paper I and the spectral images discussed by
\citet{Federman.2024.ApJ.966.41}
show several distinct components of the bipolar nebula associated with the B335 protostar.
We refer to \citet{vanDishoek.2025.AA...699A.361.JOYS}, their Figure 1, for a schematic drawing
of a typical protostellar outflow and the identification of its various components.
A wide reflection nebula seen primarily at continuum wavelengths is probably limited by
the cone angle of the continuum radiation emanating from the immediate surroundings of the
protostar.
The variability of these components, and the fact that the variability is not
uniform but shows structure indicating shadow effects originating from near the protostar, all point
in the direction of the nebula being defined by the limited solid angle of illumination
emanating from the protostar and the disk surrounding it.
The column density maps demonstrate that there is plenty of dust outside of the reflection
nebula that could potentially scatter light, but that this dust is simply not being illuminated
from the protostar.
It is also clear from the comparison of line images by
\citet{Federman.2024.ApJ.966.41} (their Figure 3) that emission of shock-excited H$_2$
traces an outflow cone that is narrower than the walls of the outflow cavity reflection nebula.

Based on near-infrared images, 
\citet{Hodapp.1998.ApJ.500L.183.B335} discovered and 
\citet{Galfalk.2007.A&A.475.281.B335.outflows} measured the proper motion of the large, more distant shock fronts in B335. To avoid confusion between 
their naming terminology, IR 3, 4, and 5, from the scheme used by
\citet{Hodapp.2024.AJ.167.102.B335} (Paper I) and in this paper, they are identified
again in Figure\ \ref{cloudshine-image}.
For these, we could not obtain a new proper motion measurement
because the second epoch image did not cover these shocks. 

We show a difference image of two Spitzer Space Telescope channel 2
images, obtained in 2004 and 2016, to give an overview of the larger-scale outflow in Figure~\ref{Shock-PM} (top panel) and to qualitatively confirm the large proper motion of the shock fronts
in the jet, in particular A, B, IR5, F, and H). The two Spitzer
images, taken with the same instrument 12 years apart, represent
a very consistent data set. However, apparent changes in the shape
of individual shock fronts make it an oversimplification to just
quote a single proper motion value for each shock front. 
The proper
motion away from the protostar is apparent in the differential image.
This statement
includes the large bow shock 5 in \citet{Galfalk.2007.A&A.475.281.B335.outflows}, ``IR5'' in 
\citet{Hodapp.2024.AJ.167.102.B335}, for which \citet{Galfalk.2007.A&A.475.281.B335.outflows} 
has measured a high proper motion (245 km\,s$^{-1}$), similar
to the jet shock fronts that we will discuss here.
A noteworthy exception are the shocks IR3 and IR4, where we do not
see the same differential (black-white) pattern. These complicated
shocks do not move with anything close to the speed of the jet in B335. In the Spitzer data, 
they are consistent with being stationary. \citet{Galfalk.2007.A&A.475.281.B335.outflows} lists
25 km\,s$^{-1}$ for IR3 and 16 km\,s$^{-1}$ for IR4a.
We conclude that these are shocks in the much lower velocity disk wind or an entrained lower
velocity wind that is carving
out the cavity, possibly shocking against the stationary cavity walls. 

\begin{figure*}[h]
\begin{center}
	\includegraphics[angle=0.,scale=0.50]{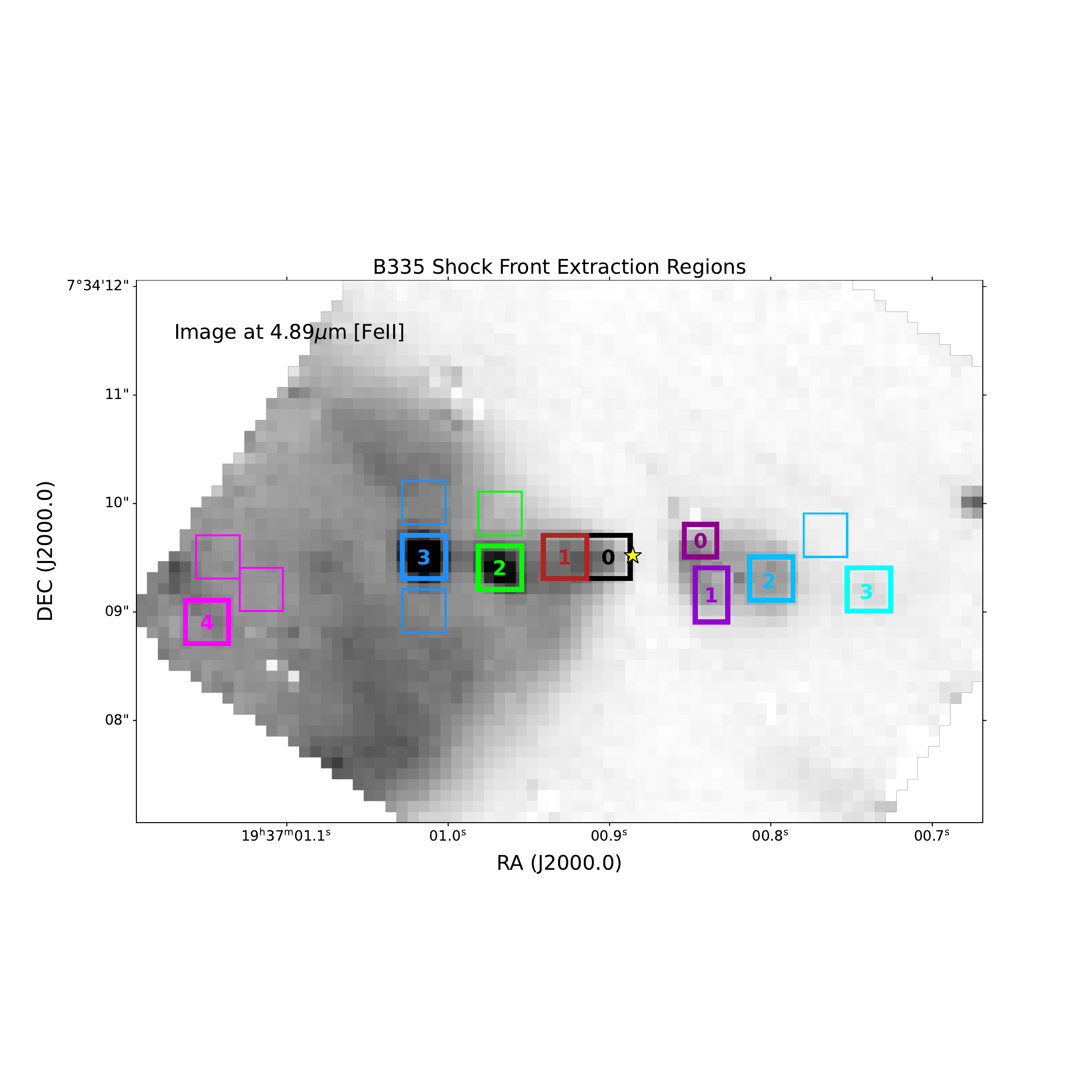}
	\caption{Positions of the regions for NIRSpec IFU datacube spectrum extraction
    of shock fronts.
                 The color coding corresponds to the colors of the spectra in 
                 Figure~\ref{All-Shock-Spectra}.
                 The image is the data cube slice at a wavelength of 4.89 $\mu$m, centered on an [\ion{Fe}{2}] emission line, to emphasize high-excitation
                 shock fronts.
                 Thick outlines and labels indicate the extraction regions for shock-excited emission features,
                 thin outlines are the reflection-only ``cavity background'' regions used for scattered light subtraction. The position of the protostar at the
                 epoch of the NIRSpec observations is indicated by a yellow star symbol.\label{Shock-extraction-regions}}
	\end{center}
\end{figure*}

Near the central axis of the outflow, a narrow jet indicated primarily
by the high-excitation atomic line of [\ion{Fe}{2}] has been identified by
\citet{Federman.2024.ApJ.966.41},
their Figure\ 3.
This central jet was studied in more detail by
\citet{Federman.2026.ApJ...998..282.MIRIjet} on the basis of additional MIRI MRS IFU spectra
in [\ion{Fe}{2}], [\ion{Ni}{2}], [\ion{Ne}{2}], and [\ion{Ar}{2}]. They find
that the blue- and red-shifted jets are at a  168.5\arcdeg relative to each other,
not at 180\arcdeg as simple symmetry would predict. Their finding is consistent
with the appearance of other, more distant shock fronts in our images, e.g. in
Figure~\ref{Shock-PM}, where shocks 3W and 6W are not axisymmetric to shocks 3E and 6E,
and their proper motion vectors are not anti-parallel.
This is similar to the bending of the jet of HH 30 found by
\citet{Estalella.2012.AJ.144.61.HH30} and modeled by interaction in a binary
star system.
For other shocks, most notably 4E and 4W,
this misalignment is less pronounced, which could be explained by the finding by
\citet{Federman.2026.ApJ...998..282.MIRIjet} that the B335 jet axis ``wiggles''
on spatial
and therefore time scales shorter than that of the overall bending.
Some of the jet knots are also prominent in Br$\alpha$.
The strongest shock front in the jet, 3E, is also detected in H$_2$, but the other
jet knots in [\ion{Fe}{2}] are not prominent in H$_2$ emission.
Table I, lists the measured proper motions of the shock fronts that we could
study in our images.

\begin{deluxetable*}{llllllll}
\tabletypesize{\scriptsize}
\tablecaption{B335 Shock Fronts\\}
\tablewidth{0pt}
\vspace{0.5cm}
\tablehead{
\colhead{Identification} & \colhead{RA} & \colhead{Dec} & \colhead{PM mas yr$^{-1}$} & \colhead{PM [km\,s$^{-1}$]} & \colhead{PA [deg]} & \colhead{Kin. Age [yr]} & \colhead{Eject Year}}
\startdata
HH119 JWST 7E & 294.26104 & 7.56845 & 206 & 161 & 90 & 130 & 1983\\
HH119 JWST 6E & 294.25722 & 7.56895 & 204 & 160 & 99 & 62 & 1961\\
HH119 JWST 5E & 294.25660 & 7.56875 & 208 & 163 & 93 & 51 & 1972\\
HH119 JWST 4Ea & 294.25506 & 7.56929 & 200 & 156 & 88 & 24 & 1999\\
HH119 JWST 4Eb & 294.25498 & 7.56919 & 210 & 164 & 91 & 22 & 2001\\
HH119 JWST 4Ec & 294.25486 & 7.56910 & 167 & 131 & 104 & 25 & 1998\\
HH119 JWST 3E & 294.25433 & 7.56927 & 266 & 208 & 90 & 8.6 & 2014\\
HH119 JWST 2E & 294.25413 & 7.56926 &  &  &  &  & \\
HH119 JWST 0E & 294.25378 & 7.56927 & 42 & 33 & 70 &  &  \\
HH119 JWST 0Wn & 294.25359 & 7.56931 & 12 & 10 & 121 &  &  \\
HH119 JWST 3W & 294.25306 & 7.56920 & 290 & 227 & -107 & 8 & 2015\\
HH119 JWST 4W & 294.25225 & 7.56897 & 187 & 146 & -90 & 28 & 1995\\
HH119 JWST 6W & 294.24930 & 7.56795 & 183 & 143 & -107 & 93  & 1930\\
\enddata
\end{deluxetable*}

Shock front 2E, emitting almost exclusively in ionic lines, appears faint
in the F444W filter images, and its proper motion is therefore poorly
determined.
The bright, compact shock front 3E has proper motion of 208 km\,s$^{-1}$
and a radial velocity  of -74 km\,s$^{-1}$ (Figure \ref{CO-radial-velocity})
relative to the systemic velocity represented by the average of 0E, 0Wn, and 0Ws.
For the next shock in the blue-shifted lobe, shock 4E, we can distinguish three spatial
components, listed separately in Table I. Only the most southern of these components (4Ec)
is included in the field of the NIRSpec data cube.

From the proper motion and the distance from the position of the protostar,
we have determined the kinematic ages of the shock front prior to the epoch
of the first NIRCam images in 2023,
and their ejection date.
For the bright shock front 3E, we have
indicated this kinematic age by an arrow in Figure \ref{NEOWISE-lightcurve}, indicating that its
launch coincides with the early phase of the photometric outburst that is,
at present, in its declining phase.

From their position-velocity diagrams based on CO emission observations
taken in 2017
with ALMA, 
\citet{Bjerkeli.2019.AA.631.64.B335.ALMA.kinematics}
found a high-velocity feature named the ``molecular bullet'' with a very
short kinematic age of only 1.7 yr, i.e., launch in early 2016, again
coincident with the early phase of the present phase of increased accretion
luminosity. Based on the same data set,
\citet{Kim.2024.ApJ.961.108.COoutflow}
analyzed this high velocity component in more detail and put it in relation
to the on-going infrared outburst. 
The high-velocity blue-shifted (-30 km\,s$^{-1}$) in the 2017 ALMA data is
almost certainly related to shock front 3E observed by us.
Shock 3E, being part of a larger jet of shock fronts, must be an
internal shock front in that fast jet, where younger, newly ejected and faster 
jet components run into, and shock against, older, slower jet gas.

\subsection{Changes in Shock Excitation Conditions Along the Jet}

\begin{figure*}[h]
\begin{center}
        \includegraphics[angle=0.,scale=0.30]{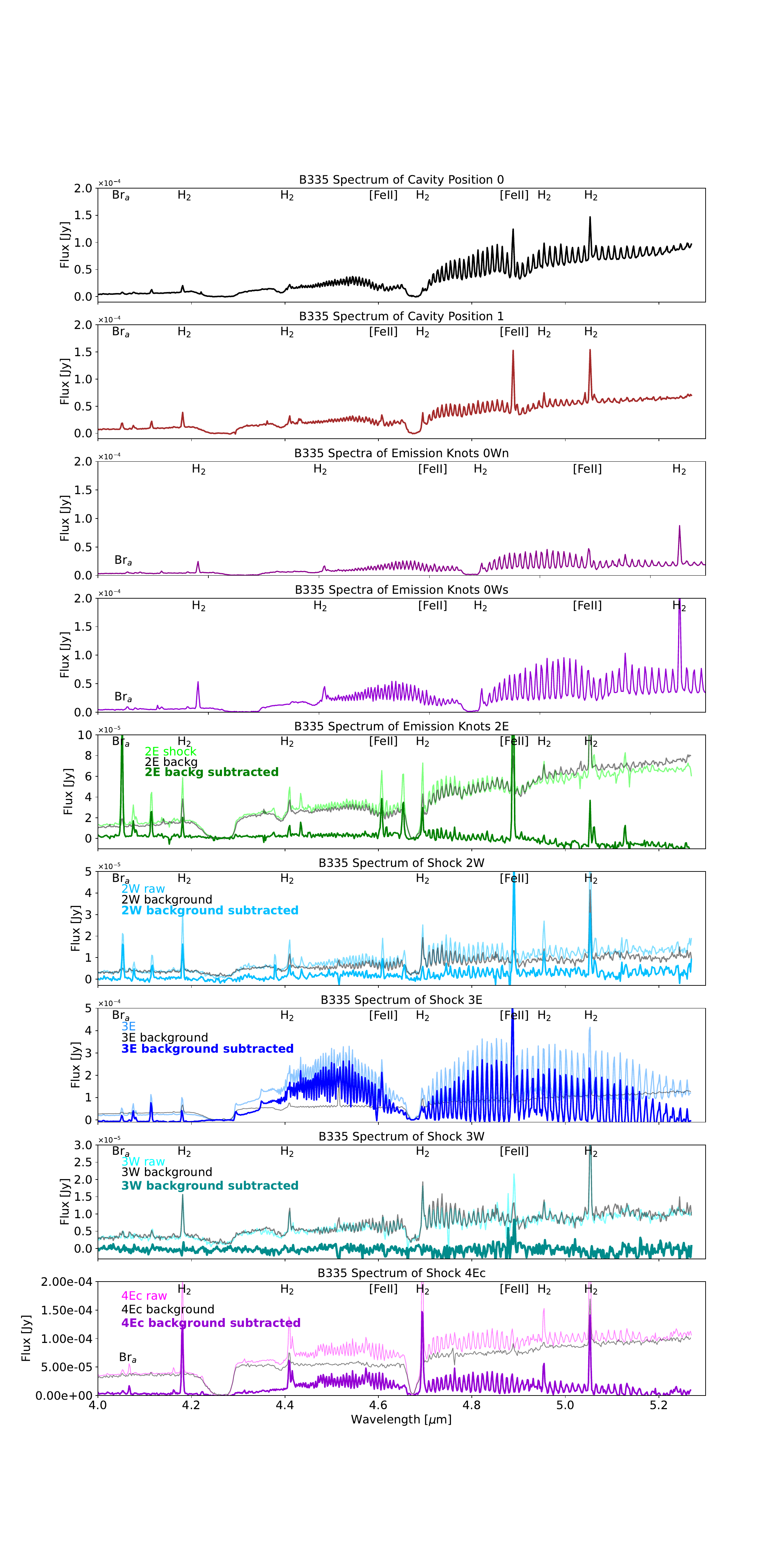}
        \caption{The spectra extracted from the 2022 NIRSpec IFU data cube. The color coding of
    the spectra corresponds to the colors of the extraction apertures in Figure \ref{Shock-extraction-regions}. The spectra are arranged in order of increasing distance from the protostar.
                 \label{All-Shock-Spectra}}

        \end{center}
\end{figure*}

In the following, we discuss individual shock fronts
in order of distance from the protostar.
Figure \ref{All-Shock-Spectra} shows all the spectra extracted from the
NIRSpec data cube.

The first set of apertures for spectral extraction, shown in Figure\ \ref{Shock-extraction-regions},
are the shock-excited emission knots previously identified
in Paper 1. For shocks 2E, 3E, and 4E, we made an effort to subtract flux contributions from the background
and foreground cavity emission. The cavity-background boxes are outlined by thin lines in the
same color as the shock extraction box. For the other shocks, 0E, and 0Wn and 0Ws, this procedure
was not possible because they are so close to the protostar that no suitable background region
could be identified.\

For the extraction of line fluxes, and for the image cutouts in Figure~\ref{lines-v-distance},
the continuum was calculated from a linear fit through the minima between the four adjacent CO lines,
and subtracted from the spectra. We are using the strong H$_2$ (0,0) S(8) line at 5.053~$\mu$m and
the \ion{Fe}{2} line at 4.889~$\mu$m because they are close together and extinction corrections between
them are negligible for our purposes. At the spectral resolution of the NIRSpec data, both this [\ion{Fe}{2}] and H$_2$ line are blended together with a CO emission line. We therefore subtracted the average
line flux of the four adjacent CO lines from the raw integrated [\ion{Fe}{2}] and H$_2$ line flux,
to arrive at the line flux free of the contribution from the blended CO line. We are using this average
of neighboring CO lines as a measurement of the CO emission line strength at the [\ion{Fe}{2}] and H$_2$ lines and thus display two data points for the CO emission in Figure~\ref{lines-v-distance}.

A detailed analysis of the excitation conditions in each shock front is beyond the scope of this paper, and is being worked on for separate publications \citet{Kim.2026.inpreparation.B335CO} and
\citet{LeGouellec.2027.in.prep.COexcitation}.
We only comment here on the
relative strength of emission lines in the four youngest shock fronts 0E through 4E in the eastern, blueshifted lobe of the B335 outflow on the basis of the 2022 NIRSpec data set.

The origin of molecules observed toward Class 0 jets is still debated.
Models invoking ejection from the stellar magnetosphere and inner disk edge
\citep{Shu.1994.ApJ...429..781.magdrivenoutflow}
\citep{Zanni.2013.AA...550A..99.magoutflows}
are expected to be dust-free given that the launching radius lies inside the sublimation radius.
In turn, the small dust fraction does not provide a sufficiently high FUV shielding
to hinder the photodissociation of molecules by the irradiation emanating from the accretion shock.
Several studies have explored the survival of molecules such as H2O, CO, and SiO in such dust-poor irradiated jets
\citep{Glassgold.1991.ApJ...373..254.moleculeformation},
\citep{Raga.2005.RMxAA..41..137.H2formation}
and concluded that molecule formation is not likely.
More recently,
\citet{Tabone.2020.AA...636A..60.H2formation}
explored this irradiated chemistry in dust-poor protostellar jets and found molecular formation routes
involving new gas phase formation routes of H$_2$, and self-shielding of atomic species within internal shocks.
A second scenario proposed that some parts of jets are launched beyond the sublimation radius via
MHD disk winds where dust can shield molecules from the FUV field, and allow faster H$_2$ reformation
on the dust grains' surface \citep{Panoglou.2012.AA...538.2.moleculesurvival}.
Finally, a last scenario was developed by
\citet{Raga.1993.AA...278..267.outflows},
\citet{Raga.2005.RMxAA..41..137.H2formation},
\citet{Tabone.2018.AA...614A.119.pulsatingjet}, and
\citet{Rabenanahary.2022.AA...664A.118.outflowfromnarrowjet}
that molecules can be entrained by jets,
where jet internal shocks would sweep up surrounding material along the
different shock working surfaces propagating along the interaction layers
with the envelope, or low-velocity outflowing material.
In the context of the observed photometric outburst of B335 caused by an increase in
the accretion rate, it appears possible that the fast, dust free, atomic and ionized jet
is a transient phenomenon in B335, and that the outflow is dominated by a more steady,
slower disk wind that is molecular and contains dust.
This is the scenarion in which we will interpret the observed line fluxes and proper motions
observed in the B335 outflow and jet.

\subsubsection{Shock 0E}
The emission knot closest to the position of the protostar in the eastern,
blue-shifted outflow lobe is shock 0E (Paper I). 
In the difference image of the 2023 and 2024 images, 
a position shift is noticeable in knot 0E and listed in Table 1; 
however, this proper motion is much smaller
than the motion in the other emission knots further downstream in the jet. 
The spectrum of 0E can be decomposed into a very steep continuum spectrum,
and an emission line spectrum dominated by CO line emission.
The CO lines of shock 0E have similar radial velocity as 0Wn and 0Ws, the related knots 
in the other outflow lobe. Together, these three knots closest to the protostar
have a very different, lower, radial velocity than shock 3E, the brightest and youngest
shock in the central jet. They also have slower proper motion than shock 3E.
Shock 0E is clearly not part of the jet ejected
from the protostar.
The protostar is not directly visible in the NIRCam and NIRSpec 
wavelength range out to 5.0 $\mu$m. 
Shock 0E appears faint due to the very high extinction in the protostellar disk seen from the side.
The difference images of H$_2$ and CO emission in shock 0E shown in Figure\ \ref{lines-v-distance} shows
that H$_2$ emission extends in a broad paraboloid cone east of the protostar position, while CO is concentrated
at the apex of this cone and of the [\ion{Fe}{2}] jet. This CO emission knot is the closest feature to the protostar
position detected in this NIRSpec data set. However, it is not coincident with the position
of the protostar from ALMA observations, reduced to the epoch of the NIRSpec data, that is indicated
by a star symbol in Figure\ \ref{lines-v-distance}. 
The extraction box for the 0E spectra 
in Figure~\ref{All-Shock-Spectra} encompases this CO-dominated region close to the
protostar. The spectrum, shown in black, is dominated by the emission of CO gas, with
H$_2$ and [\ion{Fe}{2}] lines only roughly as strong as the typical CO line.
\citet{Federman.2024.ApJ.966.41}
have shown an extracted spectrum of the innermost region of the eastern outflow, larger than, but encompassing
the shock defined as 0E here.
The region 0E is distinct from the other shock fronts as it shows much less proper motion (Table I).
We therefore interpret it as a combination of scattered light from the protostellar source itself, explaining the strong continuum, and shocks from
a much slower disk wind.
Shock 0E shows CO emission west of the starting point of the jet seen in Br$\alpha$ and [\ion{Fe}{2}] and the apex of the outflow cavity outlined in H$_2$ emission.
This CO line flux is the closest to the position
of the protostar observable at NIRSpec wavelengths, but it is not direct light from the protostar itself but rather scattered light.
Even at MIRI wavelengths \citet{Federman.2026.ApJ...998..282.MIRIjet} did not detect direct light
from the protostar.

The launch region of the jet traced in [\ion{Fe}{2}] is much
narrower than the inner regions of the outflow cavity.
The small difference image of H$_2$ and [\ion{Fe}{2}] in Figure\ \ref{lines-v-distance}
shows the distinct [\ion{Fe}{2}] jet that emerges from the protostar,
while H$_2$ emission is outlining the wider cone of the cavity walls, probably
excited by shocks from a much slower disk wind.
The wide outflow cavity seen in
H$_2$ emission is better seen in the wider fields shown in
\ref{CO-emission-map} top left panel that we will discuss in Section 3.7.3 in the
context of CO gas emission.
In the spectral images shown by \citet{Federman.2024.ApJ.966.41}, the CO emission is concentrated
in a cone of roughly the same width as the H$_2$ emission, but does not extend as far down along
the outflow axis. 
\citet{Rubinstein.2024.ApJ.974.112.B335.IFU} published an analysis of the CO excitation
conditions based on population diagrams for a larger sample of protostellar objects, including
the innermost outflow lobes of B335, one of them being close to our cutout definition
of 0E, the other encompassing 0Wn and 0Ws. They identify two temperature components (900K and 1700K) based on the $\nu$ 1-0 transition
and a higher temperature component of 6300K based on the $\nu$ 2-1 transition.

\subsubsection{Knots 0Wn and 0Ws}

To the west of the protostar position, there are two emission
knots detected in the F444W filter.
The southern of these knots, 0Ws, does not
show a clear signature of proper motion in the difference image,
and follows the rest of the outflow cavity reflection nebula in
being fainter in 2024 than in 2023.
This emission region shows complex morphological changes between
our two epochs of imaging observations. 
Interpreted naively, these changes appear to be
a strong apparent proper motion
in a general direction towards SE, different from any of the other
shock fronts. The radial velocity of this feature is close to the
systemic velocity represented by shock 0E and very distinct from that of the jet shock fronts.
With the information available, we cannot clearly identify this
as a shock front, and favor an interpretation as a reflection nebula
subject to locally changing illumination conditions.
For this reason we do not list shock 0Ws in Table 1 and we do not
show a proper motion vector in Figure~\ref{Shock-PM}. 
The northern knot (0Wn) also does not show motion, but has increased
in brightness (i.e., appears white in the difference image).

We should point out that both 0Wn and 0Ws are much further away
from the protostar than 0E. In fact, their separation in NS direction is
similar to the opening of the outflow cavity at the same distance from
the protostar in the eastern direction. 
The 0Ws region shows shock-excited features in its spectrum, 
including CO emission, similar to that seen in the 0E and 0Wn region.

\subsubsection{Shocks 1E and 2E}
The ratio image Figure\ \ref{Shock-PM}
shows the typical dark-bright pattern of a moving shock front at the positions
of shock 1E and 2E.
Knot 1E, identified in paper I, appeared less defined in 2024
and a proper motion could not be measured with confidence.
We will not discuss the questionable shock 1E any further here.

Shock 2E (Figure\ \ref{All-Shock-Spectra}) only shows very weak CO emission, in
distinct contrast to the strong emission in shock 3E discussed below.
However, shock 2E shows many emission lines of ionic species in the
MIRI spectra presented by
\citet{Federman.2026.ApJ...998..282.MIRIjet}
and appears actually more highly excited that 3E.
In the NIRCam F444W images, shock 2E appears faint, since only the \ion{Fe}{2} lines
contribute substantially to the flux. Therefore, and because of the complex flux
distribution between the protostar and shock 3E, the proper motion of shock 2E is
only poorly measured, and for this reason, we do not include this measurement in Table 1.

\subsubsection{Shocks 2W and 3W}
The two shock fronts 2W and 3W are only faintly indicated in the continuum image.
In the western, redshifted outflow lobe, shock 2W itself, 
shown in Figure\ \ref{All-Shock-Spectra}, after subtraction of a adjacent cavity background flux, shows very little CO emission
and is dominated by H$_2$ and [\ion{Fe}{2}] line emission.
In the spectral images shown by
\citet{Federman.2024.ApJ.966.41} shocks 2W and 3W appear as the western part of the
[\ion{Fe}{2}] jet.
While the symmetric distribution around
the protostar position and the similar proper motions in the east (blueshifted) and west (redshifted)
outflow lobe strongly suggest that corresponding shocks were ejected at the same time,
their excitation conditions are clearly not identical. While their spectra are similar
to the minor shock 2E in the eastern lobe, they differ substantially from the
shock front 3E, discussed in the following subsection, where CO emission dominates. 
In particular, shock 3W, which is symmetric to the strong CO emitting shock 3E,
only shows some [\ion{Fe}{2}] emission, after cavity background subtraction.

\subsubsection{The Strong Shock 3E}

Shock 3E (Figure\ \ref{All-Shock-Spectra}) shows emission lines from shock-excited H$_2$ and
[\ion{Fe}{2}], but also very strong CO emission. Shock 3E, and also 2E, have some H$_2$ emission, but it is relatively weak compared to the
shocks further downwind.
We have subtracted spectra from two adjacent positions north and south of shock 3E, indicated in Figure \ref{Shock-extraction-regions},
representing the cavity foreground and background.
The continuum contribution in the spectrum of 3E originates
from the foreground and background in the outflow cavity.
Shock 3E in Figure\ \ref{All-Shock-Spectra} itself is a
pure emission line spectrum.

\begin{figure*}[h]
\begin{center}
	\includegraphics[angle=0.,scale=0.50]{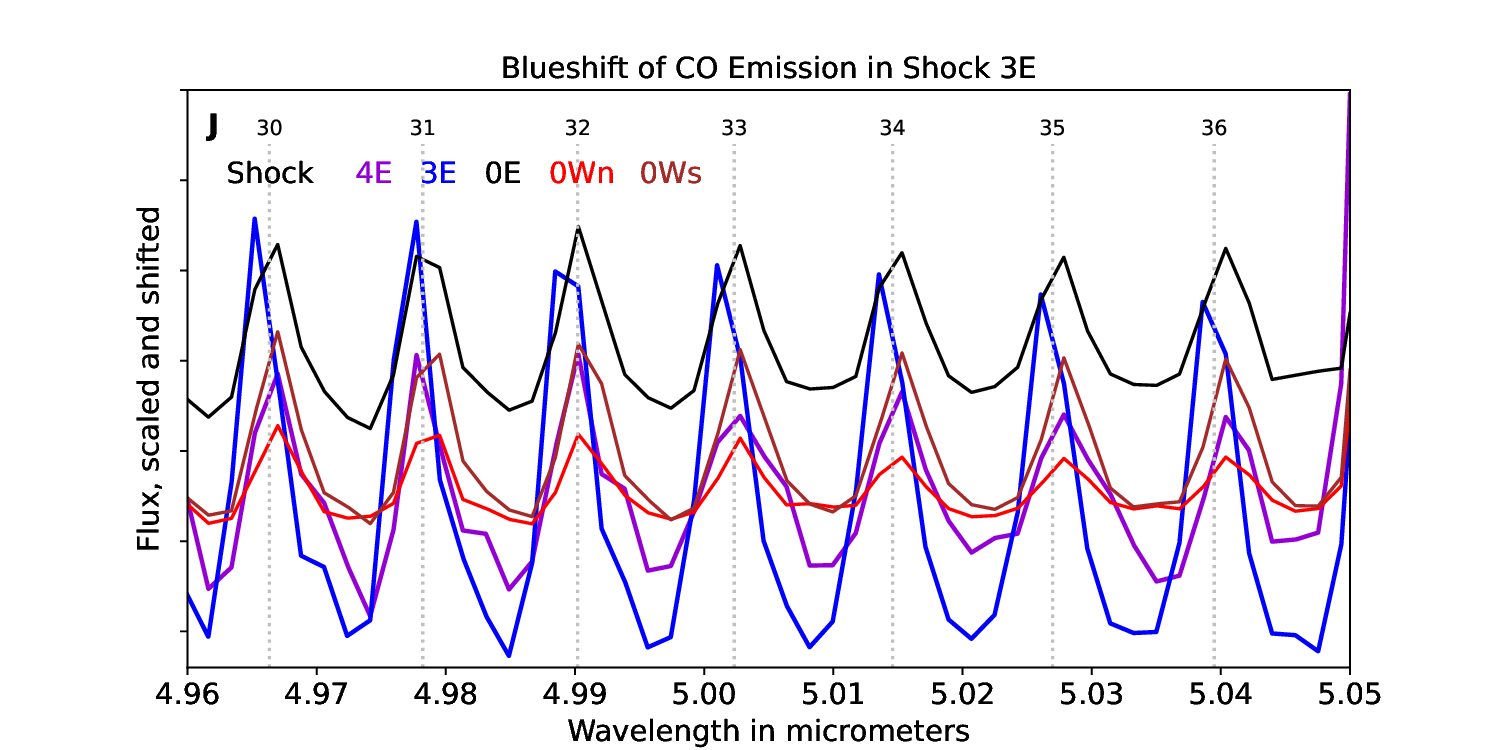}
	\caption{For different integration apertures, this figure shows spectra over a small wavelength range to illustrate the particular blue-shifted velocity of shock 3E. The spectra are in units
of flux, but are both scaled and shifted for clarity. 
The CO emission lines in shock 3E (blue) are blue-shifted compared to the lines in
the regions close to the protostar 0E (black), 0Wn (red), and 0Ws (brown). The older shock 4E is
slightly blue-shifted relative to the shocks near to the protostar, but less so than shock 3E.
The emission lines of shock 3E are blueshifted by a radial
velocity of -74 $\pm$ 3 km\,s$^{-1}$.
                 \label{CO-radial-velocity}}
	\end{center}
\end{figure*}

\begin{figure*}
\begin{center}
        \includegraphics[angle=0.,scale=0.38]{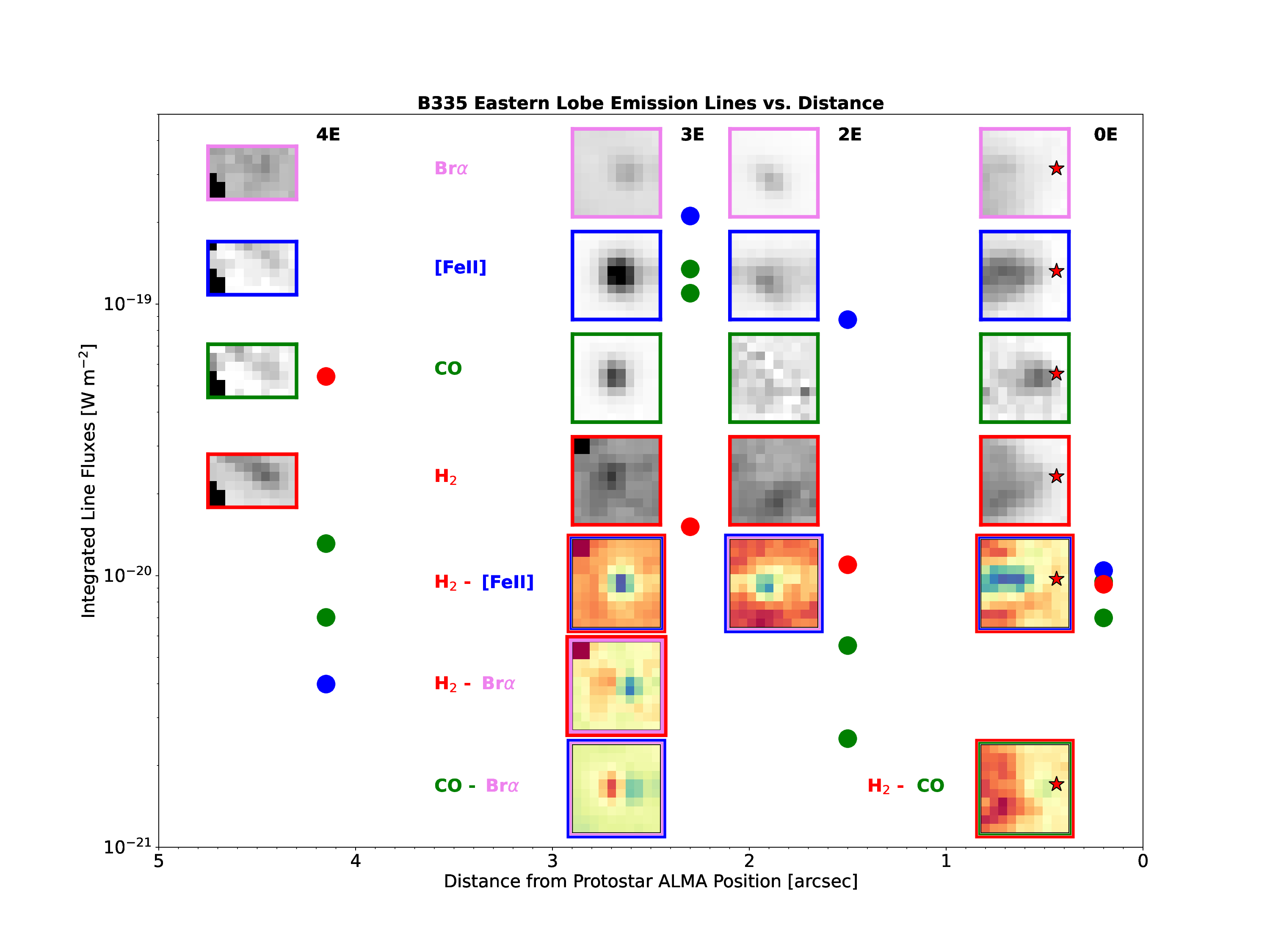}
        \caption{This figure shows both the line flux as large circles and the spatial distribution
    of the line flux as small image cutouts, for the four innermost shock regions in B335: 0E, 2E, 3E, and 4E. The shock front 4E was not covered in its entirety by the NIRSpec IFU data cube, so we only show the southern part of this shock. The color coding, both for the line flux symbols and for the image frames, is
    [\ion{Fe}{2}] (blue), CO (green) adjacent to the [\ion{Fe}{2}] and H$_2$ lines, and H$_2$ (red) lines.
The x-axis shows the distance from the ALMA position of the protostar, from right to left, so that the inserted cutout images
have the proper orientation relative to the axis. The inserted cutouts are integrated intensities over the these lines after subtraction of the continuum and of overlapping CO lines. Their vertical arrangement is in order of wavelength, and has no relationship to the flux scale. We also include Br $\alpha$ at 4.052~$\mu$m as image cutouts, but do not plot line fluxes, since its wavelength is shorter than those of the other lines and
extinction corrections would be substantial and are not precisely known.
We show some difference images in colored intensity scaling to document that high-excitation atomic
and ionic lines in the jet are emitted at different locations than relatively low-excitation molecular lines showing the structure of excitation conditions within each shock front.
                  \label{lines-v-distance}}
        \end{center}
\end{figure*}

Shock 3E is a very strong shock. [\ion{Fe}{2}] emission is strong, and so is CO,
indicating both an ionized component and high density in the molecular component of the shock.
In shock 3E, the position of the shock varies with the excitation of the emission lines.

The maximum of Br$\alpha$ is closest to the protostar.
The [\ion{Fe}{2}] emission is upstream from the H$_2$ emission, as is demonstrated in the color-coded difference image (Figure\ \ref{lines-v-distance}).
Emission from H$_2$ is relatively weak, peaks farthest downstream, and is more extended than [\ion{Fe}{2}].
It is probably excited by UV radiation from that shock, before being dissociated by it..
The [\ion{Fe}{2}] emission in such a strong shock is further enhanced by grain sputtering in the high-temperature environment of the shock, increasing the Fe abundance
in the gas phase.
We interpret 3E, which has a kinematic age corresponding to the onset of the present photometric outburst, as the shock front between the hot and fast
jet generated in this event and slower material ejected prior to that. The slower, older outflow material is largely molecular and was probably
ejected by a disk wind in the quiescent phase prior to the present accrection event.

The displacement of the
atomic and ionic lines from the molecular lines suggests that this newly ejected wind is shocking
against a slower, primarily molecular jet component. The actual shock produces UV radiation
that excites the molecular, primarily CO, emission ahead of this hot shock.
The location of the H$_2$ and CO emission regions downstream from the \ion{Fe}{2} emission
is not consistent with these molecules reforming behind the shock.

In Figure \ref{CO-radial-velocity}, we illustrate that shock 3E is unique in showing a strong blue-shift in its CO emission lines.
We computed the radial velocity relative to the average velocity of the inner emission regions on either side of the
protostar, 0E, 0Wn, and 0Ws, assuming that
this average is close to the radial velocity of the protostar. We used two wavelength ranges free of other strong emission lines
to determine the radial velocity by cross-correlation of the marginally resolved CO emission line pattern: 
4.77$\mu$m, to 4.88$\mu$m and 4.96$\mu$m to 5.048$\mu$m.
The error was calculated from the scatter of radial velocities calculated from
sub-sections of these wavelength ranges.
Relative to this average radial velocity, shock 3E has a
radial velocity of -74 $\pm$ 3 km\,s$^{-1}$.

In Figure~\ref{CO-radial-velocity}
the CO emission lines in shock 4E (violet lines) appear less blue-shifted than those of shock 3E (blue lines)
and correlation with the spectra of 0E, 0Ws, and 0Wn gives a radial velocity
of -28.5 $\pm$ 2.5 km\,s$^{-1}$ for 4Ec, the southern component in shock 4E (Table I), 
very different from 3E indeed.
Given that the radial velocity of shock 3E is peculiar in comparison with
the other shock fronts, the jet inclination calculated from this and the
proper motion gives a value that is inconsistent with 
\citet{Hirano.1988.ApJ.327L.69.B335.inclination}.
A possible explanation that will be explored in a separate paper on
the CO excitation conditions is 
that the $^{12}$CO ro-vib lines of shock 3E
could be optically thick and suffer from self absorption effects, 
so that the measured radial velocity
is dominated by the surface of the shock facing the observer. In shock 3E, like in all
other internal shock fronts in the jet, gas is being squeezed perpendicular to
the jet axis, expanding the shock front sideways. This is clearly demonstrated by
the morphological changes going from shock fronts 3E to 4E, 5E, 6E out to IR5.
The high blueshifted value of the 3E radial velocity probably just represents
the surface facing the observer of a laterally expanding shock front.
Therefore, the radial velocity of the bulk of gas in shock 3E is probably lower than
the measured value.
We have measured the ellipticity of the 3E shock front on the F444W images in 2023 with 0.20 and 2024 with 0.18.
This measurement does not support the notion of a sideways expansion of the shock front, but
the epoch difference of one year is probably insufficient to show this effect.
The average FWHM of the jet perpendicular to the jet axis in 2023 and 2024 is 50$\pm$4 AU giving
a rough estimate of the width of the jet launched during the current eruptive event.
This paper is deliberately limited to a qualitative discussion of the shock spectra.
A detailed analysis of the excitation conditions in the B335 jet, in particular shock 3E,
will be published by \citet{Kim.2026.inpreparation.B335CO} and a new, higher spectral resolution
NIRSpec IFU data set is being 
analyzed and will be published by \citet{LeGouellec.2027.in.prep.COexcitation}.

\subsubsection{Shock 4E}
For shock front 4E, only the southern component 4Ec was included in the
NIRSpec data cube from program 1802.
For shock front 4E, the cavity background was measured in a separate rectangular
aperture from the aperture integrating over the 4Ec component. The result is very
similar than for 3E: the cavity background (black line in Figure \ref{All-Shock-Spectra}) shows essentially a continuum and no CO emission,
and the spectrum of the shock is practically a pure emission line spectrum.
This spectrum confirms the much fainter
CO emission, compared to shock front 3E, and the very strong H$_2$ emission lines.

Shock 4E has very similar proper motion as shock 3E, which we interpret as the
bulk motion of the central jet in the outflow, with the shocks being the
result of perturbations in this flow.
However, the measured radial velocity is only marginally different for the average of the
regions near the protostar, i.e., 0E, 0Wn, and 0Ws. Shock 4E has a relative radial velocity
of -14 $\pm$ 7 km\,s$^{-1}$.
[\ion{Fe}{2}] is very weak. 4E is a low velocity internal shock downwind from the point of interaction (shock 3E) of the strong, fast, and hot wind generated
in the recent photometric event.
Shock 4E is much wider than 3E or 2E, possibly indicating that it is generated by
a wider wind than the well collimated jet producing 3E.
As already discussed in Section 3.4, we have one Spitzer photometry data point indicating
that in 2004, B335 was brighter than the minimum brightness observed prior to the present
accretion event, indicating a prior outburst. It is plausible that this prior outburst
has ejected the wind that shocked into slower material to produce shocks 4a, b, and c,
whose ejection dates is Table I are 2000 $\pm$ 2.

\subsubsection{Synoptic Discussion of the Excitation along the Eastern Jet}

In a synoptic overview of the B335 jet and outflow, we need to explain the following facts:
Aside from 0E,0Wn, and 0Ws,
the measured proper motions are quite uniform. 3E is the best measurement, 2E is difficult because
it is faint on the images, and 4E is difficult because of its complex structure.
All shocks discussed here are internal shocks within the outflowing material, and their internal shock
velocity is much lower than the proper motion.
In the cavity-background-subtracted spectra in Figure
\ref{All-Shock-Spectra},
there is a pattern of decreasing excitation going from the youngest shock fronts, 2E and 3E, towards 4E.
Hodapp et al. 2024 (Paper I), based on slitless spectroscopy, already pointed out that
shock 3E shows H$_2$, CO and [\ion{Fe}{2}] emission, the older shocks further downwind (4E, 6E, and 7E) show predominately
H$_2$ emission.
The subtraction of the continuum foreground and background of the outflow cavity
is an unavoidable side effect of the spectrum extraction in slitless spectroscopy used in Paper I
so the results in Paper I are directly comparable to the background-subtracted spectra presented here.

It is particularly interesting to discuss the different chemical composition
of the shocks 2E and 3E, that in spite of being relatively close to each other (~0.8" = 132 AU)
exhibits different atomic and molecular emission line species.
Notably, [NeII], [NeIII], and [ArII] emission lines
(at 6.99 $\mu$m, 12.81 $\mu$m, and 15.56 $\mu$m, respectively)
are bright toward 2E in the JWST MIRI data of B335
(Program ID 1802, \citep{Federman.2026.ApJ...998..282.MIRIjet})
and dim toward 3E.
The [NeII], [NeIII], and [ArII] lines are $\approx$2-3, $\approx$2, and $\approx$3-4 times brighter in 3E than in 2E, respectively.
The ratio [NeIII]/[NeII] $\approx$ 0.05 could be explained by a UV irradiated environments,
\citep{Hollenbach.2009.ApJ...703.1203.linediagnostics}, with the UV originating in
the accretion disk of the protostar.
Alternatively, the line ratios can be explained
with a very high internal shock velocity following
\citet{Hollenbach.1989.ApJ.342.306}.
Comparing the [NeII]@12.81um and SI@25.25um line intensities with the shock model grid of
\citet{Hollenbach.1989.ApJ.342.306}
indicates a pre-shock density of $\approx$10$^4$ cm$^{-3}$ and a shock velocity of $\approx$90 km s$^{-1}$
(see also
\citet{Tychoniec.2024.AA...687A..36.TMC1}, \citet{CarattioGaratti.2024.AA...691A.134.HH211}, and
\citet{vanDishoek.2025.AA...699A.361.JOYS})
for recent usage of this grid to constrain protostellar jet shock parameters from JWST IFU data).

The identified shock front proper motions do not measure the velocity
of the ionized jet ejected during the current accretion burst. What we see is the velocity
of internal shock fronts where the velocity of the jet changes abruptly. In case of 2E and 3E,
our working assumption is that 3E is the forward shock where the current jet is decelerated to
the velocity of the pre-outburst molecular material. 
Due to its faintness in the F444W images, the proper motion of shock 2E could not be
measured with confidence and was not included in Table I. Future proper motion measurements
of shock 2E will be able to decide whether 2E is a forward or reverse shock,
where the
jet hits the jet material already decelerated, in shock 3E, to nearly the velocity of the molecular
pre-outburst outflow. If 2E is a forward shock, this would imply that the jet has increased its
velocity during the course of the ongoing accretion event.

The stringent difference between the 2E and 3E shocks is that no molecules are detected toward 2E
while 3E exhibits bright H2 and CO ro-vibrational emission lines,
despite originating from a similar launching region, where the initial dust content is expected to be comparable and small.
In contrast, the lack of molecular emission in 2E may reflect either an absence
of dust and molecular material or conditions unfavorable for molecule survival.
The formation of H$_2$ on dust grains is the most efficient chemical pathway
and is essential for subsequent CO formation
\citep{Hollenbach.1971.ApJ...163..155.H2formation},
\citep{Thielens.1985.ApJ...291..722.photodissociation}
We note, however, that re-formation of H$_2$ behind shocks takes much longer than
the very short timescales of only a few years involved in the very recent ejection
of the 2E and 3E shocks in B335. Following \citet{Neufeld.1989.ApJ...340..869.reformation}
timescales for H$_2$ re-formation range from 10$^2$ to 10$^5$ years, depending on density,
but are longer than a few years for any density below 10$^7$ cm$^{-3}$, in cold molecular
clouds.

Alternatively, we consider the scenario that the molecular material that shock
3E runs into has been ejected in the quiescent phase prior to the present accretion
burst under conditions that allowed the survival of H$_2$ and CO molecules in the wind.
The fast, ionized jet launched during the accretion burst is now running into this
pre-existing molecular material.
The faster atomic and ionic material ejected during the outburst catches up with slower molecular material ejected in the quiescent phase
of the protostar prior to the present outburst,
producing a two-shock structure:
a forward shock (3E) driven into the slower material ahead and producing H$_2$ and CO emission in the molecular material by UV excitation,
and, possibly, a reverse shock where the ionic gas from the jet hits the gas decelerated in the forward shock, producing a shock in mostly ionized
gas, as described in the dual shock model by \citet{Raga.1990.ApJ...364..601.jetsvariablesources}.
Observational evidence for such reverse shocks was found
in HH34 by \citet{Reipurth.1992.AA...257..693.HH34} and then confirmed
with HST imaging by \citet{Reiputh.2002.AJ....123..362.HH34HST}.
The Mach disk in HH34, consisting of knots D, E, and F in their terminology,
is $\approx$8\arcsec behind the tip of the bow shock, with HH34 assumed to
be at the distance of 414$\pm$7 pc measured by
\citet{Menten.2007.AA.474.515.Oriondistance} for the Orion GMC.
More evidence for reverse shocks was found in HH 1-2 by
\citet{Bally2002-HH1-2} and in HH47 and HH34 by \citet{Hartigan.2011.ApJ...736...29.Machdisks}.
Based on Keck I telescope imaging in 1996,
\citet{Hodapp.1998.ApJ.500L.183.B335}
had identified the double shock front near the B335 protostar (now called shock 6E), as a bow-shock and Mach disk system.
We note that in Paper I, we had discussed the evolution of shock front 6E that changed its appearance
from a double shock observed in 1996 to a single shock in 2023, where the two shocks had apparently
merged. In the scenario of Rage et al. 1990, such an eventual merging of the front and back shock
is plausible. We expect that shocks 3E and 2E will similarly merge in the decades to come.

Shock 4E, whose emission date is earlier than the present photometric outburst, shows even lower
excitation, and is dominated by molecular emission. In the light curve, the oldest Spitzer data
point from 2003 indicates that the B335 protostar underwent another outburst, and this outburst
can plausibly be associated with shock 4E. We don't know the amplitude or duration os this earlier
outburst, and therefore do not know whether it was as large and long as the present one.
The fact that molecular material is present behind shock 4E and ahead of shock 3E would
indicate that the wind ejected during this earlier event was largely molecular.

In the western, redshifted outflow lobe, the excitation of the shock fronts is
remarkably different than in the eastern, blueshifted lobe. In particular, the
western shock 3W, having the same ejection date as the strongest shock 3E in the east,
only shows some [\ion{Fe}{2}] emission, indicative of a high-excitation ionized environment, but no molecular emission from H$_2$ and CO.

The fact that we see in the spectra of 3E, 4E and 6E a consistent decrease of CO emission
with the distance separating the shock from the protostar suggests that
such decrease is related to either the dynamical age of the shock, or the distance from the protostar, or both.
With the age of the shock front increases its sideways expansion,
which decreases the momentum flux deposited per working surface area,
and thus the equivalent shock velocity decreases and with it the shock kinetic energy.
This results in cooler and less dense shocks with time.
However, CO fundamental lines needs a large density (critical density of $\approx$10$^{11}$  cm$^{-3}$)
to be excited by collisions, or a strong radiation field to IR-pump the vibrational $\nu \geq 1$ levels. Evidence for IR-pumped CO lines in S68N was presented by 
\citet{LeGouellec.2025.ApJ...985..225.S68N}.
We just showed that the density decreases with distance and so does the shock temperature, and thus IR radiation continuum.
Adding to that the fact that the protostar-induced IR irradiation decreases
at least with the square of the distance separating the shock from the protostar,
we have several physical arguments explaining why emission of CO fundamental lines
decreases with distance from the protostar.
And finally, the pre-shock density
decreases with distance from the protostar within the cavity,
since the outflow angle is conical and further expansion occurs in the cavity.

\subsection{The Outflow Cavity Nebula}

The B335 protostar is not directly detected \citep{Stutz.2008.ApJ.687.389.B335.Spitzer} 
at wavelengths up to 5.3$\mu$m that we discuss here.
In addition to the individual shock fronts discussed above, B335 is a bipolar nebula, with its
eastern, blue-shifted lobe being more prominent. In the context of the variability of the B335 
bipolar nebula, we have already established that the rapid variability and the local shadow effects indicate that a substantial
fraction of the bipolar nebula is reflected light. 
The extracted line images in 
\citet{Federman.2024.ApJ.966.41}
show that H$_2$ emission lines outline an outflow cavity that is broader than the
jet, but narrower than the continuum reflection nebula.

\begin{figure*}[h]
\begin{center}
	\includegraphics[angle=0.,scale=0.40]{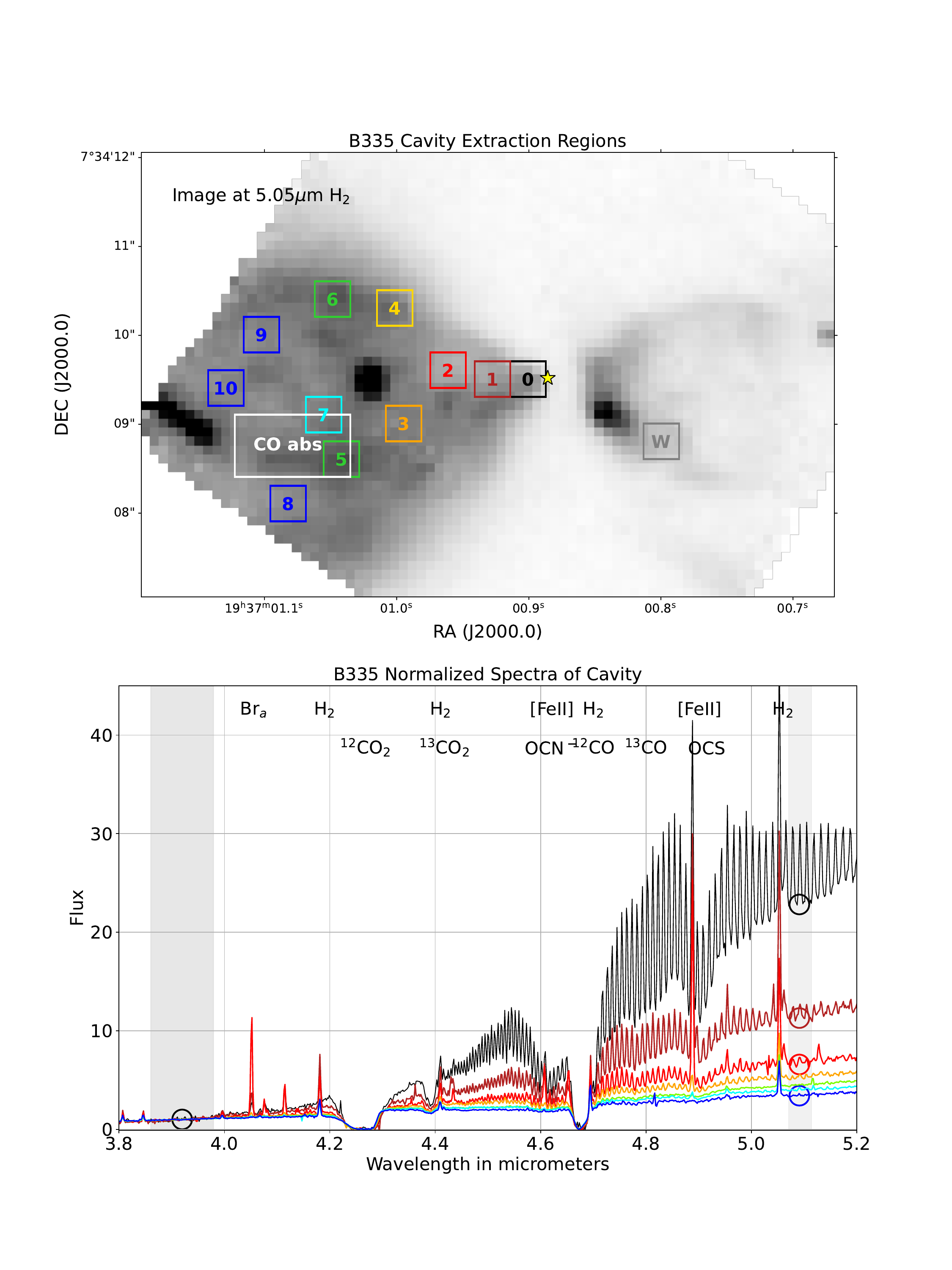}
	\caption{Top Panel: Extraction apertures in the reflection nebula other than shock fronts.
                 The image is the NIRSpec IFU data cube slice at a wavelength of 5.053 $\mu$m, centered on an H$_2$ emission line.
                 The extraction boxes are color coded (blue-green-yellow-red-brown-black) in a sequence suggesting increasing extinction, consistent with the coloring of the spectra in the bottom panel. The regions in yellowish, greenish, and blue were
                 averaged for these spectra in to improve the S/N ratio. The white rectangle indicates the extraction region where we found CO gas absorption. All other regions show CO in emission. The position of the protostar at the
                 epoch of the NIRSpec observations is indicated by a yellow star symbol.
                 Bottom Panel: The spectra extracted from the reflection nebula 
                 positions outside of known jet
                 shock fronts were normalized in an emission line free region 
                 from 3.86 to 3.98 $\mu$m, indicated by
                 an open circle.
                 The spectra show strong, saturated ice absorption features of $^{12}$CO$_2$ and $^{12}$CO,
                 and the non-saturated feature of $^{13}$CO$_2$. 
                 Closest to the protostar, in particular in Knot 0E,
                 the spectrum is dominated by strong CO gas emission. 
                 The overall slope of the spectra indicates rapidly increasing
                 extinction close to the protostar.
                 \label{Cavity-extraction-regions-and-spectra}}
\end{center}
\end{figure*}

\begin{figure*}[h]
\begin{center}
	\includegraphics[angle=0.,scale=1.0]{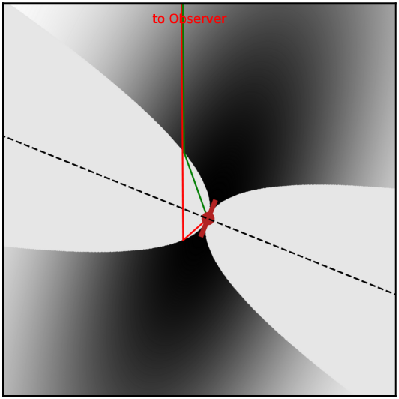}
	\caption{A schematic cross-section illustrating the scattering geometry for light scattered
             on the outflow cavity walls, and then suffering extinction in the outer regions
             of the flattened molecular core surrounding the protostar. The direction toward the observer is vertically to the top. The scattered light in the blue-shifted
             cavity does not pass through the equatorial plane of this flattened molecular core, nor through
             any protoplanetary disk surrounding the protostar, the latter being schematically indicated in brown color.
             The jet axis of the outflow near the symmetry axis of the outflow cavities is indicated as a dashed line.
             \label{B335-geometry-sketch}}
	\end{center}
\end{figure*}

Based on the results by \citet{Hirano.1988.ApJ.327L.69.B335.inclination} the B335 outflow axis is close to the plane of the 
sky (inclination $<$ 10$\arcdeg$), and the protostar is therefore seen through a nearly edge-on disk as we illustrate
in the schematic cross-sectional sketch in Figure~\ref{B335-geometry-sketch}.
In order to quantify the rapid increase of the extinction close to the protostar,
we have selected a set of extraction apertures in the blue-shifted outflow cavity
that are not associated with any shock fronts or shock-excited emission knots and
identify them in Figure\ \ref{Cavity-extraction-regions-and-spectra}
superposed on an image extracted from the NIRSPEC data cube, in colors matching those
used for the spectra in the bottom panel of this figure.
Position C0 is identical to shock front 0E, which we have determined not to be part of the
jet, on the basis of its low proper motion and different radial velocity.
Positions C1 and C2 are at the same distance from the protostar as shock fronts 1E and 2E, but
are north of those shock fronts and in general, are away from the jet axis.
Cavity positions C1 and C2 still show CO bandhead emission, but with decreasing strength.
Cavity positons C3 and C4 (yellow), are away from the jet axis, at distances slightly closer to the
protostar than shock 3E. Cavity positions C5, C6, and C7 (green) are removed from individual shock fronts
at a distance larger than shock 3E. The combined spectral traces in yellow show very little CO emission,
while the green averaged spectrum shows essentially none. All these cavity positions show H$_2$ emission
lines, with diminishing strength with increasing distance.

At the positions closest to the protostar, we observe dominant CO emission in the spectrum of
position 0E. In 1E and 2E, the CO lines are roughly equally strong in the jet than in adjacent
``background'' positions, which essentially is our finding that those shocks themselves 
are free of molecular emission.
The CO lines at these positions are not emitted in the jet, but in a
wider outflow cone of similar opening angle to the H$_2$ emission cone.
With the low (R$\approx$1000) spectral resolution, we cannot cleanly separate CO line emission, or
CO line absorption, from the continuum. Using the minima between the CO lines as a proxy for
the continuum, we qualitatively find that
in those positions (1E and 2E),
the CO emission appears fainter relative to the continuum than in 0E.
The continuum is mostly
reflected light from near the protostar, based on the shadow effects and the variability.
A strong contribution from local thermal continuum emission
from dust heated in or near the region of H$_2$ shock emission could not explain the rapid
variability of the flux.
We will discuss CO gas emission and absorption in more detail in Section 3.7.3.

\subsubsection{Extinction in front of the Reflection Nebula}

The NIRSpec IFU spectra integrated over these extraction
apertures are shown in Figure\ \ref{Cavity-extraction-regions-and-spectra}. Since we are interested in the shape of the continuum
spectrum, and not in the absolute flux, we normalize all spectra to unity at an
effective wavelength of 3.92 $\mu$m. For clarity, the spectra are color coded in the
same way as the apertures in Figure\ \ref{Cavity-extraction-regions-and-spectra}, 
in a wavelength sequence of color going from violet
for the most distant to brown and black for the closest extraction apertures.
It is clear from Figure\ \ref{Cavity-extraction-regions-and-spectra} that the spectra are much steeper closer to the protostar.
The spectra closest to the protostar show strong CO emission around 5 $\mu$m on top
of a continuum spectrum. We use the minimum flux between a set of five CO emission lines
at an effective wavelength of 5.09 $\mu$m as a measure of the continuum flux in this region.
The continuum points extracted this way are shown as open circles in Figure \ref{Cavity-extraction-regions-and-spectra}.
From the flux ratio between 5.09 and 3.92 $\mu$m we compute the A$_V$ using the
infrared extinction law by
\citet{Wang.2019.ApJ.877.116.Extinction}. The results are nominal extinction values
that incorrectly assume a flat spectrum of the protostar illumination. However, it is a
reasonable approximation that the spectral energy distribution of the light emerging
from 
the protostar and then illuminating the outflow cavity is the same for all positions in the cavity and therefore is the same each of the extraction
apertures. 
Relative extinction values between the different apertures can therefore be determined
with better certainty.
We measure an increase of extinction by A$_V$ = 250 mag from the most distant extraction aperture 3\farcs5 from
the protostar to the closest, 0\farcs2 from it.
A caveat and the reason for stating this as an approximation is that we have evidence
for local and time-variable absorption effects from the time variability of the cavity flux shown in
Figures \ref{Shock-PM} and \ref{Variation-Crosscut}.

\clearpage

\begin{figure*}[h]
\begin{center}
	\includegraphics[angle=0.,scale=0.30]{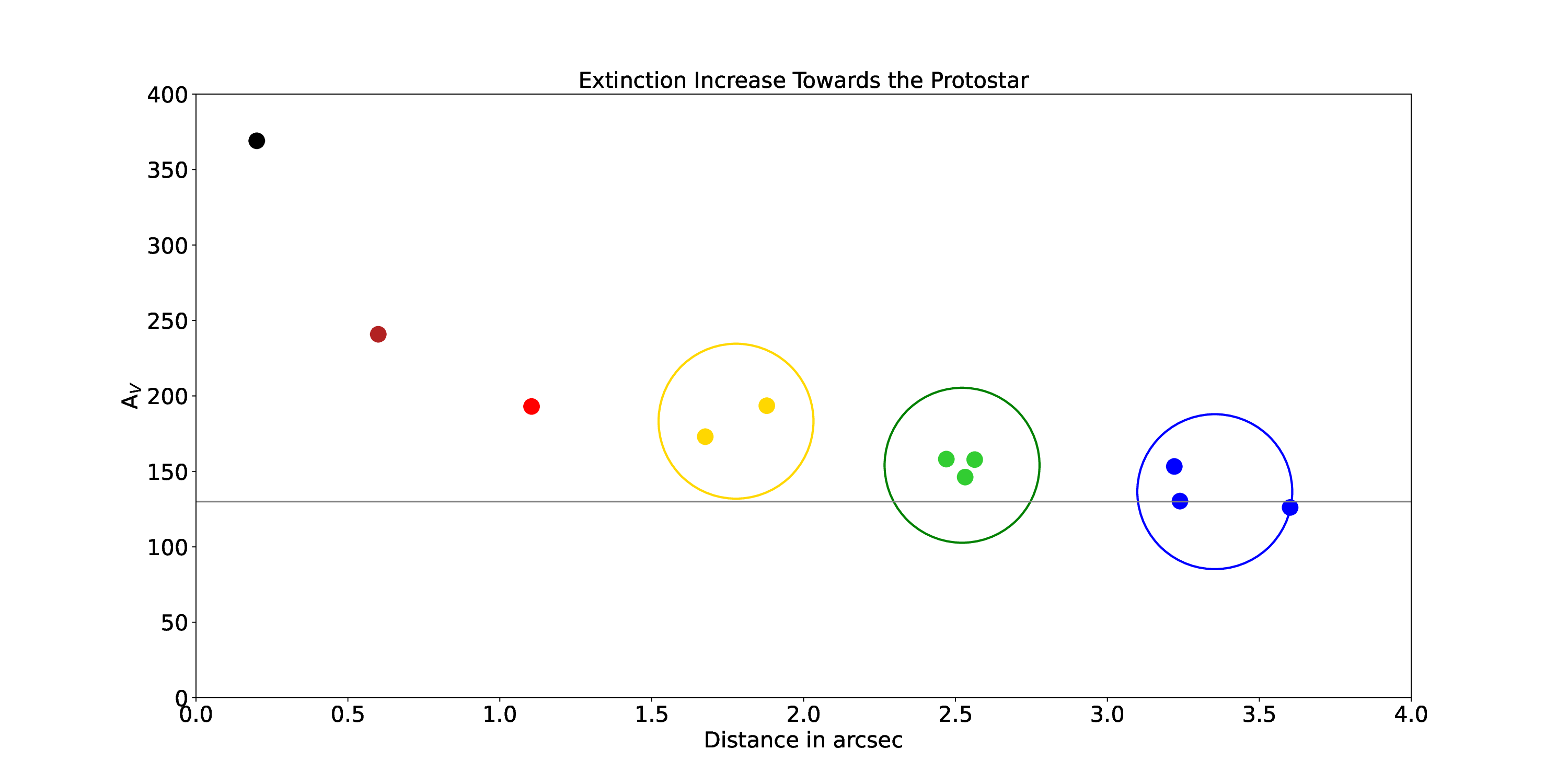}
	\caption{Plot of the A$_V$ in the selected apertures in the blue-shifted 
                 outflow cavity walls (Figure \ref{Cavity-extraction-regions-and-spectra}), dependent  on the distance from the protostar.
                 The extinction was determined from the normalized spectra in Figure \ref{Cavity-extraction-regions-and-spectra}
                 between 
                 the shaded intervals
                 at effective wavelengths 3.92 and 5.09 $\mu$m
                 indicated by open black circles. At the long wavelength point, the minima in the
                 CO line forest were assumed to represent the continuum.
                 While the illuminating spectrum from the protostar is unknown, the assumption
                 is that every point in the reflection nebula is illuminated by the same spectrum, i.e.,
                 that the continuum extinction in the outflow cavity is negligible.
                 Therefore, while the absolute value of the extinction is uncertain, the increase with 
                 position towards the protostar can be measured. The line at a level of A$_V$ = 130
                 indicates the baseline at a distance of 3\farcs5 from the protostar.
                 Therefore, we measure an increase of extinction of about 200 magnitudes of
                 A$_V$ from that position to the closest observed feature near the protostar.
                 We also indicate the groups of positions that were averaged for the analysis of the
                 ice features by large circles, with their color corresponding to the ice absorption
                 spectra in Figure \ref{Norm-ice-spectra}.
                 \label{Distance-AV}}
	\end{center}
\end{figure*}

We have extracted continuum spectra of the cavity positions defined in Figure \ref{Cavity-extraction-regions-and-spectra}
by using the minima between the individual CO emission lines as a measure of the 
continuum. We are going to refer to this spectrum as the CO-line-minimum spectrum.
This procedure uses only a fraction of the data points in the
original spectra, and therefore is undersampling narrow emission lines.
However, the broad absorption features of ices are adequately sampled.
We extracted this continuum and ice absorption spectrum for each of the
cavity positions. For clarity of presentation, and illustrated in Figure \ref{Distance-AV}, 
we have averaged the
spectra in cavity positions 3 and 4, the result being shown in red in Figure \ref{Norm-ice-spectra}
cavity positions 5, 6, and 7, the result being shown in green,
and positions 8, 9, and 10, the result being shown in blue.
For wavelengths below 4.4 $\mu$m, we show the normalized spectra
in their original sampling, since there are only a few well separated
emission lines in this wavelength range.
The cavity position numbers correspond to increasing distance from
the protostar and the color coding in Figures \ref{Cavity-extraction-regions-and-spectra}, \ref{Distance-AV}, \ref{Norm-ice-spectra}, 
and \ref{Tau-ice-vs-distance} is intended to suggest
increasing extinction. All the spectra in Figure \ref{Norm-ice-spectra} were normalized by
the median of the CO-line-minimum spectrum between 4.45 and 4.55 $\mu$m.
We again note the trend of decreasing slope of the continuum with distance
from the protostar. We also note that all ice absorption features shown here
become more shallow with increasing distance.

\subsubsection{Ice Absorption}

\begin{figure*}[h]
\begin{center}
	\includegraphics[angle=0.,scale=0.30]{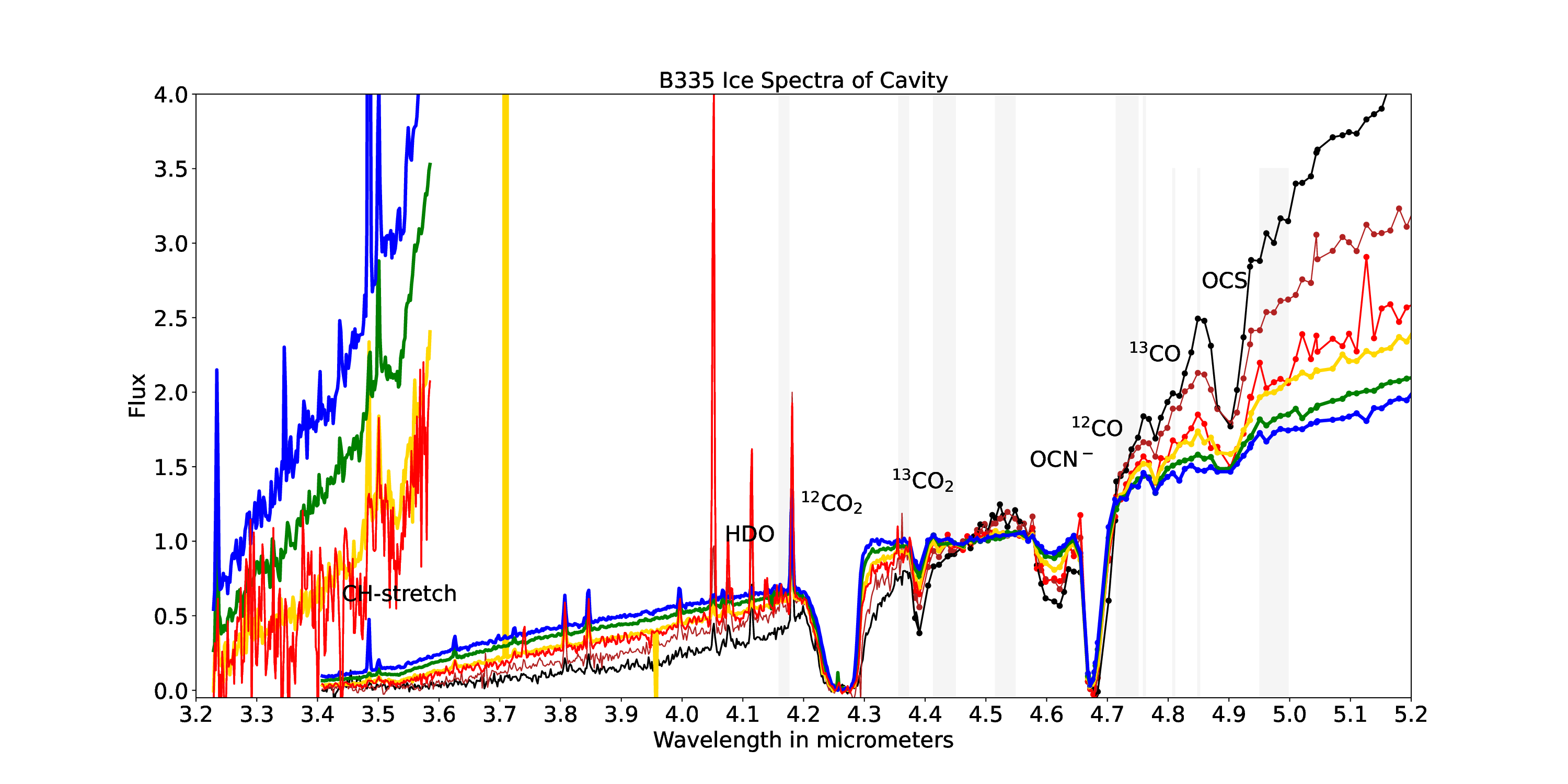}
	\caption{The spectra extracted from the reflection nebula positions 
    (Figure \ref{Cavity-extraction-regions-and-spectra}) outside of known
shock fronts were normalized in an emission line free region from 4.45 to 4.47 $\mu$m.
For wavelengths longer than 4.4~$\mu$m, where CO gas emission bands are strong, the continuum
spectrum was approximated by interpolating through the minima between CO lines, leading effectively
to a reduced spectral resolution that was, however, still adquate to show the broad ice absorption
features.
Averaged spectra from several extraction positions are indicated
by thicker lines. 
The band of H$_2$O ($\sim$3.0\,$\mu$m) is saturated and marks the lower limit of
useful data. At the short wavelength end, we show the same data expanded by
a factor 20 to bring out the faint signature of the C-H stretch at 3.53 $\mu$m. CO$_2$ ($\sim$4.27\,$\mu$m) ice is saturated at all positions, as is the CO ice band ($\sim$4.67\,$\mu$m).
The weaker ice features of OCN$^-$ ($\sim$4.60\,$\mu$m), OCS ($\sim$4.90\,$\mu$m), $^{13}$CO$_2$ ($\sim$4.38\,$\mu$m), and $^{13}$CO ($\sim$4.78\,$\mu$m) are not saturated.
\label{Norm-ice-spectra}}
	\end{center}
\end{figure*}

Figure \ref{Norm-ice-spectra} shows that the absorption features of
$^{12}$CO$_2$ and $^{12}$CO ice are strong and saturated in all
outflow cavity positions. $^{13}$CO$_2$, OCN$^-$ (cyanate anion) and
OCS (carbonyl sulfide) ices are convincingly detected and their depths
increase towards the protostar.  $^{13}$CO ice is weakly detected.
The broad absorption of HDO ice at 4.1$\mu$m
\citep{Slavicinska.2024.AA.688A.29.HDO} is tentatively detected 
as a weak, broad depression relative to the interpolated continuum in the spectrum of
region 0E, the position closest to the protostar, and only there.
The continuum
emission declines strongly at shorter wavelengths, and that portion in
Figure \ref{Norm-ice-spectra} (enhanced by a factor of 20 for clarity)
shows that the absorption feature of the C-H stretch mode is detected
in the three cavity positions farther away from the protostar
($>$1\arcsec). In the context of YSOs, this C-H stretch signature is
commonly identified as CH$_3$OH, laboratory spectra of which were
shown in \citet{Dawes.2016.PCCP.18.1245}, but our data are of
insufficient quality to support this specific identification.
Finally, while the NIRSpec data cubes cover the wavelength range of
the broad H$_2$O ice feature near 3.0 $\mu$m, no usable data were
obtained of this feature due to the high column density and saturation
of the absorption as well as the faint continuum.

For all these ice features, we computed optical depth spectra relative
to a local continuum and determined integrated optical depths (in
units of cm$^{-1}$). The resulting integrals were divided by the
integrated laboratory band strengths ($A$) to give column densities in
units of molecules cm$^{-2}$.  We used the $A$-values from
\citet{Bouilloud.2015.MNRAS.451.2145.waterice.A} for $^{13}$CO$_2$.
For OCN$^-$ we used $A$ = $1.51\times10^{-16}$ from
\citet{Gerakines.2025.MNRAS.537.2918.OCN-}, and for OCS
$1.20\times10^{-16}$ from \citet{Slavicinska.2025.OCS}.

The NIRSpec data cube was used to produce ice band optical depth maps
across the entire outflow cavity (Figure~\ref{Ice-map}) and also to
study the ice band profiles at specific positions in more detail
(Figure \ref{Norm-ice-spectra}). The spatial distribution of the ice
band optical depths as well as the continuum extinction are shown in
Figure~\ref{Ice-map} on a per-pixel basis. We omitted the area of the
strong shock 3E from the discussion because it is dominated by gas
emission lines and the derived optical depths are unreliable. Two
general trends are visible. First, the ice absorption and dust
extinction strongly increase towards the protostar, as the line of
sight passes through denser regions in the flattened molecular core,
as schematically illustrated in Figure~\ref{B335-geometry-sketch}. And
second, it is apparent that the absorption is concentrated outside of
the outflow cavity. To highlight this spatial relationship, we show an
image of the shock-excited H$_2$ emission generated at the interface
between the outflow cavity and the ambient molecular core material. We
conclude that the outflow cavity shows less absorption both in
continuum dust extinction and in specific ice features compared to the
surrounding molecular core, indicating a reduced dust density in the
outflow. This is consistent with the results in Figures~\ref{H2O-AV-map}
from the photometric H$_2$O ice column density
maps on larger spatial scales, and the appearance of the outflow
cavity in the scattered light (coreshine) images in
\ref{cloudshine-image}. Put together, three largely independent
methods indicate that the outflow cavity has lower dust extinction and
lower absorption in all our observed ice features: background star
photometry, scattered light imaging at large distances, and scattered
light spectroscopy on short distances from the protostar.  We believe
this is due to the higher temperature of the outflowing gas, compared
to the ambient molecular core material, both from its launch
conditions near the protostar and continual heating from shocks. Under
such elevated temperatures, dust, and certainly its ice mantles, are
destroyed during outflow launch, and will not recondense in the
outflow cavity.

The absorption feature of $^{12}$CO$_2$ near 4.27 $\mu$m is saturated
in all the spectra recorded here (Figure \ref{Norm-ice-spectra}), but
the long-wavelength wing strength depends on the distance from the
protostar and therefore on total extinction.  The position closest to
the protostar appears to show some evidence of the scattering peak
short of the saturated $^{12}$CO$_2$ feature as well.  These wings of
the $^{12}$CO$_2$ ice feature have been modeled by
\citet{Dartois.2022.AA.666A.153.CO2ice.profiles} as scattering by dust
mixtures containing grains of up to several micrometers in size, and
thus they are evidence of grain growth in dense environments.  Both
the short and long-wavelength wings have been observed in other
protostellar sources. Of the sources discussed by
\citet{Dartois.2022.AA.666A.153.CO2ice.profiles}, NGC7538 IRS9 has the
strongest features, but these are not as pronounced as in the B335
lines of sight.

The $^{12}$CO ice feature near 4.67 $\mu$m is saturated closest to the
protostar, but appears to just come out of saturation for the less
obscured lines of sight (Figure \ref{Norm-ice-spectra}). From the
scattering models in \citet{Dartois.2022.AA.666A.153.CO2ice.profiles},
asymmetries between the short and long wavelength wings of the main
absorption are expected here as well. However, blending with other ice
features (OCN$^-$, CO mixed with polar molecules) complicates
detections of these possible scattering asymmetries.

The $^{13}$CO$_2$ ice feature is not saturated in any of our spectra,
and its depth is strongly dependent on the distance to the central
star.  Based on the same data,
\citet{Brunken.2024.AA.685A.27B.B335.13CO2ice} presented a detailed
analysis of the $^{13}$CO$_2$ feature in a larger aperture centered on
the protostar position, i.e., comprising features 0E, 0Wn, and 0Ws in
our terminology. They find that the strongest absorption component is
from $^{13}$CO$_2$ mixed with H$_2$O, followed by $^{13}$CO$_2$ mixed
with CH$_3$OH, with minor contributions from $^{13}$CO$_2$ mixed with
CO of two different temperatures (15K and 25K), and finally a very
small contribution from pure $^{13}$CO$_2$ at 80K. Given the viewing
geometry where the protostar itself is behind a close to edge-on disk,
it is not surprising to find contributions from ices with a
substantial range of temperatures.

\begin{figure*}[h]
\begin{center}
	\includegraphics[angle=0.,scale=0.30]{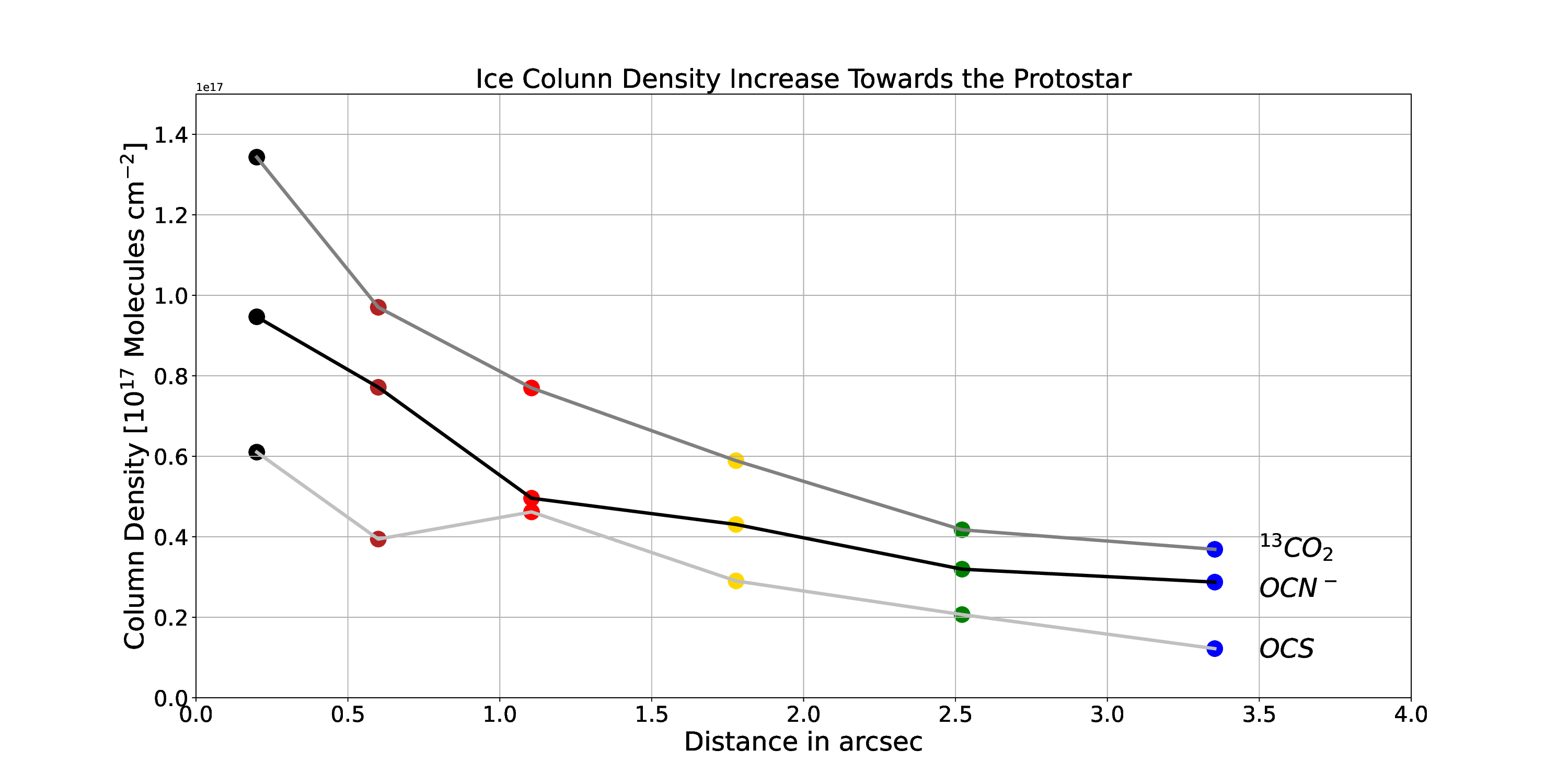}
	\caption{Ice column density vs. distance from protostar for three ice species
    $^{13}$CO$_2$, OCN$^-$, and OCS that are not saturated in the B335 cavity and have
    sufficient strength for a significant detection. 
\label{Tau-ice-vs-distance}}
	\end{center}
\end{figure*}

\begin{figure*}[h]
\begin{center}
	\includegraphics[angle=0.,scale=0.40]{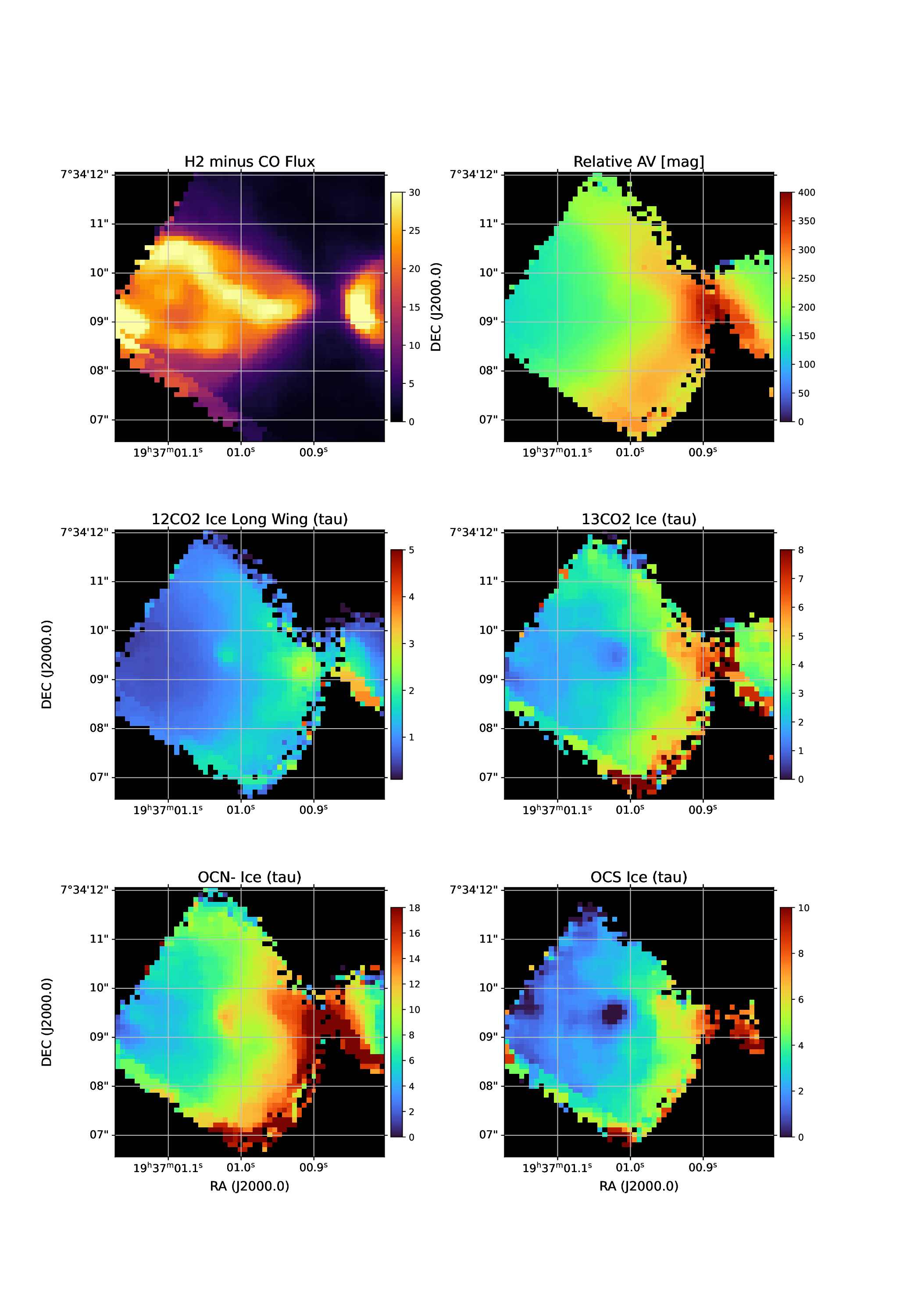}
	\caption{Maps of the dust absorption near the B335 protostar based on the NIRSpec IFU
    data cube. The top left panel is a flux image of shock-excited H$_2$ emission, corrected for
    CO gas emission. It shows the shocked interface between the outflow cone and the surrounding
    molecular cloud material and is used for comparison with the dust absorption maps.
    The top right panel shows the continuum extinction in the scattering path in units of A$_V$, 
    with an unknown contribution from the color of the protostar light, as explained in the text.
    The lower four panels show the distribution of the optical depth in the scattering path for
    the $^{12}$CO$_2$ long-wavelength wing, the $^{12}$CO$_2$ feature, and the OCN- and OCS features.
    In all panels, the position of the strong shock 3E is standing out. Due to the very strong line
    emission in this shock, the data at this position are not reliable. Near the limits of the outflow
    cavity, the fluxes are very low and the extracted ice optical depths are therefore very uncertain.
    In particular, while we show the map in its entirety, we do not believe features on the jet axis
    can be trusted.
    All absorption features show a general trend of more absorption closer to the protostar, and 
    stronger absorption in an envelope outside the H$_2$ emission cone, and less absorption in this
    cone itself.
\label{Ice-map}}
	\end{center}
\end{figure*}

As is shown in Appendix C, the strong detection of the OCS and OCN$^-$
ice absorption bands against the scattered continuum emission allows
for a detailed study of the ice band profiles across the B335 outflow
cavity. The analysis of the CO ice band is limited by it being
saturated. The OCS band has a Gaussian peak position and FWHM of 4.900
$\pm$ 0.008 $\mu$m and 0.04 $\pm$ 0.01 $\mu$m, respectively, for all
extracted positions across the cavities. This is comparable to the
bulk of the massive YSO targets studied in Figure~13 of
\citet{Boogert.2022.ApJ.941.32B.OCN}, and consistent with OCS embedded
in CH$_3$OH-rich ices, and proton-irradiated H$_2$S- or
SO$_2$-containing ices. While the column densities increase towards
the protostar (Figures \ref{Norm-ice-spectra} and
\ref{Tau-ice-vs-distance}), the OCN$^-$/OCS column density ratio is
constant across the outflow cavities, with a mean of 4.0 and a
standard deviation of 0.4 (see also
Figure~\ref{Tau-ice-vs-distance}). This is $\sim$50\% lower compared
to massive YSOs \citep{Boogert.2022.ApJ.941.32B.OCN}.  Overall, other
than column densities, the ice characteristics seem to depend little
on the varying conditions in the B335 environment, and mildly at most
when compared to deeply embedded massive YSOs with orders of magnitude
larger luminosities.  An astrochemical discussion of the origin of
these similarities and differences is beyond the scope of this paper.
The lack of strong signs of ice processing by the
protostellar radiation (such as strong crystalline $^{13}$CO$_2$ ice
and a weak feature of solid CO, the most volatile ice) is likely
because most absorption is seen against light scattered out of the
outflow cavities and then passes through substantial column densities of dust
and ice from the dense envelope (Figure~\ref{B335-geometry-sketch}).

\subsubsection{CO Gas Emission and Absorption}
\begin{figure*}[h]
\begin{center}
	\includegraphics[angle=0.,scale=0.50]{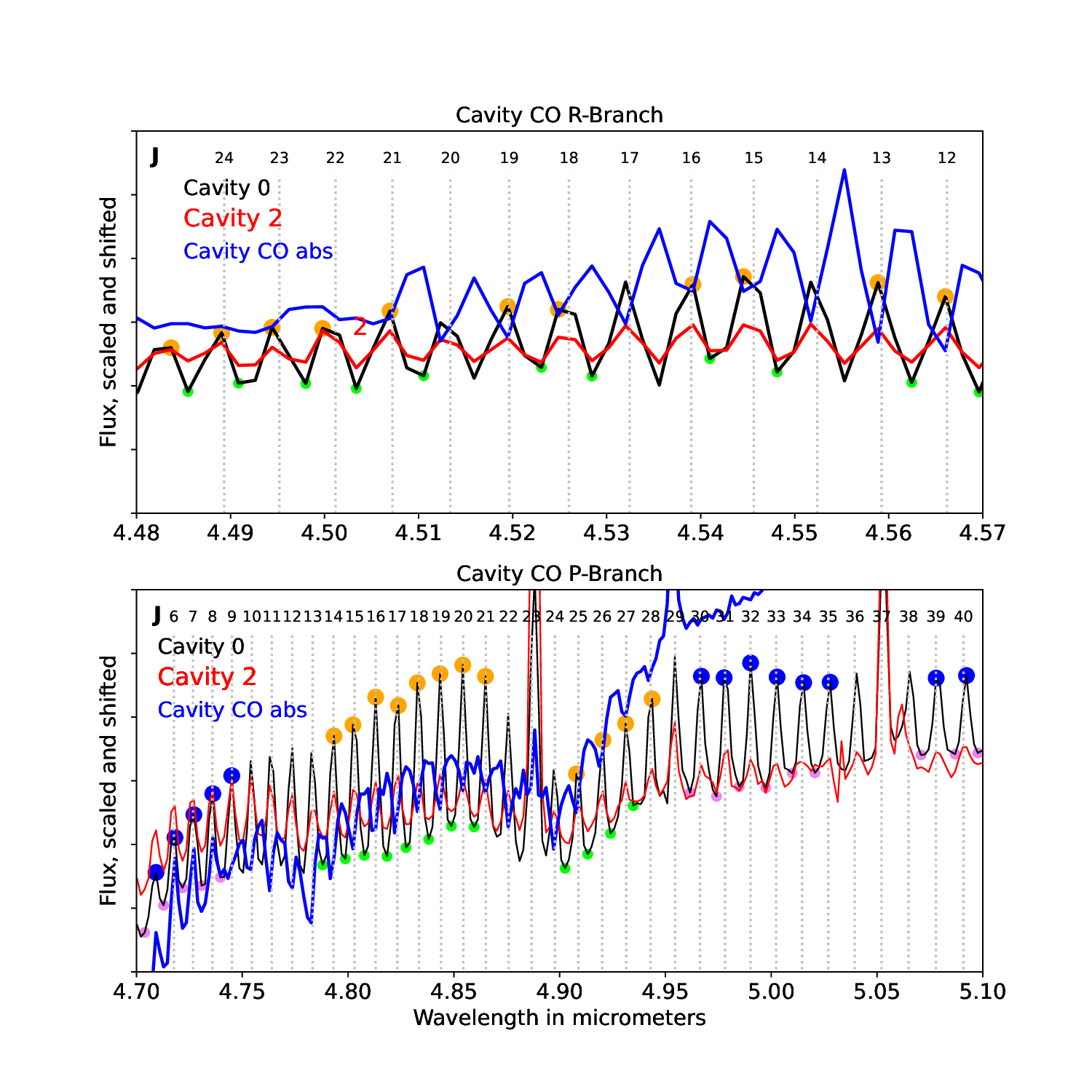}
	\caption{For different integration apertures, this figure shows spectra over a small wavelength range to illustrate both the CO emission
and CO absorption in different regions of the outflow cavity. The spectra are in units
of flux, but are both scaled and shifted for clarity. The CO line laboratory wavelengths
are labelled with their J-number and indicated by dotted lines.
Top Panel: CO R-branch emission and absorption.
Bottom Panel: CO P-branch emission and absorption. 
Large orange circles indicate CO lines where absorption is observed in the ``CO abs'' position and
that were used in the mapping of the absorption. In the bottom panel, blue circles indicate
CO lines where no absorption was observed.
\label{CO-emission-absorption}}
	\end{center}
\end{figure*}

\begin{figure*}[h]
\begin{center}
	\includegraphics[angle=0.,scale=0.50]{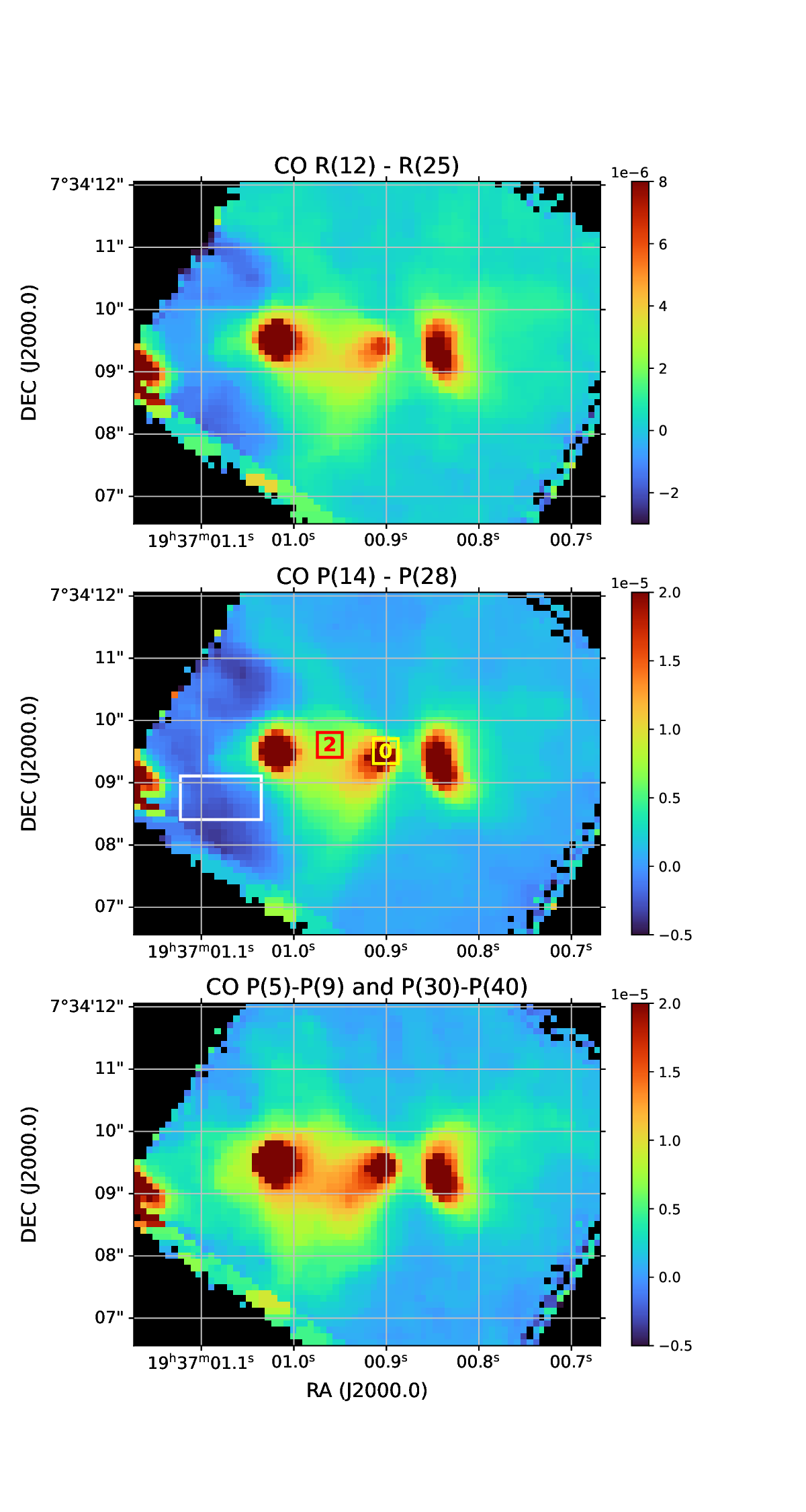}
	\caption{Maps of CO emission in the R and P branches. Red to brown indicates position difference
    signal, i.e., emission in CO lines. Violet to blue indicates negative difference signal, i.e., absorption.
    The white rectangle indicates the region where the spectra in Figure \ref{CO-emission-absorption} were
    extracted.
    \label{CO-emission-map}}
	\end{center}
\end{figure*}

All the positions in the B335 outflow cavity show CO gas lines, in general with diminishing
amplitude with increasing distance from the protostar. The spectra shown here were obtained
with the NIRSpec G395M medium resolution grating, resulting in a spectra resolution of just
above 1000 in the spectral region discussed here. 
At higher spectral resolutions, e.g., in the observations of GSS~30 by
\citet{Pontoppidan.2002.AA.393.585.GSS30}, CO emission lines appear as narrow lines above
a well-defined continuum.
In the NIRSpec B335 data, as a consequence of this relatively low
spectral resolution, the CO line spectrum appears as a near symmetrical oscillation of the
flux, with emission and absorption not being immediately distinguishable.
Wavelengths and identifications of the CO lines were taken from 
the HITRAN/HITEMP database, which uses the data from
\citet{Goorvitch.1994.ApJS.95.535.COlines}.
The radial velocity of the B335 molecular core relative to LSR is 8.34 km\,s$^{-1}$
\citep{Yen.2015.ApJ.799.193.B335systemic8.34}. With
the kinematic correction applicable to the NIRSPEC observations, lines appear blueshifted
by $\approx$20 km\,s$^{-1}$ relative to the laboratory wavelengths indicated in
Figure~\ref{CO-emission-absorption}. 
We compare the CO gas spectrum of two regions
close to the protostar (0E and 2E) with the average of a larger area away from the jet
axis, at a distance between that of shocks 3E and 4E, indicated by a white rectangle
in Figure~\ref{CO-emission-absorption}. In this area, we note a phase shift
in the R-branch (short wavelengths) lines, indicating absorption. In the P-branch, the 
situation is more complex, with lines of J-values between 13 and 28 appearing in absorption,
while the other J values are in emission, i.e., in phase with the emission from regions
0E and 2E (black and red in Figure~\ref{CO-emission-absorption}). We map the distribution
of CO line absorption in Figure~\ref{CO-emission-map}. Within the field
of view of the NIRSpec data, the absorption
is confined to the more distant regions of the blue-shifted lobe, but within the outflow
cavity defined by shock-excited H$_2$ emission, shown by 
\citet{Federman.2024.ApJ.966.41}.

\citet{Federman.2024.ApJ.966.41} report that CO emission in B335 is comparatively weak, but they do not explicitly describe the presence of spatially extended CO absorption. 
They do describe CO absorption in two other protostellar sources in their sample, HOPS 370 and,
in particular, IRAS 20126, where they clearly see CO absorption by a distinct spatial feature.
In our data, we find that the CO fundamental-band lines transition from net emission near the protostar to net absorption at projected distances of 2.0$\arcsec$ along the cavity walls, and $\approx$3$\arcsec$ on
the jet axis to the east. 
This is similar to what is seen
in the sample of
low-mass young stellar objects (YSOs) studied by \citet{Herczeg.2011.AA...533A.112.COinYSO}
with dense envelopes (SED Class I).
Ten object show CO emission and six objects (HH 100 IRS, IRS 63, WL 12,
Elias 29, Elias 32, and TMC 1A) show wind absorption in the
CO 1-0 lines. 
They find extended CO emission in a few objects, best seen in GSS 30, which is morphologically similar
to B335 
\citep{Chen.2007.AA...475..277.GSS30AO}, 
though less deeply embedded.

The CO absorption region seen in Figure \ref{CO-emission-map} is confined to the outflow cavity.
The strong shock fronts from 0E, 3E, and 4W show CO emission, but outside of these shocks, the
CO gas goes from emission to absorption at approximately the distance of shock 3E from the protostar. We cannot clearly distinguish whether the CO gas emission occurs in the
immediate vicinity of the protostar and is then scattered in the outflow cavity or its walls, or whether
the CO is emitted in the densest and warm regions of the outflow cavity.
The fact that CO lines do not extend as far into the outflow cavity as the continuum,
and that they indeed go into absorption indicates that CO lines are more strongly absorbed
in the outflow cavity than continuum radiation.
The CO absorption requires a region of the outflow wind where the temperature
is low enough that emission does not dominate, but absorption of the continuum and line flux
emitted further upwind does occur. This is different from the situation in IRAS 20126 discussed
by \citet{Federman.2024.ApJ.966.41} where the absorption apparently occurs in a separate object
in front of the outflow cavity. The spectral resolution of R$\approx$1000 of the spectra does not
allow a clean detection of the absorption lines relative to the continuum. Despite this limitation
the fact that we can detect CO R-branch absorption at J-values up to J=22 allows a rough estimate
that the gas temperature in the absorbing gas is about 1000 K. In their study of the excitation 
conditions of CO emission near the B335 protostar, at positions equivalent to our positions
0E, 0Wn, and 0Ws, \citet{Rubinstein.2024.ApJ.974.112.B335.IFU} find multiple temperature components,
the lowest on being close to 1000K. Their result is consistent with the warm absorbing CO gas
being present in the outflow cavity.

We have now obtained new NIRSpec IFU data with the G395H grating, that have about double the
spectral resolution of the data analyzed here, and we will, in the near future, revisit the
CO gas emission and absorption with the goal of studying the temporal evolution of the gas emission.
\clearpage

\section{Summary and Conclusions}

We have analyzed two epochs of JWST NIRCam images of the isolated globule B335 and the single
protostar it contains. We have also presented new analysis of public JWST NIRSpec IFU data of the region 
closest to the protostar.
Based on NEOWISE photometry of the integrated light in the reflection nebula, 
the B335 protostar underwent
a photometric outburst around 2015 and is at present in the declining brightness phase of
this outburst, but has not yet returned to its quiescent brightness. In addition to the overall 
decline of brightness of the blue-shifted, eastern lobe of the reflection nebula, shadow effects
from obscuring clumps in the path of the illuminating radiation from the protostar are projected
into the reflection nebula. From the fact that substantial variations in these shadow effects
are seen in the one year epoch difference between our two images, we conlude that those obscuring clumps must be of order one AU from the protostar. From the amplitude of the variations, we estimate that
those clumps have optical depths of order unity.

We have presented maps of the column densities of H$_2$O ice, continuum extinction, and broad-band extinction based on
NIRCam photometry of background stars and confirm the overall flattened structure of the B335 core, and the
carving out of an outflow cavity that contains less dust and ice than the undisturbed molecular core.
A study of the scattered interstellar radiation field, the coreshine, qualitatively confirms
these findings and extends them to regions of lower density.

A set of shock-excited emission knots
near the center of the outflow cavity outlines a fast jet emanating from the protostar.
Based on our astrometry of the individual shock fronts, their kinematic ages are short, of order of a few decades. 
The kinematic age of shock 3E, the brightest and most compact shock, is 
consistent with it having been launched in the early phases of the presently on-going photometric outburst. 
This shock front 3E is remarkable for its strong CO gas emission.
The excitation levels of the shock fronts along the fast jet indicate that the present outburst of
the protostar has produced a fast, hot, largely ionized jet. Shock 3E is the shock front
where this new jet is shocking against previously ejected slower molecular gas leading to highly excited
emission and a distinct blue-shifted radial velocity, 
probably due to lateral ejection of material from the
shock. Shocks further downwind
(4E and older) show rapidly decreasing levels of CO emission, and become dominated by H$_2$ emission.
The kinematic ejection date of the complex shock front 4E is consistent with
its ejection in a previous photometric outburst which is tentatively indicated by one Spitzer observation.
The emission knots closest to the ALMA position of the protostar, in particular 0E, are not part of the fast jet, but are stationary shocks or slower moving disk wind features.

We have studied the spectra at selected positions in the blue-shifted outflow cavity that are not
part of the jet. 
The continuum slope indicates rapid increase of the extinction toward the
protostar, with an increase of approximately A$_V$ = 200 mag in the inner $4\arcsec$  towards the protostar, which establishes a lower limit for the total extinction towards the protostar.
In the 4–5 $\mu$m region, these scattered light spectra are dominated by ice features:
$^{12}$CO$_2$ and $^{12}$CO are in deep saturation, and both show evidence of short and long-wavelength wings
in their profile, indicating a contribution from larger grains. The rarer isotopologues $^{13}$CO$_2$ and $^{13}$CO are detected without saturation, as are OCN$^-$ and OCS. The features of $^{13}$CO$_2$, OCN$^-$, and OCS show strongly increasing optical depth towards the protostar, but no evidence of further chemical evolution closer to the protostar.

We have mapped the emission of gaseous CO in the bipolar nebula associated with B335 and find
that close to the protostar, CO is in emission, with low radial velocities, consistent with the
systemic velocity of the B335 molecular core. We have found CO absorption at distances
from the protostar beyond the position of Shock 3E, indicating absorption by warm gas
within the outflow cavity.

\begin{acknowledgments}

We thank the referee for thoughtful comments that helped to improve this paper.
We thank Fengwu Sun for the use of his astrometrically re-calibrated NIRCam images.
We thank Neal Evans and Steven Federman for communicating their new
results on the distance of B335 prior to publication.
D.J.\ is supported by NRC Canada and by an NSERC Discovery Grant.
This project was supported by NASA through the JWST/NIRCam project, contract No. NAS5-02105 (M. Rieke, University of Arizona, PI).
This work is largely based on observations made with the NASA/ESA/CSA James Webb Space Telescope. 
V.J.M.L.G. acknowledges support by the Spanish program Unidad de Excelencia María de Maeztu CEX2020-001058-M, financed by MCIN/AEI/10.13039/501100011033, and by the MaX-CSIC Excellence Award MaX4-SOMMA-ICE. 
V.J.M.L.G. acknowledges support by the European Research Council (ERC) under the European Union’s Horizon 2020 research and innovation program (grant agreement No. 101098309 - PEBBLES).

The data were obtained under GTO program 1187 and GO program 1802 and downloaded from the Mikulski Archive for Space Telescopes (MAST) at the Space Telescope Science Institute, which is operated by the Association of Universities for Research in Astronomy, Inc., under NASA contract NAS 5-03127 for JWST.
This publication makes use of data products from the Near-Earth Object Wide-field Infrared Survey Explorer (NEOWISE), which is a joint project of the Jet Propulsion Laboratory/California Institute of Technology and the University of California, Los Angeles. NEOWISE is funded by the National Aeronautics and Space Administration.
This research has made use of the NASA/IPAC Infrared Science Archive, which is funded by the National Aeronautics and Space Administration and operated by the California Institute of Technology.
This work is based in part on observations made with the Spitzer Space Telescope, which was operated by the Jet Propulsion Laboratory, California Institute of Technology under a contract with NASA.
ALMA is a partnership of ESO (representing its member states), NSF (USA) and NINS (Japan), 
together with NRC (Canada), MOST and ASIAA (Taiwan), and KASI (Republic of Korea), 
in cooperation with the Republic of Chile. 
The Joint ALMA Observatory is operated by ESO, AUI/NRAO and NAOJ.
The National Radio Astronomy Observatory is a facility of the National Science Foundation 
operated under cooperative agreement by Associated Universities, Inc.
This work is based in part on observations made with the Spitzer Space Telescope, which was operated by the Jet Propulsion Laboratory, California Institute of Technology under a contract with NASA.
\end{acknowledgments}

\vspace{5mm}
\facilities{JWST (NIRCam, NIRSpec), MAST, IRSA, WISE, NEOWISE, Spitzer, ALMA}

\appendix
\section{Position and Proper Motion of the Protostar from ALMA Data}

We took the proper motion of the B335 protostar into account, since any
shock features of the outflow should be measured relative to the protostar, that is
not directly detected at NIRCam wavelengths.
Since B335 is a well-studied object, numerous observations are available publicly.
We have selected the highest spatial resolution continuum data publicly available in mid 2024 from
the ALMA Science Archive. 
The data used here are from
\citet{Maury.2018.MNRAS.477.2760.B335.ALMA}
\citet{Bjerkeli.2019.AA.631.64.B335.ALMA.kinematics},
\citet{Imai.2019.ApJ.873L.21.B335.ALMA},
\citet{Okoda.2022.ApJ.935.136.ALMA.position}
and other unpublished ALMA data included in the following list:\\

\noindent
uid://A001/X11f/X7b PI name: Yen, Hsi-Wei 2014-09-02\\
uid://A001/X145/X2b3 PI name: Ruiz, Maria Teresa 2015-08-27\\
uid://A001/X2fe/X21 PI name: Evans, Neal 2016-07-17\\
uid://A001/X2c9/X17 PI name: Yen, Hsi-Wei 2016-07-24\\
uid://A001/X1284/X2059 PI name: Bjerkeli, Per 2017-10-08\\
uid://A001/X133d/X2eed PI name: Imai, Muneaki 2019-06-23\\
uid://A001/X1590/X1e95 PI name: Jorgensen, Jes 2022-08-20\\
uid://A001/X1590/X1e91 PI name: Jorgensen, Jes 2022-08-20\\
uid://A001/X2d20/X2f92 PI name: Jorgensen, Jes 2023-04-12\\
uid://A001/X2df7/X600 PI name: Plunkett, Adele 2023-04-20\\
uid://A001/X2df7/X736 PI name: Yang, Yao-Lun 2023-04-21\\
uid://A001/X2d20/X1c3b PI name: Baek, Giseon 2023-05-25\\

These data were taken at a variety of frequencies, and with different
configurations of ALMA, and therefore have a variety of angular resolutions. Since, to the
best of current knowledge, the B335 protostar is an isolated single object at sub-mm and mm
wavelengths, we use all these measurements to determine the proper motion of the protostar.
Different centroiding methods implemented in various python routines tend to give slightly
different results. This has been studied and documented for the case of NIRISS data by
\citet{Goudfrooij.2022.STScI.doc.centroid}.
We use both the momentum based classical centroid algorithm as implemented in \texttt{photutils.aperture.ApertureStats}
and the 2D-Gaussian fitting as implemented in centroid\_sources with $centroid\_func=centroid\_2dg$.

We have included one older position measurement of the B335 protostar from
\citet{Reipurth.2002.AJ.124.1045.B335.VLA}
from VLA configuration A observations at 3.6 cm wavelength, 8h of exposure time
estimated position error 0\farcs05, beam for B335 was 0\farcs32 x 0\farcs28. 
The source in B335 was identified as extended, interpreted as evidence for a compact thermal radio jet.
The 2001 VLA position fits very well with the later, shorter wavelength ALMA observations, despite
the wavelength difference and probably the difference in the excitation mechanism of the emission.

\begin{figure*}[h]
\begin{center}
	\includegraphics[angle=0.,scale=0.70]{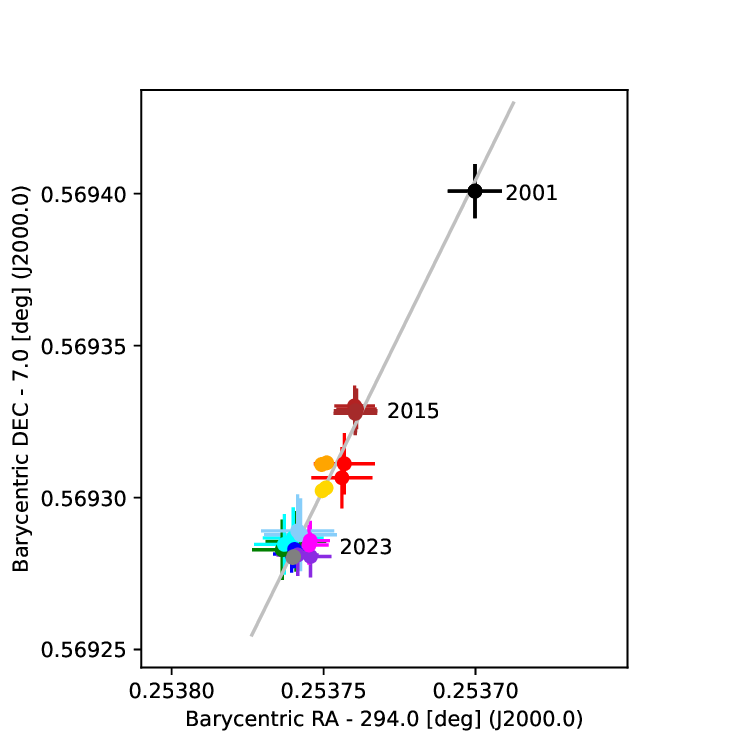}
	\caption{Proper motion of the B335 protostar from public ALMA data and
one published VLA data point.\label{ALMA-PM}}
	\end{center}
\end{figure*}

All ALMA positions and the VLA positions were reduced to barycentric coordinates using the
distance of 165 pc. The parallax is small at $\approx$ 7 mas, but is of similar magnitude as
the annual proper motion.
From the fits of the VLA and ALMA positions of the B335 protostar shown in Figure \ref{ALMA-PM}, we get
the ICRS position in degrees (J2000.0):

\noindent

\begin{flushleft}
$
\mathrm{RA}(\mathrm{MJD})\mathrm{[deg]}  =
\left( 294.2537586 \pm 1.1\times10^{-6} \right)\,\mathrm{[deg]}
+
\left( 7.1 \pm 0.5 \right)\times10^{-9}\,\mathrm{[deg\,day^{-1}]}\,
(\mathrm{MJD}-60000)\mathrm{[day]}
$
\end{flushleft}

\begin{flushleft}
$
\mathrm{DEC}(\mathrm{MJD})\mathrm{[deg]} =
\left( 7.5692829 \pm 1.7\times10^{-6} \right)\,\mathrm{[deg]}
-
\left( 15.0 \pm 0.7 \right)\times10^{-9}\,\mathrm{[deg\,day^{-1}]}\,
(\mathrm{MJD}-60000)\mathrm{[day]}
$
\end{flushleft}

\noindent
RA PM on sky [mas/yr] =  9.3 $\pm$ 0.6, 
DEC PM in [mas/yr] =  -19.7 $\pm$ 0.9\\

\noindent
The measured projected velocity of the B335 protostar is 17$\pm$1 kms$^{-1}$ at a 
position angle of 155 $\pm$ 2\arcdeg.

\section{H$_2$O Ice Column Density from Background Star Photometry}
We obtained aperture photometry of most stars in our imaging field around
B335. We compiled a list of star coordinates initially created by daofind.
The list was manually curated to eliminate spurious detections in shock fronts
or diffraction spikes, to eliminate close pairs of stars, and most extragalactic
objects that could be visually identified as such. 

In the shortest wavelength filters (F070W and F090W), our field in B335
appears blank. Therefore, every star in the field has at least one filter where
it is not saturated and a precise position could be measured.
The optical depth of the broad absorption feature of H$_2$O ice can be approximately
measured using filter photometry. The ideal filters for this purpose would have been
the NIRCam filters F250M, F300M, and F360M. However, since our program was primarily focused on
obtaining slitless grism spectroscopy, we only obtained direct images in the F300M filter
for H$_2$O ice photometry, and in the F277W and F356W filters that were also used for
the WFSS. 

\begin{figure*}[h]
\begin{center}
	\includegraphics[angle=0.,scale=0.50]{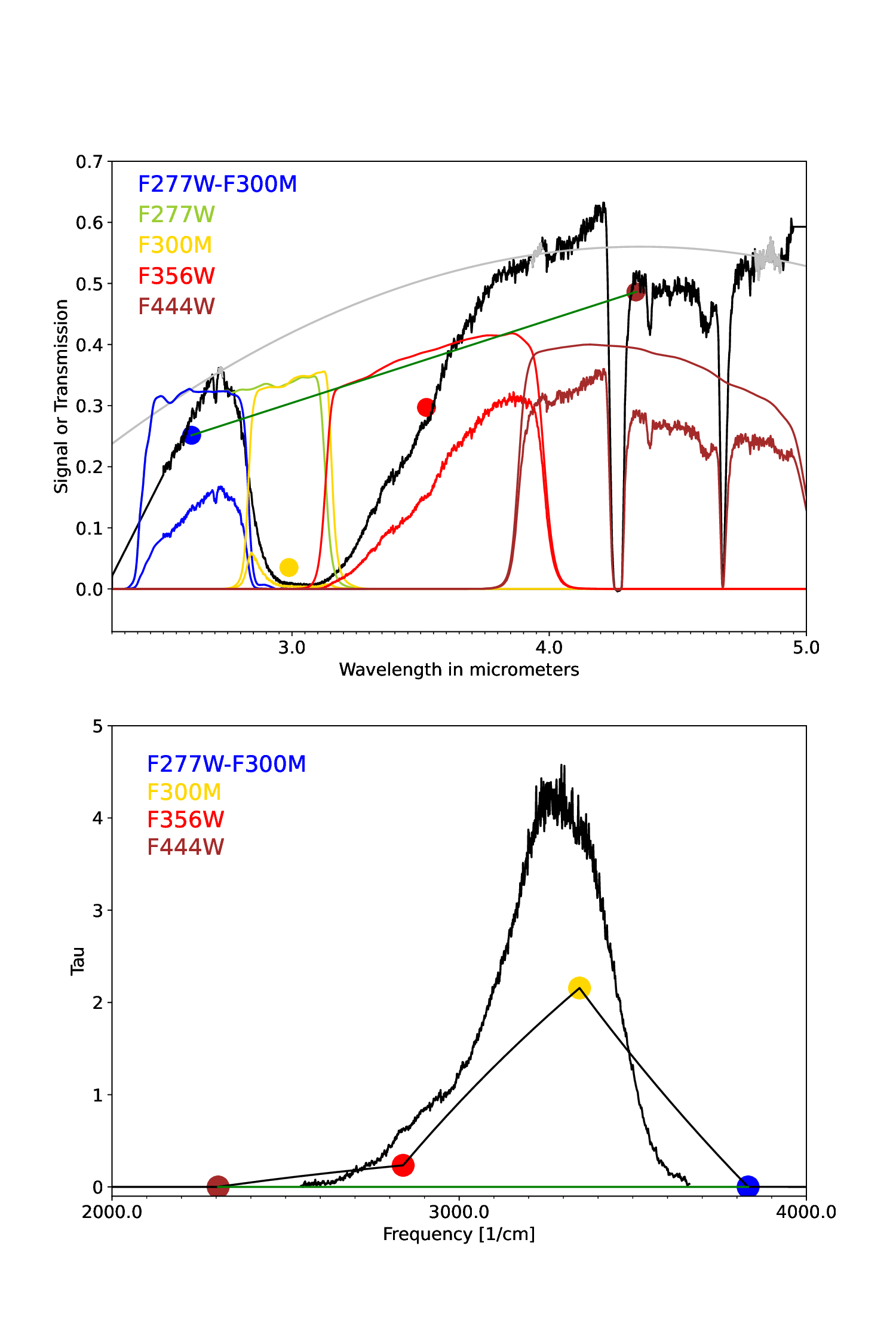}
	\caption{Top: spectrum of NIRS38 from 
\citet{McClure.2023.NatAs.7.431.NIR38} shown as a black line. The continuum regions for
the polynomial continuum fit are shown in grey. The transmission curves of the NIRCam filters are the smooth thin colored lines. The product of the spectrum and the
transmission are shown in the same color as the filter transmission curves. Large filled circles
indicate the effective wavelengths of the filters, and the spectrum signal integrated over the filter
transmission. The synthetic filter (F277W-F300M) has an effective wavelength of 2.61$\mu$m.
The green line is the linear interpolation of the continuum between the F277-F300M and F444W synthetic photometry points. Bottom: 
Optical depth $\tau$ vs. spatical frequency from the spectrum and from the linear interpolation of the synthetic photometry. Filled circles denote the individual filters, as in the top panel.\label{synthetic-photometry}}
	\end{center}
\end{figure*}

\begin{figure*}[h]
\begin{center}
	\includegraphics[angle=0.,scale=0.50]{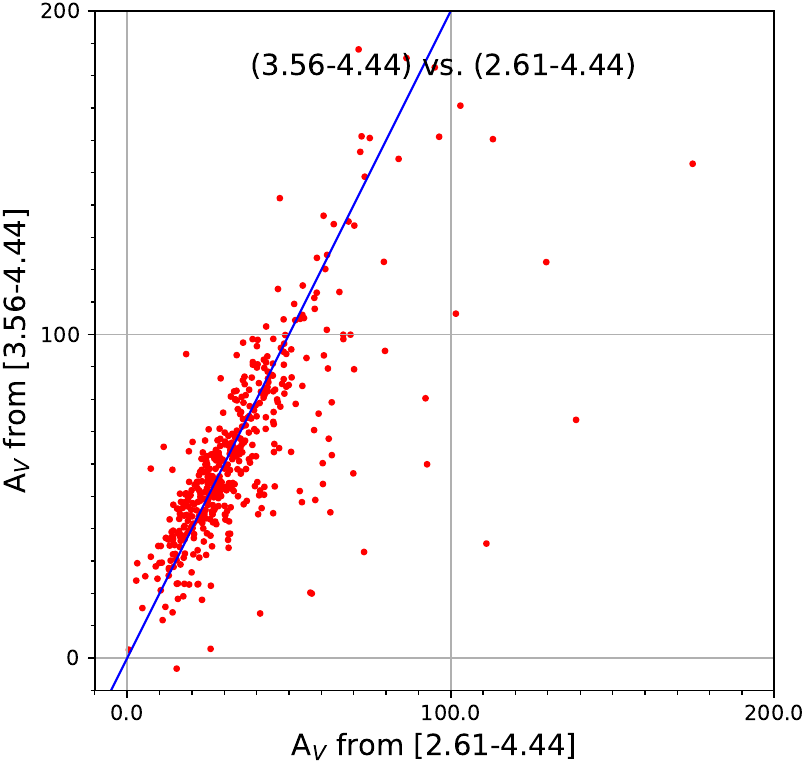}
	\caption{
    A$_V$ extinction measured from the F356W-F444W color versus extinction measured from the synthetic 2.61$\mu$m - F444W color.
    F356W-F444W, being affected by the long-wavelength wing of the H$_2$O ice feature, overestimates the extinction by a factor of 2.0, shown as the blue line. \label{AV-comparison}}
	\end{center}
\end{figure*}

We performed aperture photometry with an aperture of 3 pixels radius in those
filters on combined (...i2d) images to approximately measure the column density of H$_2$O ice in B335.
The images were destriped by subtracting the median of the bottom 1/4 of data values, effectively a robust minimum, along each image row.
Aperture photometry was done with an aperture of 0\farcsec189, using photutils.
The sky brightness and some extended emission was locally subtracted using an annulus in the range 
of 0\farcsec378 $<$ r $<$ 0\farcsec629.
The aperture correction was determined for each filter by doing
the same aperture photometry, including sky annulus subtraction, on
synthetic images of the psf, created by STPSF (formerly WebbPSF).
While our program did observe B335 in the F200W filter, we found that bandpass
of little use for mapping the extinction close to the protostar since the central
region of B335 appeared opaque at that relatively short wavelength.

We used the following method to approximately measure the H$_2$O ice column density
from the available photometry, illustrated in Figure\ \ref{synthetic-photometry}.
The top panel shows a high signal-to-noise spectrum of a heavily obscured background star behind
the Chameleon I molecular cloud 
from \citet{McClure.2023.NatAs.7.431.NIR38}. 
We also show the transmission curves of all the NIRCam filters used for 
that photometric column density measurement: F227W, F300M, F356W, and F444W.
The F277W filter covers continuum in the lower part of its transmission range, but
its upper half includes the H$_2$O ice absorption band.
The F300M filter is narrower than the F277W filter, and covers the center of the H$_2$O ice band.  
As Figure \ref{synthetic-photometry} (top panel) shows, 
a synthetic photometric value using the difference of the transmission curves
of F277W (green transmision curve) minus F300M (yellow transmission curve) gives a better 
approximation of the continuum shortward of the H$_2$O
ice feature, and is shown as the blue transmission curve with an effective wavelength of
2.61 $\mu$m. 
The F300M filter (yellow) covers the center of the H$_2$O ice feature, and the F356W filter 
(red transmision curve) covers
its long-wavelength wing. 

\begin{equation}
f_{\mathrm{syn},261}
=
f_{277W}\,
\frac{\int T_{277W}(\lambda)\, d\lambda}
     {\int T_{\mathrm{syn},261}(\lambda)\, d\lambda}
-
f_{300M}\,
\frac{\int T_{300M}(\lambda)\, d\lambda}
     {\int T_{\mathrm{syn},261}(\lambda)\, d\lambda}
\end{equation}
where 
$f_{\mathrm{filter}}$ is the flux measured in a filter and
$T_{\mathrm{filter}}$ is the system transmission for the filter and
\begin{equation}
T_{\mathrm{syn},261}(\lambda)
=
\max\!\left[\,T_{277W}(\lambda)-T_{300M}(\lambda),\,0\,\right]
\end{equation}

The F444W filter (brown transmission curve) covers spectral regions outside of the H$_2$O ice
feature, but is affected by the narrow, though deep, absorption bands of CO$_2$ and CO.
This filter therefore underestimates the flux in the continuum.

As an approximation, we use the synthetic ``F277-F300M'' flux and the F444W flux as
continuum points, we linearly interpolate between those two points as an estimate of the
continuum, and compute the optical depth $\tau$ as the log$_{10}$ of the flux ratio of the
interpolated continuum and the measured flux at the effective wavelengths of the F300M
and F356W filters. 
The resulting values of $\tau$ are shown in the bottom part of Figure\ \ref{synthetic-photometry}.
We then integrate this linearly interpolated $\tau$ spectrum over frequency in units of
cm$^{-1}$. 

For comparison and calibration, we also determine $\tau$ from the example spectrum,
using a second order polynomial fit to the three continuum regions free of ice features
and indicated in grey color in Figure\ \ref{synthetic-photometry} (top). The $\tau$ measured from the spectrum is
the blue spectrum in Figure\ \ref{synthetic-photometry} (bottom).
The ratio of the spectral to the photometric $\tau$ measurement in this one example
is 1.44, and we use this as our best estimate for the calibration factor to go from the
photometric $\tau$ to the true spectroscopic $\tau$.

\section{Measuring Extinction from the Photometry}
Our goal was to map the extinction in the B335 cloud as completely as possible,
without suffering ''blind spots'' simply from the lack of detectable background stars
in the area of very high extinction close to the protostar.
The best measure of the continuum extinction, i.e., the extinction outside of the
H$_2$O ice band that dominates this region of the spectrum, is to measure the color
excess between the synthetic F277W - F300M flux, and the F444W filter. The latter is
not affected by H$_2$O ice absorption, but is affected by the narrower CO$_2$ and CO ice
absorption bands. As our Figure \ref{H2O-AV-map} shows, the extinction  map so derived
is still blind to the high extinction levels immediately near the protostar.
We have therefore decided to also use the color excess F356W-F444W as a measure of
extinction, even those the F356W filter covers the long-wavelength wing of the
H$_2$O ice band. Figure \ref{AV-comparison} plots the extinction of individual stars derived from this
F356W-F444W color excess against the H$_2$O-free extinction derived using the synthetic 
value at 2.61$\mu$m. We find that the F356W-F444W color overestimates the extinction
by a factor of 2.0 compared to the H$_2$O-ice-free color because H$_2$O ice absorption
correlates with continuum extinction, as shown in Figure \ref{AV-comparison}.
Since the F356W filter data
reach much deeper than the F277W and F300M data, the F356W-F444W extinction map avoids
the blind spot near the protostar and properly shows the strong concentration of
extinction near it.
\section{Component Analysis of Ice Absorption Features}

Ice absorption and scattering bands are detected throughout the outflow cones of B335. We analyze the set of 11 sight-lines extracted within the rectangular apertures identified in Figure
\ref{Cavity-extraction-regions-and-spectra}.  The 3.0\,$\mu$m band of H$_2$O ices is severely saturated in all sight-lines and therefore
not shown in Figure~\ref{Norm-ice-spectra}. The strong C--O stretching modes of CO$_2$ and CO at 4.25 and 4.67\,$\mu$m, respectively, are detected everywhere. The CO$_2$ band is always saturated, and the CO ice band is saturated or nearly saturated in its core. The weaker 4.38, 4.60, and 4.90\,$\mu$m features, attributed to $^{13}$CO$_2$, OCN$^-$, and OCS ices, are detected in all sight-lines, while the $^{13}$CO ice band at 4.78\,$\mu$m is seen in a few. Finally, the C--H stretch mode at 3.53\,$\mu$m is visible in a number of sight-lines, while in others it is limited by the low S/N of the continuum at those wavelengths.

Forests of emission lines of the ro-vibrational P- and R-branch transitions of gas-phase CO are visible (Figure~\ref{Cavity-extraction-regions-and-spectra}) in the sight-lines close to the protostar. For the purpose of the ice component analysis, these were removed from the flux spectra following the procedure described in \citet{Boogert.2022.ApJ.941.32B.OCN}. Line fluxes were measured with Gaussian fits to the most isolated, highest quality emission lines, from which rotation diagrams were constructed. Linear fits to the observed data points in the rotation diagrams were then used to determine the strengths of all transitions, which were subsequently subtracted from the observed flux spectra. This yields spectra in which the ice absorption features are much less contaminated by CO lines, but preserve the full spectral sampling of the data, allowing for more accurate band profile and column density measurements for the spectra at the cavity apertures 0, 1, 2, and 3. The remaining apertures perhaps show weak CO emission lines as well, but their effect on the ice bands is small. Finally, any hydrogen emission lines were masked if they overlap with the ice features.

\begin{figure*}[h]
\begin{center}
    \includegraphics[angle=180.,scale=0.5]{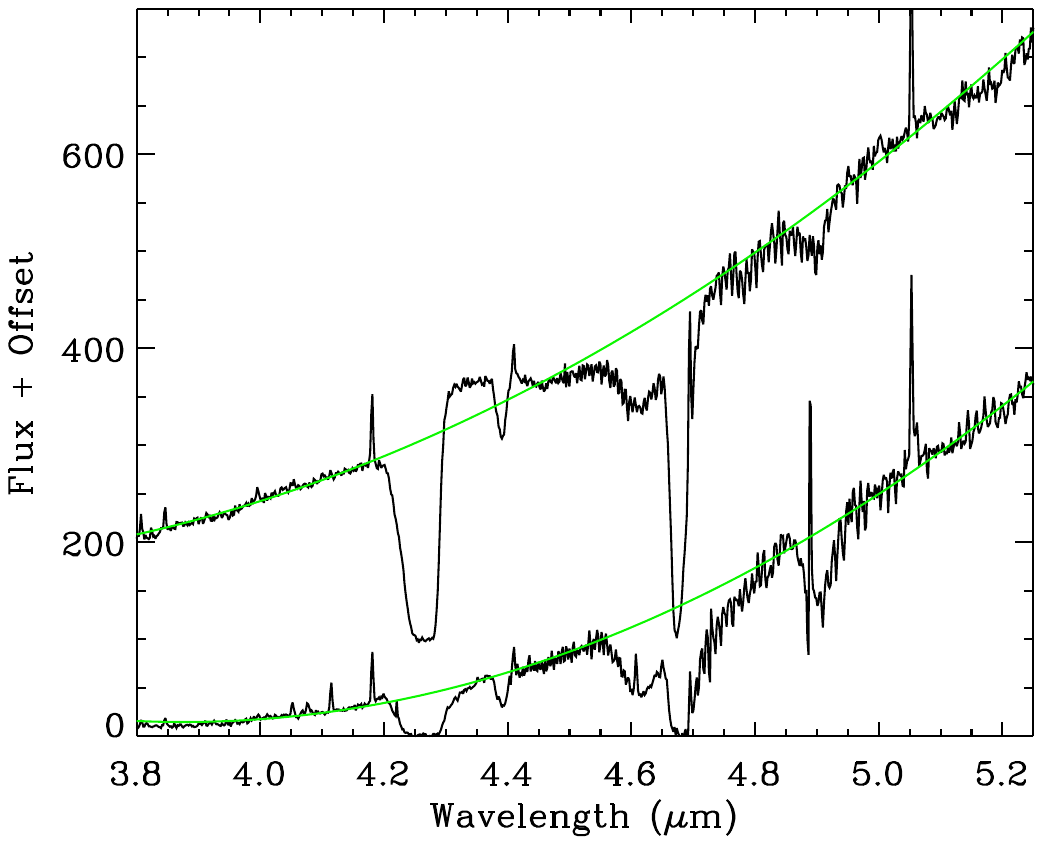}
	\caption{Polynomial baseline (green) used to derive the optical depth spectra for the ice features. The spectrum for position 8 in the B335 cavity is shown in the top figure, offset for clarity, and position 0 is shown at the bottom. Position 8, like most positions at larger distances from the central star, shows a distinct jump across the 4.25 $\mu$m CO$_2$ ice feature. We have assumed that this is caused by scattering on the ice feature and have excluded it from the baseline fits. Note that position 0 does not show this jump, while it increases for sight-lines further away from the protostar.
\label{Adwin-3}}
	\end{center}
\end{figure*}

In order to analyze the profiles of the ice features and derive column densities, we derive local baselines with low-order polynomials. As is clear in Figures~\ref{Norm-ice-spectra} and \ref{Adwin-3}, cavity 
position~0
shows a smoothly rising continuum level, that is well fitted with a second-order polynomial in the 3.8--5.3\,$\mu$m wavelength range
(green line). 
Most other sight-lines show a jump in the continuum level across the 4.27\,$\mu$m CO$_2$ ice band, 
shown in Figure 
\ref{Adwin-3} 
for position 8. 
This jump is up at longer wavelengths, which is reversed from that seen toward JWST/NIRCam and NIRSpec observations of the Chamaeleon background stars published in 
\citet{McClure.2023.NatAs.7.431.NIR38}
and \citet{Dartois.2024.NatAs.8.359.IceAge.models}. For these background stars the jump is attributed to scattering by large ($\sim$0.9\,$\mu$m) sized grains. The reversed jump seen in the B335 cavity is reproduced in some of the higher inclination circumstellar disk models presented in 
\citet{Dartois.2022.AA.666A.153.CO2ice.profiles} (their Figures~10 and 11).
Such a jump is also distinctly present in the \textit{AKARI} spectrum of the edge-on disk surrounding the low-mass YSO IRAS~04301+2247 
\citep[their Figure~3]{Aikawa.2012.AA.538.57.AKARI}. 
This disk, like that of B335, is nearly completely edge-on. It is remarkable that the strength of the scattering jump relative to the observed continuum increases for sight-lines further away from the protostar. We speculate that this is due to a larger column of grains in the foreground for sight-lines closer to the star, less affected by scattering in the disk geometry. For the purpose of the ice analysis, we chose to fit a second-order polynomial to all sight-lines, excluding the long-wavelength side of the CO$_2$ band from the fits (Figure \ref{Adwin-3}). This preserves scattering effects on the ice bands in the optical depth spectra.

\begin{figure*}[h]
\begin{center}
	\includegraphics[angle=180.,scale=0.50]{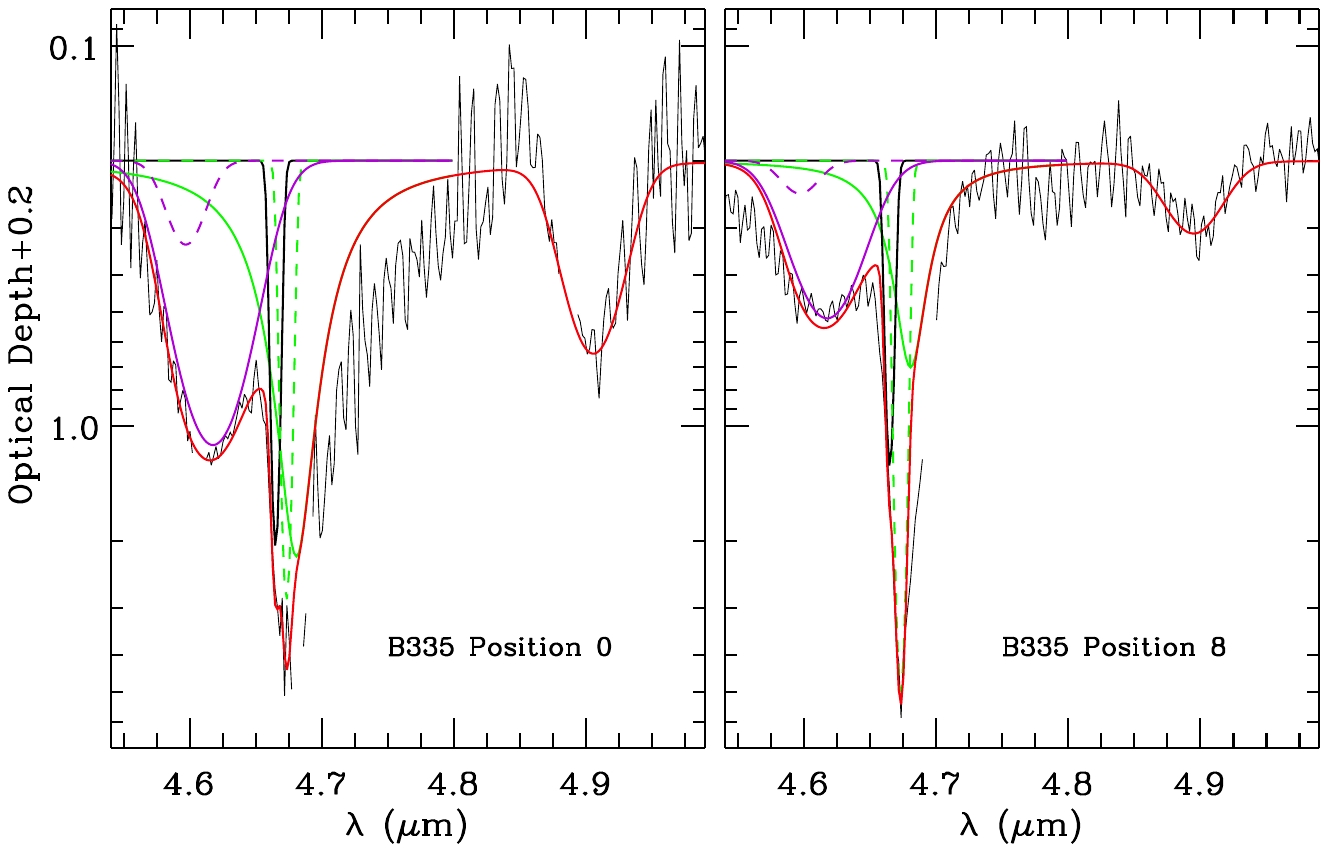}
	\caption{Optical depth spectra of two sight-lines 
    (position 0 on the left and position 8 on the right)
    within the B335 outflow cavity. Gaussian and Lorentzian fits to the CO ice bands (green: CO$_{\rm blue}$, green dashed: CO$_{\rm apolar}$, black: CO$_{\rm polar}$) are shown, and Gaussian fits to the 2165.7 and 2175.4\,cm$^{-1}$ components of the OCN$^-$ band (solid purple and dashed purple, respectively). The OCS band at 4.90\,$\mu$m is fitted with a single Gaussian. The red line is the sum of all components.
\label{Adwin-4}}
	\end{center}
\end{figure*}

In order to put the B335 ice bands in context, we fit the OCN$^-$, CO, and OCS ice bands with functions that were used in previous work, following the approach of \citet{Boogert.2022.ApJ.941.32B.OCN}. The OCN$^-$ band was found to consist of two components, centered on 2165.7 and 2175.4\,cm$^{-1}$ (4.6174 and 4.5969\,$\mu$m; \citealt{Broekhuizen.2005.AA.441.249.OCN}). The 2165.7\,cm$^{-1}$ component is consistent with solid OCN$^-$ in polar environments, while the other component relates perhaps to CO bonding to the grain surfaces or to OCN$^-$ in apolar environments \citep{Oberg.2011.ApJ.740.109O.Spitzer}. We adopt the same decomposition procedure, by fitting two Gaussians, one at 2165.7\,cm$^{-1}$ (FWHM = 26\,cm$^{-1}$) and one at 2175.4\,cm$^{-1}$ (FWHM = 15\,cm$^{-1}$). The fitting was done simultaneously with those of the CO ice feature, since they overlap. The CO ice feature was found to consist of three components \citep{Pontoppidan.2003.AA.408.981P.CO}. The narrow, central component of CO in an apolar environment (CO$_{\rm apolar}$) was fitted using a Gaussian function (peak wavelength 4.6731\,$\mu$m, FWHM = 0.0076\,$\mu$m). The broader component of CO in polar environments (CO$_{\rm polar}$) was fitted with a Lorentzian (4.6806\,$\mu$m, FWHM = 0.0232\,$\mu$m), and a narrow short-wavelength component, perhaps due to CO in a CO$_2$ environment (CO$_{\rm blue}$), was fitted with a Gaussian (4.6648\,$\mu$m, FWHM = 0.0065\,$\mu$m). Thus, the only variables in these fits are the peak optical depths. The OCS feature at $\sim$4.90\,$\mu$m, however, was fitted with an unconstrained Gaussian function. Examples of these fits are shown in Figure \ref{Adwin-4}.

Due to the peak optical depths of $>4$, fits to this CO ice band---in particular that of the CO$_{\rm apolar}$ component---are compromised by saturation effects. In addition, the presence of a hydrogen emission line near 4.7\,$\mu$m limits the use of the long-wavelength wing to determine the strength of the CO$_{\rm polar}$ component. We find that, when fitting the extended wing, visible up to $\sim$4.75\,$\mu$m, with the CO$_{\rm polar}$ Lorentzian component, a significant fraction of the OCN$^-$ band is consumed. Considering the overlap with gas-phase lines and possible scattering effects, we have limited the depth of the fitted CO$_{\rm polar}$ component to minimize the effects on the OCN$^-$ band. Other than this, reasonable fits are obtained for the OCN$^-$ band, showing a dominance of the long-wavelength component attributed to OCN$^-$ in polar ice environments, in agreement with previous work on low- and high-mass YSOs (e.g., \citealt{Broekhuizen.2005.AA.441.249.OCN,Oberg.2011.ApJ.740.109O.Spitzer,Boogert.2022.ApJ.941.32B.OCN}).

The OCS band has a Gaussian peak position and FWHM of $4.900\pm0.008$\,$\mu$m and $0.04\pm0.01$\,$\mu$m, respectively, for all extracted positions across the B335 outflow cavities. This is comparable to the bulk of the massive YSO targets studied in \citet[their Figure~13]{Boogert.2022.ApJ.941.32B.OCN}, and consistent with OCS embedded in CH$_3$OH-rich ices, and proton-irradiated H$_2$S- or SO$_2$-containing ices. 
On the other hand, while the OCN$^-$ /OCS column density ratio is constant across the outflow cavities, with a mean of 4.0 and a standard deviation of 0.4 (see also Figure 21)., this is $\approx$50\% lower compared to massive YSOs \citet[their Figure~13]{Boogert.2022.ApJ.941.32B.OCN}.   An astrochemical  discussion on the origin of this difference is beyond the scope of this paper.

\end{document}